\documentclass{aa}  
\usepackage[english]{babel}
\usepackage{graphicx}
\usepackage[authoryear]{natbib}
\usepackage[maxfloats=40]{morefloats}
\usepackage{txfonts}
\usepackage{color}
\newcommand{\Hi}{\ion{H}{I}}
\newcommand{\Cn}{\ion{C}{}}
\newcommand{\Ci}{\ion{C}{I}}
\newcommand{\Nn}{\ion{N}{}}
\newcommand{\Ni}{\ion{N}{I}}
\newcommand{\On}{\ion{O}{}}
\newcommand{\Oi}{\ion{O}{I}}
\newcommand{\Nai}{\ion{Na}{I}}
\newcommand{\Mgn}{\ion{Mg}{}}
\newcommand{\Mgi}{\ion{Mg}{I}}
\newcommand{\Aln}{\ion{Al}{}}

\newcommand{\Sin}{\ion{Si}{}}
\newcommand{\Sii}{\ion{Si}{I}}
\newcommand{\Siii}{\ion{Si}{II}}
\newcommand{\Ki}{\ion{K}{I}}
\newcommand{\Can}{\ion{Ca}{}}
\newcommand{\Cai}{\ion{Ca}{I}}
\newcommand{\Caii}{\ion{Ca}{II}}
\newcommand{\Scn}{\ion{Sc}{}}
\newcommand{\Sci}{\ion{Sc}{I}}
\newcommand{\Scii}{\ion{Sc}{II}}
\newcommand{\Tin}{\ion{Ti}{}}
\newcommand{\Tii}{\ion{Ti}{I}}
\newcommand{\Tiii}{\ion{Ti}{II}}
\newcommand{\Vn}{\ion{V}{}}
\newcommand{\Vi}{\ion{V}{I}}
\newcommand{\Crn}{\ion{Cr}{}}
\newcommand{\Cri}{\ion{Cr}{I}}
\newcommand{\Mni}{\ion{Mn}{I}}
\newcommand{\Fen}{\ion{Fe}{}}
\newcommand{\Fei}{\ion{Fe}{I}}
\newcommand{\Feii}{\ion{Fe}{II}}
\newcommand{\Coi}{\ion{Co}{I}}
\newcommand{\Nin}{\ion{Ni}{}}

\newcommand{\Baii}{\ion{Ba}{II}}
\newcommand\temp{$T_{\fontsize{6}{6}\selectfont \mbox{eff}}$}\normalfont
\newcommand\logg{$\log g$}
\newcommand\loggf{$\log gf$}
\newcommand\met{[M/H]}
\newcommand\Space{SP\_Ace}
\newcommand\kmsec{km s$^{-1}$}

\newcommand\Vrot{$V_{rot}$}
\begin{document}
   \title{\Space: a new code to derive stellar parameters and elemental abundances}

   \author{
	C. Boeche and E.K. Grebel
	}          

   \offprints{corrado@ari.uni-heidelberg.de}

   \institute{
Astronomisches Rechen-Institut, Zentrum f\"ur Astronomie der Universit\"at Heidelberg, M\"onchhofstr. 12-14, D-69120
Heidelberg, Germany
}


 
  \abstract
  {Ongoing and future massive spectroscopic surveys will collect large
numbers ($10^6$-$10^7$) of stellar spectra that need to be analyzed. Highly automated 
software is needed to derive stellar parameters and chemical abundances from these
spectra.}
  {We developed a new method of estimating the stellar parameters \temp, \logg,
\met, and elemental abundances. This method was implemented in a new code,
\Space\ (Stellar Parameters And Chemical abundances Estimator).
This is a highly automated code suitable for analyzing the spectra of large spectroscopic
surveys with low or medium spectral resolution (R=2\,000-20\,000). 
}
  {After the astrophysical calibration of the oscillator strengths of 4643 absorption
lines covering the wavelength ranges 5212-6860\AA\ and 8400-8924\AA, 
we constructed a library that contains the equivalent widths (EW) of these
lines for a grid of stellar parameters. The EWs of each line are fit
by a polynomial function that describes the EW of the line as a function
of the stellar parameters. The coefficients of these polynomial functions are
stored in a library called the ``$GCOG$ library". \Space, a code written in FORTRAN95,
uses the $GCOG$ library to compute the EWs of the
lines, constructs models of spectra as a function of the stellar parameters
and abundances, and searches for the
model that minimizes the $\chi^2$ deviation when compared to the observed spectrum. The code has
been tested on synthetic and real spectra for a wide range of signal-to-noise and
spectral resolutions.}
  {\Space\ derives stellar parameters such as \temp, \logg, \met, and chemical 
abundances of up to ten elements for low to medium resolution spectra of
FGK-type  stars with precision comparable to
the one usually obtained with spectra of higher resolution. Systematic errors in
stellar parameters and chemical abundances are presented and identified with tests
on synthetic and real spectra.  Stochastic errors are automatically
estimated by the code for all the parameters. A simple Web front end of \Space\ can be found at
http://dc.g-vo.org/SP\_ACE, while the source code will be published soon.}
   {}

   \keywords{Methods: data analysis -- Atomic data -- Stars: fundamental parameters -- Stars: abundances
-- Techniques: spectroscopic -- surveys
               }
   \titlerunning{\Space: a code to estimate \temp, \logg, \met, and [El/H]}
    \authorrunning{Boeche \& Grebel}

   \maketitle
%

\section{Introduction}\label{intro_grad}
The physical, chemical, and kinematic information carried by the stellar
spectra are fundamental for understanding how the Milky Way formed and evolved.
The increasing demand for stellar spectra by astronomers engaged in Galactic
archaeology led to large spectroscopic surveys that could be carried out thanks
to the availability of efficient multi-object spectrographs,
to the fast growth of the data storage capability, and the 
computational power of modern computers.
Past, present, and future surveys (such as the RAdial Velocity Experiment, RAVE, \citealp[Steinmetz
at al.][]{rave}; the Sloan Extension for Galactic
Understanding and Exploration, SEGUE, \citealp[Yanny et
al.,][]{yanny}; the Large Sky Area Multi-Object Fiber Spectroscopic 
Telescope, LAMOST, \citealp[Zhao et al.][]{zhao}; The Apache Point
Observatory Galactic Evolution Experiment, APOGEE, \citealp[Allende Prieto
et al.][]{apogee}; the Galactic Archaeology with
HERMES-GALAH Survey, \citealp[Zucker et al.][]{zucker}; the Gaia-ESO Public
Spectroscopic Survey, \citealp[Gilmore et al.][]{gilmore}; the 4-Metre
multi-Object Spectroscopic Telescope, 4MOST, \citealp[de Jong et
al.][]{dejong}; Gaia, \citealp[Perryman et al.][]{perryman},
\citealp[Lindegren et al.][]{lindegren}) delivered and
will deliver millions of stellar spectra that need to be analyzed to
derive stellar parameters (effective temperatures \temp, gravity \logg,
metallicity \met) and elemental abundances. The analysis of such an amount of data
is a challenge that can be addressed by its automation.
Today there is considerable effort to develop software for this purpose. 

Some software packages implement the classical spectral analysis by measuring equivalent
widths (EW) of isolated, well known absorption lines and by deriving
stellar parameters from the excitation equilibrium and ionization balance
(such as the Fast Automatic Moog Analysis, FAMA, \citealp[Magrini et
al.][]{magrini}; GALA, \citealp[Mucciarelli et al.][]{mucciarelli};
ARES \citealp[Sousa et al.][]{sousa}). 
These programs are particularly oriented to
high-resolution, high signal-to-noise (S/N) spectra, for which isolated lines
can be recognized, and their EW can be reliably measured thanks to a safe
continuum placement. 
Other methods are based on grids of synthetic spectra, but they
differ in the ``line-fitting" or ``full-spectrum-fitting" approach, i.e., 
by fitting isolated absorption lines one-by-one (e.g., MyGIsFOS, 
\citealp[Sbordone et al.][]{sbordone}; Stellar Parameters
Determination Software, SPADES, \citealp[Posbic et al.][]{posbic}), or 
by fitting full spectral ranges (e.g., the MATrix Inversion for Spectral SynthEsis, MATISSE,
\citealp[Recio-Blanco et al.][]{recio-blanco}; neural networks, among others
{\it statnet} by \citealp[Bailer-Jones][]{bailer-jones}; 
FERRE, \citealp[Allende Prieto et al.][]{allendeprieto}).
Spectroscopy Made Easy (SME, \citealp[Valenti \& Piskunov][]{valenti})
distinguishes itself from the other codes since it
synthesizes on-the-fly single absorption lines or parts of the 
spectrum to be matched with the observed ones.\\
The line-fitting analysis can derive stellar
parameters and chemical abundances with the drawback of neglecting the
significant amount of information carried by the unused part of the 
observed spectrum. This penalizes the line-fitting approach to low S/N, low
metallicity, and low resolution spectra, in which the number of usable lines
may be too small to carry out this analysis.
On the other hand, the full-spectrum-fitting approach cannot deliver
chemical abundances
because the grid of synthetic spectra needed to cover the
whole parameter and chemical space (and account for
many elemental abundances) would be too big to be handled
\footnote{The ASPCAP pipeline can derive
abundances for up to 15 elements with a technique that can be classified as
a line-fitting approach. See Elia Garcia Perez et al.
\cite{eliagarciaperez}.}.
Other techniques can use real spectra as templates
(``The Cannon", \citealp[Ness et al.][]{ness}; ULySS, \citealp[Koleva et
al.][]{koleva}) with the advantage of overcoming the systematic errors that
stem from the synthetic spectra (due to our incomplete knowledge of atomic parameters
and stellar atmospheres) but share with the previous techniques the
challenge to collect a number of templates 
large enough to uniformly cover the stellar parameter and chemical space.\\

The accuracy of the parameters derived with any of the methods proposed so
far (including this work) depends on two fundamental pillars: the
precision and accuracy of i) the atomic parameters of the absorption lines
employed and ii) the reliability of the stellar atmosphere models. In both
these areas, significant
progress has been made in recent years, and because of their importance, they deserve
further support. The recent praiseworthy efforts to supply laboratory
oscillator strengths (\citealp[Ruffoni et al.][]{ruffoni_ges},
but see also other works cited later on) cover a number of lines that may meet
the needs of the classical spectral analysis (i.e., the line-fitting
approach), but these are too few for the needs of a full-spectrum-fitting analysis.
On the other hand, the 3D stellar atmosphere modeling
(\citealp[Asplund][]{asplund2005}; \citealp[Freytag et
al.][]{freytag}; \citealp[Magic et al.][]{magic}) show that realistic
model atmospheres can reproduce the observed spectra with great accuracy.
However, the computational power required today to analyze wide spectral
ranges with these tools is prohibitive. 

To this lively field that is rich in new ideas, we contribute with a software called
\Space\ (Stellar Parameters And Chemical abundances Estimator) that
implements a new method of performing stellar spectral analysis.  \Space\ is based on a
method born from the experience of the RAVE chemical pipeline \cite[Boeche et
al.][]{boeche11} developed to derive elemental abundances from the
spectra of the RAVE survey (Steinmetz et al. \citealp{rave}; Kordopatis
et al. \citealp{kordopatisDR4}). 
The RAVE chemical pipeline relies on stellar parameters that must be 
provided by other sources, and it only derives chemical abundances.
\Space\ extends the RAVE chemical pipeline's foundations and performs 
an independent, complete spectral analysis. Although \Space\ employs a
full-spectrum-fitting approach, it derives stellar parameters as much as
chemical elemental abundances for FGK-type stars. Unlike other codes
dedicated to stellar parameter estimation, \Space\ does not rely on a
library of synthetic spectra, or measure the EW of
absorption lines, but it makes use of functions that describe how
the EW of the lines changes in the parameter and chemical space.
In the next section we explain the general concepts on which \Space\ is
based. 

\section{Method}\label{sec_method}
The usual methods employed to estimate stellar parameters from spectra are i) 
to directly compare the observed spectrum with the synthetic one to find
the best match and ii) to measure the EWs of the absorption lines of the
observed spectrum from which the stellar parameters are inferred.
In both cases the spectrum must be synthesized, and the stellar parameters of the
synthetic spectrum are varied until the spectrum (first case) or
the line's EWs (second case) match the observed ones. Any spectrum synthesis 
depends on a stellar atmosphere model,
that represents the physical conditions in the stellar atmosphere to 
the best of our knowledge.
For this reason stellar parameters and chemical abundances obtained from spectral
analysis are indirect measurements, and we say that they are derived (and not
measured).
Regardless of the method employed, to estimate the stellar parameters from spectra 
we must construct a spectrum model and compare it to the observed
spectrum. \Space\ makes no exception: it constructs a spectrum model
and compares it with the observed spectrum with a simple $\chi^2$ analysis. 
Its peculiarity is the novel way to construct the spectrum model, which is
not a direct synthesis. At first glance this method may look cumbersome, but it 
eventually gives consistent advantages that we describe below.\\

Consider a stellar spectrum of low to medium spectral resolution\footnote{In
the following the spectral resolution at wavelength $\lambda$ is defined as
R$=\frac{\lambda}{\Delta \lambda}$, where $\Delta \lambda$ is the
Full-Width-Half-Maximum (FWHM) of the instrumental profile.}
($R\sim2\,000-20\,000$) with a known instrumental profile. An absorption line can be fit
with a Voigt profile of known FWHM and strength (i.e., EW).
We start from the na\"ive idea that a normalized spectrum can be reproduced by
subtracting Voigt profiles of appropriate wavelengths, FWHMs,
and EWs from a constant function equal to one (representing the normalized
continuum). Under the weak line approximation
the spectrum so constructed would reproduce the observed spectrum
with fair precision\footnote{We here want just outline the main idea.
To generalize the method the weak line approximation
must be removed, and this is discussed in Sec.~\ref{sec_ewlibrary}.}. 
To construct a full spectrum in this way we 
need to know the EWs of the lines at the wanted \temp, \logg, and
abundance [El/H]\footnote{We define chemical 
abundance of a generic element ``El"
as [El/H]=$\log\frac{N(El)}{N(H)}-\log\frac{N(El)_{\odot}}{N(H)_{\odot}}$ where
$N$ is the number of particle per unit volume.}
of the generic element ``El" the lines belong to.
For this purpose we synthesize the lines for a grid of stellar parameters \temp,
\logg, and chemical abundance [El/H]\footnote{The microturbulence employed
is a function of \temp\ and \logg\ as clarified in Appendix \ref{appx_microt}.}, 
measure the EWs at such points, 
and store them into a library that we called the EW library.
The EW library contains all the information that describes the
strength of the lines in the stellar parameter and chemical space.
So defined, the EW of an absorption line is a function of the stellar parameters that we call 
General Curve-Of-Growth (GCOG) to remember that it
is the generalization of the well known Curve-Of-Growth (COG) function
(which can be obtained from the GCOG by fixing the parameters \temp\ and
\logg, and leave the abundance [El/H] as free variable).
By using the EW library we can construct spectrum models with
stellar parameters and abundances corresponding to grid
points of the library.
To overcome the discreteness of the grid in the parameter space, we use
continuous functions that fit the EWs of the lines in the parameter and
chemical space. This can be done with polynomial functions that we
call ``polynomial GCOGs" and that we store in the ``GCOG library". 
The advantage of this method is that
we just need to vary the parameters \temp, \logg, and abundances [El/H] in
the polynomial GCOGs to vary the strength of the lines and construct
spectrum models for any stellar parameters and abundances until we find the
one that matches the observed spectrum best\footnote{The difference to
the RAVE chemical pipeline (Boeche et al.  \citealp{boeche11}) is that this
one takes \temp\ and \logg\ as external input and uses polynomial COGs to
only derive chemical abundances.}.  This method is implemented in the code
that we call \Space.

To achieve this result, three steps are necessary: i) to build a line list of
absorption lines that must be as much complete as possible (possibly all the
lines visible in stellar spectra) ii) to build an EW library where the EWs of
every absorption line are stored as a function of \temp, \logg, and [El/H], and 
iii) to use the EW library to fit the polynomial GCOGs and store their
coefficients in the GCOG library that is employed by the code \Space\ to construct the
spectrum model. These steps are outlined in the next sections.\\

\section{The line list}\label{sec_lines_list}
To build the EW library we need a list of atomic and molecular absorption
lines and their physical parameters.
These physical parameters are: wavelength, atom or molecule
identification, oscillator strength ($f$, often expressed as logarithm
\loggf, where $g$ is the statistical weight), 
excitation potential ($\chi$), 
van der Waals damping constant $C6$, and dissociation energy $D_0$ (only for
molecules).
The atomic line list was taken from the Vienna Atomic Line Database (VALD, 
Kupka et al. \citealp{kupka}), with the option that selects the lines with 
expected strengths larger than 1\% in at least one of the normalized 
spectra of the Sun, Arcturus, and Procyon\footnote{The VALD web interface
lists the lines which strengths are larger than 1\% of the normalized
flux of a synthetic spectrum which stellar parameters correspond to the
nearest point of the stellar parameters grid 
to the stellar parameters provided by the user. For instance, for
the Sun the closest grid point is at \temp=5750~K, \logg=4.5, \met=0~dex.}. 
Afterwards, the EWs of these lines were 
re-computed with the code MOOG \cite[Sneden][]{sneden} 
and only the lines with EW$>$1m\AA\ in at least 
one of these stars were included in the line list.
The molecular line list was taken from Kurucz \cite{kurucz} and selected
with the same procedure employed for the atomic lines. 
In the present work the line list covers the wavelength intervals
5212-6860\AA\ and 8400-8924\AA. We chose the first interval because
it is commonly covered by optical spectra, while the second
interval (the \Caii\ triplet region) becomes particularly important because
of Gaia spectral coverage. Extentions of the line list to other wavelength
ranges can be done in the future.

\subsection{The atomic lines}\label{sec_atomic_lines}
The wavelengths and excitation potentials were adopted from the VALD database. 
The oscillator strengths are discussed in Sec.\ref{sec_gf_calibration}.
The van der Waals damping constants $C6$ were taken from the VALD database
when such values are available.
When VALD does not provide the damping constants,
we adopted the Uns\" old approximation (computed by MOOG) multiplied by
the enhancement factor E$_{\gamma}$ following the recipe of Edvardsson et
al. \cite{edvardsson} and Chen et al.  \cite{chen}.  For the neutral iron
lines \Fei, this recipe assigns E$_{\gamma}=1.2$ for lines with
$\chi\le2.6$eV and E$_{\gamma}=1.4$ for $\chi>2.6$eV (Simmons \& Blackwell
\citealp{blackwell}), whereas for the ionized \Feii\ lines E$_{\gamma}=2.5$
(Holweger et al.  \citealp{holweger90}). For \Ki, \Tii, and \Vi\
E$_{\gamma}=1.5$ (Chen et al.  \citealp{chen}), for \Nai\ E$_{\gamma}=2.1$
(Holweger et al. \citealp{holweger71}), for \Cai\ E$_{\gamma}=1.8$ (Oneill 
\& Smith \citealp{oneill}), for \Baii\ E$_{\gamma}=3.0$ (Holweger \&
Mueller \citealp{holweger74}). For any other element,
E$_{\gamma}=2.5$ (Maeckle et al. \citealp{maeckle}).\\
Precise damping constants were computed by Barklem et
al. \cite{barklem00} and Barklem \& Aspelund-Johansson \cite{barklem05}.
Such values are contained in the MOOG data files and, by setting
the MOOG keyword ``{\it damping}=1" we imposed to use the Barklem values
when they are available.
There are few cases for which there are no Barklem damping
constants and for which the enhancement factor E$_{\gamma}$ does not apply.
These are:\\
\begin{itemize}
\item The strong and broad lines of \Hi.
Our synthesis with MOOG under Local Thermodynamic Equilibrium (LTE)
assumptions and one dimensional (1D) stellar atmosphere models (see
Sec.\ref{sec_spectra_models} for details) renders a too weak H$\alpha$ line
at the line core, whereas the synthetic Paschen \Hi\ lines 
in the near infrared are too strong at the tip of the line, and 
too weak in the wings with respect to 
the observed lines. For all these lines we adopted the
Uns\"old approximation and calibrate the \loggf s by hand to improve the
fit. However, the match between synthetic and observed \Hi\ lines remains 
unsatisfactory and the lines at 6562.797\AA, 8467.258\AA, 8502.487\AA,
8598.396\AA, 8665.022\AA, 8750.476\AA, and 8862.787\AA\
are neglected during the \Space\ estimation process. The other Paschen \Hi\ lines
in the interval 8400-8924\AA\ are so weak in our standard stars that 
they can be neglected in the \temp\ and \logg\ range considered.
\item For \Sii\ the damping constants reported in VALD appear always too
small. In fact, the \Sii\ lines observed in real spectra are always broader
than the lines synthesized with the VALD damping constant. Also the
value E$_{\gamma}=2.5$ suggested by Holweger \cite{holweger73} 
appears too small\footnote{This is in contrast with Wedemeyer
\cite{wedemeyer} who found that the Uns\" old approximation can well
describe the wings of silicon lines on the Sun.}. 
After some tests, we adopted E$_{\gamma}=4.5$ 
which improves the match of the wings in many (but not all) \Sii\
lines.
\item For the \Mgi\ lines 8712.682\AA, 8717.815\AA, 
and 8736.016\AA\  we adopted E$_{\gamma}=6.0$ in order to
match better their broad wings. These lines are 
multiplets treated as one line (as explained in the following).
\end{itemize}

The need for the adjustments of the E$_{\gamma}$ just reported was
recognized during the firsts attempts to calibrate the \loggf\ and
applied before the final calibration procedure (
described later in Sec.~\ref{sec_spectra_models}).
In the line list there are multiplets where the lines are so close that
they are physically blended. Because lines of multiplets have the same $\chi$, 
these physically blended multiplets can be described (as a first approximation) 
as if they were one single line. 
For multiplets with lines closer than 0.1\AA\ we
adopted one single line with the same $\chi$ and wavelength, which 
is the average of the multiplet's wavelengths. As \loggf\ we adopt the
multiplet's largest \loggf, which is afterwards calibrated by the \loggf\
calibration routine (described in Sec.\ref{sec_gf_calibration}).\\

\subsection{The molecular lines}
Molecular lines of several species are present in the 
considered wavelength ranges. While in hot stars molecules 
have a very low probability to form and their spectral lines have negligible 
strengths, in cool stars molecular lines become important.
In the range 5212-6860\AA\ the spectra of cool stars 
show many absorption features that belong to the species CN, CH, MgH, and TiO. 
However, the very high number of molecular lines
present in the range 5212-6860\AA\ prevents us from performing a reliable 
calibration of their \loggf s (this is discussed in Sec.~\ref{sec_CN}). 
Therefore, in this work we only treat the CN molecule in the wavelength region
8400-8924\AA\ where the CN lines are sparse and
most of them can be identified one by one.
Physical parameters such as wavelengths and excitation potential are
taken from Kurucz \cite{kurucz}. 
The CN molecule dissociation energy $D_0=7.63$eV was taken from Reddy et al. \cite{reddy}.
Oscillator strengths for the CN were taken from Kurucz and afterwards
calibrated by the \loggf\ calibration routine (Sec.\ref{sec_gf_calibration}).
Molecular multiplets are treated like the atomic multiplets (see
Sec.\ref{sec_atomic_lines}).\\

\section{The \loggf\ calibration procedure}\label{sec_gf_calibration}
After the preparation of the line list outlined in
Sec.~\ref{sec_lines_list} we focus on the accuracy of the $gf$-values.\\
Most of the $gf$-values were derived from
theoretical and semi-empirical calculations (e.g., Seaton \citealp{seaton};
\citealp[Kurucz \& Peytremann][]{kurucz_gf})
which are known to have significant errors \cite[Bigot \&
Th{\'e}venin][]{bigot}. Although substantial efforts have been and are
currently being made to obtain precise $gf$-value from laboratory measurements
(from Blackwell et al. \citealp{blackwell72}, \citealp{blackwell76},
\citealp{blackwell79}, \citealp{blackwell82}, to the more recent works by Ruffoni et al.
\citealp{ruffoni}) the number of lines for which laboratory $gf$-values are
available is still small with respect to the number of lines visible in a
stellar spectrum. Besides, the lines targeted for $gf$ laboratory
measurements are the unblended ones, important for the classical 
spectral analysis. This leaves the blended lines uncovered by the laboratory
measurements. 
To improve the quality of the numerous (but inaccurate) $gf$-values provided by
the theoretical computations, some authors calibrate the oscillator
strengths by setting the $gf$-values to match the strength of the
synthetic line with the corresponding line in the Sun spectrum
(among others Gurtovenko \& Kostik \citealp{gurtovenko81},\citealp{gurtovenko82}; 
Th\'evenin \citealp{thevenin89};\citealp{thevenin90}; Borrero
et al. \citealp{borrero}), in two stars like the Sun and
Arcturus (Kirby et al. \citealp{kirby};  Boeche et al.
\citealp{boeche11}) or in three stars like the Sun, Procyon and $\epsilon$
Eri \cite[Lobel et al.][]{lobel}. Recently, Martins et al. \cite{martins}
used the spectra of three different stars (the Sun, Arcturus, and Vega)
and a statistical technique
(the cross-entropy algorithm) to recover the oscillator strengths and broadening
parameters that minimize the difference between the observed spectra and the
synthetic ones. These works employ the idea that, by using more than one
star we can disentangle lines in blends and recover their individual atomic
parameters\footnote{This idea was also employed by \citealp[Boeche et al.][]{boeche11}
for the $gf$ calibration alone, but using a less sophisticated method.}.\\
For our line list we decided to calibrate 
the oscillator strengths of any blended or isolated line on five different
stellar spectra. 
This is necessary because \Space\ employs a full-spectrum-fitting analysis, 
and the few lines with reliable oscillator strengths 
available would not be sufficient.
In the following we outline our calibration method. This method
compares the strengths of the synthetic lines with 
two (or more) stellar spectra in order to correct the $gf$ values of isolated
and blended lines. The method was first proposed in
Boeche et al. \cite{boeche11} and we report it here.\\

In the framework of 1D atmosphere models and LTE assumptions, the EW of a line is a 
function of parameters such as \temp, \logg, the abundance [El/H], the excitation 
potential $\chi$, the $gf$-value, the damping
constant, and the microturbulence $\xi$. Given the atomic parameters $\chi$
and damping constant C6, and assuming that we know with good accuracy the stellar
parameters \temp, \logg, [El/H], and $\xi$ of the stars employed, the EW of a line
can be described as a function of \loggf\ alone \\
\begin{equation}\label{eq_ew}
EW=\mathcal{F}(\log gf).
\end{equation}
Since we know the stellar parameters of the Sun,
by measuring the EW of a line in the Sun spectrum we can
determine its $gf$-value by using equation~(\ref{eq_ew}). This is the so
called ``astrophysical calibration" of the $gf$-values and it only applies
to isolated lines.\\
Now consider a blend made of two lines
$l_1$ and $l_2$. 
The equivalent width of the whole blend in the solar spectrum can be written as
\begin{equation}\label{eq_ew_oneline}
EW_{Sun}^{blend} = \mathcal{F}_{Sun}(\log gf_1,\log gf_2),\\\nonumber
\end{equation}
and the equation is underdetermined because the two oscillator strengths are
unknown. To make it determined we need to measure the EW of the
blend in another star for which we know the stellar parameters and abundances. 
Then, we can write the equation system\\
\begin{eqnarray}\label{eq_ew_twolines}
EW_{Sun}^{blend} & = & \mathcal{F}_{Sun}(\log gf_1,\log gf_2)\\\nonumber
EW_{star}^{blend} & = & \mathcal{F}_{star}(\log gf_1,\log gf_2),\\\nonumber
\end{eqnarray}
and the system is determined. In the general case of a blend composed of $n$ number
of lines, the $gf$-values can be determined by measuring the EWs of
the blends in $k$ number of different well known stars and the equation
system\\
\begin{equation}\label{eq_ew_manylines}
EW_{k}^{blend} = \mathcal{F}_{k}(\log gf_1,...,\log gf_n)\\\nonumber
\end{equation}
is determined when $k \ge n$. Degeneracy can happen when more than one line
belongs to the same element and their excitation potentials $\chi$ are
the same (like in multiplets). In this case the lines behave as one line.
If the lines are close in wavelength we can approximate them as if they were
one single line (as described in Sec.~\ref{sec_atomic_lines}). If more than
one line belongs to the same element and the $\chi$s are
not the same, the degeneracy can be broken by choosing stars with different
stellar parameters, so that the contribution of the lines to the total
EW of the blend is different in different spectra.\\

Ideally, equation system (\ref{eq_ew_manylines})
states that for any line or blend, the \loggf s can be astrophysically calibrated. In practice,
this is not fully true for at least two reasons. 
First, we can only work with a limited number of stars, which may not be large enough to solve
all the possible blends (for instance, when $n>k$, equation system~(\ref{eq_ew_manylines})
is underdetermined). 
Second, uncertainties in the
EW measurements, in the continuum correction, or in the atmospheric
parameters prevent the equality between the measured and the
synthesized EWs (which can be seen as the left- and righthand side of the equation system
(\ref{eq_ew_manylines})). In this realistic case (which is the case of this
work) we can only minimize
the residuals between the left- and the righthand side terms of the
equation system. Besides, because the analytical form of the $\mathcal{F}$ functions in
the equation system are unknown, the solution of the system relies 
on an iterative process where the EWs of the observed spectra (lefthand terms of the system) 
are compared with the EWs resulting from the synthesis of the spectra (righthand
terms of the system) and the variables \loggf s are varied
until the residuals are minimized.
A further difficulty arises when some of the elemental abundances of the stars
employed are unknown. If an element of unknown abundance has isolated lines
in the spectra of these stars, then we can derive its abundance after the \loggf\ of these
lines have been calibrated on the Sun or on another well known star.\\
In the following we outline our solution, which makes use of
the spectra of the Sun and other four stars.

\subsection{Stellar spectra and atmosphere models}\label{sec_spectra_models}
To minimize the residuals between the left- and righthand side
terms of the equation system (\ref{eq_ew_manylines}),
we used high resolution and high S/N spectra
of five stars. We chose the spectra of the Sun, Arcturus
(both from Hinkle et al., \citealp{hinkle}), Procyon,
$\epsilon$ Eri, and $\epsilon$ Vir (from Blanco-Cuaresma et al.,
\citealp{blanco-cuaresma}). These five stars belong to the Gaia FGK benchmark
stars proposed as standard stars for calibration purposes
(Blanco-Cuaresma et al., \citealp{blanco-cuaresma}; Jofr\'e et al.,
\citealp{jofre}).
The Sun and Arcturus spectra were observed with the same instrument
(the Coud\'e feed telescope and spectrograph at Kitt Peak), 
while the other spectra were observed with three different
spectrometers. The spectra of Procyon and
$\epsilon$ Vir were observed with the NARVAL spectropolarimeter with a
spectral resolution of R$\sim$81\,000 covering the full range of wavelengths
between 3000\AA\ and 11\,000\AA. For $\epsilon$ Eri, Blanco-Cuaresma et al.
only provide spectra taken with the UVES and HARPS spectrometers, which
present gaps in the wavelength coverage (at $\sim$5304-5336\AA\ for HARPS and
at $\sim$5770-5840\AA\ and 8540-8661\AA\ for UVES). 
For this reason we employed the HARPS spectrum for
the wavelength range 5712-6260\AA\ and the UVES spectrum everywhere else.
Unfortunately, the gap at 8540-8661\AA\ cannot be covered, because
no spectra from other instruments are available in this wavelength range.
This means that for this wavelength interval, $\epsilon$ Eri does not play any
role in the $\log gf$ calibration, which is only performed on the spectra of 
the other four stars.
All the spectra were re-sampled to a dispersion of 0.01\AA/pix to
match the dispersion of the synthetic spectra. In fact, the calibration
routine (described in Sec.~\ref{sec_corr_routine}) requires that both must have the same
sampling). 
Hinkle et al. provided the spectra free of telluric lines (they were
subtracted from the observed spectra), therefore
they cannot affect the strengths of the absorption lines.
Conversely, the telluric lines affect
the Blanco-Cuaresma spectra and, for the wavelength ranges
here considered, they affect mainly the range $\sim$6274-6320\AA. 
The presence of telluric lines superimposed on an absorption line can affect
its \loggf\ calibration. However, we verified that the final 
effect on the calibrated \loggf s is in general weak or negligible because the
calibration is performed simultaneously on two spectra free from telluric
lines and on other three on which the telluric lines do not lie at the same
wavelengths because of the different velocity correction $\Delta v$
applied (the spectra exibith different radial velocities).\\

To synthesize the spectra we used the code MOOG (Sneden et al.,
\citealp{sneden}) and the stellar atmosphere models 
from the ATLAS9 grid \cite[Castelli \& Kurucz][]{castelli} updated to the
2012 version\footnote{http://wwwuser.oats.inaf.it/castelli/grids.html}. 
For the solar spectrum we assumed an
effective temperature \temp=5777K, gravity \logg=4.44,  metallicity 
\met=0.00~dex. For Arcturus we assumed the stellar parameters of
Ram{\'{\i}}rez \& Allende Prieto \cite{ramirez}, while
for the other stars we adopted the stellar parameters
given in Jofr\'e et al. \cite{jofre}.
The microturbulence $\xi$ adopted is inferred during the calibration process as
explained in Sec.~\ref{sec_microt}.
All these stellar parameters are summarized in Tab.~\ref{tab_st_params}.
The elemental abundances adopted for the Sun are [El/H]=0~dex by
definition, with solar abundances adopted from
Grevesse \& Sauval \cite{grevesse}. For Arcturus we adopted the elemental
abundances given by Ram\'irez \& Allende Prieto \cite{ramirez}. For elements
for which Ram\'irez \& Allende Prieto gave no abundance we impose [El/H]=\met\ at the beginning of the
calibration and leave the possibility to change the abundance during the
calibration process. In fact, if the element has an isolated line its
\loggf\  can be calibrated on the Sun and its abundance on Arcturus can be
therefore derived. Similarly, at the beginning of the calibration process we
impose [El/H]=\met\ for the elements of the other stars and allow
possible changes of [El/H] during the process.\\

Because the instrumental resolution varies with wavelength, and because
MOOG adopts one constant FWHM per synthesized interval (used to convolve the
synthetic spectrum with the adopted instrumental profile) 
we synthesized the spectra in four pieces covering the 
wavelength ranges 5212-5712\AA, 5712-6260\AA, 6260-6860\AA, and 8400-8924\AA.
The line profile of the synthetic spectra was broadened with a Gaussian
profile (to reproduce the instrumental profile of the spectrograph) 
and a macroturbulence profile, the best matching values
of which were chosen via eye inspection for every wavelength range.
While the macroturbulence is constant across these wavelengths, the Gaussian
instrumental profile broadens with wavelength.
In the last column of Tab.~\ref{tab_st_params} we report the macroturbulence
$v_{mac}$ adopted, while in Tab.~\ref{tab_fwhm} we summarize the best matching
Gaussian FWHM chosen for the four wavelength ranges.

\begin{table}[t]
\caption[]{Effective temperature (K), gravity, metallicity (dex), micro- and 
macroturbulence (in \kmsec) adopted to synthesize the spectra of the standard stars.}
\label{tab_st_params}
\vskip 0.3cm
\centering
\begin{tabular}{l|c|ccccc}
\hline
\noalign{\smallskip}
star & sp. class & \temp & \logg & \met & $\xi$ & $v_{mac}$\\
\noalign{\smallskip}
\hline
\noalign{\smallskip}
Sun & G2V & 5777 & 4.44 & 0.00 & 1.3 & 2.5\\
Arcturus &K1.5III& 4286 & 1.66 & -0.52 & 1.7 & 5.3\\
Procyon &F5IV-V& 6554 & 3.99 & -0.04 & 2.1 & 7.5\\ 
$\epsilon$ Eri &K2Vk:& 5050 & 4.60 & -0.09 & 1.1 & 3.5\\
$\epsilon$ Vir &G8III& 4983 & 2.77 & +0.15 & 1.5 & 6.0\\
\noalign{\smallskip}
\hline
\end{tabular}

\end{table}

\begin{table}[t]
\caption[]{Instrumental FWHMs adopted for the synthetic spectra in
the four wavelength ranges.}
\label{tab_fwhm}
\vskip 0.3cm
\centering
\begin{tabular}{l|cccc}
\hline
\noalign{\smallskip}

star &  \multicolumn{4}{c}{FWHM(\AA) at} \\
     & 5212- & 5712- & 6260- & 8400\\
     & 5712\AA & 6260\AA & 6860\AA & 8924\AA\\
\noalign{\smallskip}
\hline
\noalign{\smallskip}
Sun            & 0.04 & 0.05 & 0.06 & 0.07 \\
Arcturus       & 0.04 & 0.04 & 0.04 & 0.07 \\
Procyon        & 0.05 & 0.05 & 0.06 & 0.08 \\ 
$\epsilon$ Eri & 0.05 & 0.03 & 0.07 & 0.08 \\
$\epsilon$ Vir & 0.05 & 0.05 & 0.06 & 0.07 \\
\noalign{\smallskip}
\hline
\end{tabular}

\end{table}

\begin{figure*}[t]
\begin{minipage}[t]{6cm}
\includegraphics[width=6cm,viewport=30 28 232 390]{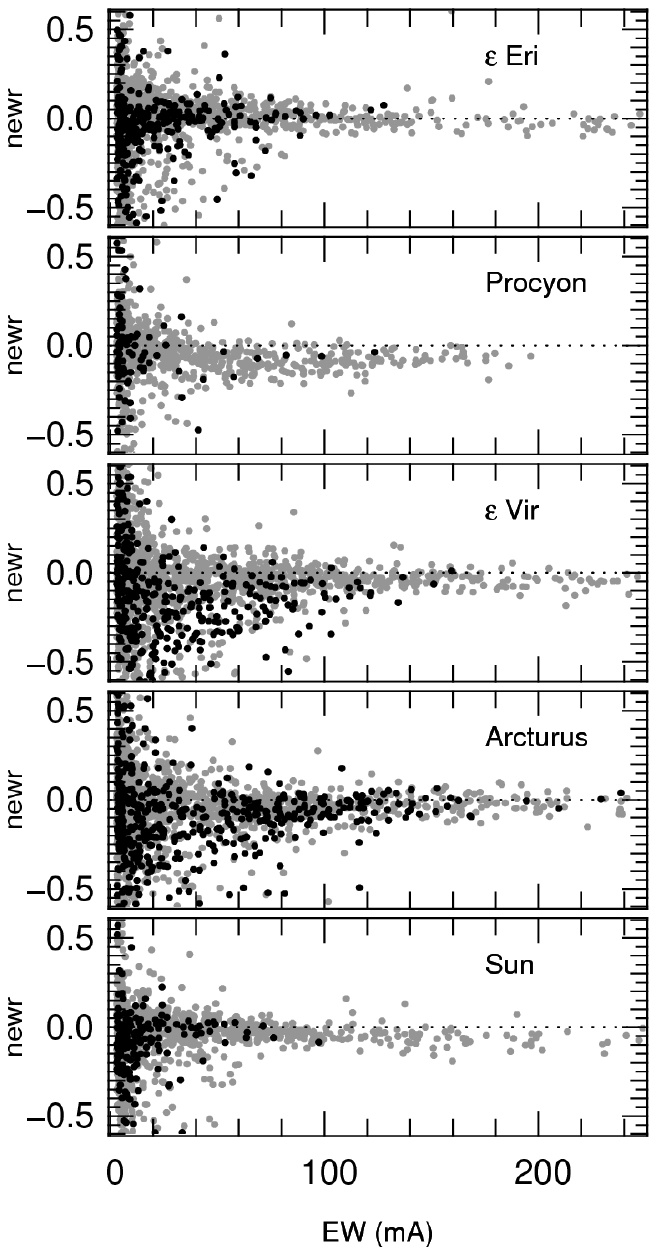}
\caption{Distributions of the $NEWR$s of the elements Fe (gray points) and Ti
(black points) as a function of their EW for the five stars before the beginning of
the calibration routine.}
\label{initial}
\end{minipage}
\hfill
\begin{minipage}[t]{6cm}
\includegraphics[width=6cm,viewport=30 28 232 390]{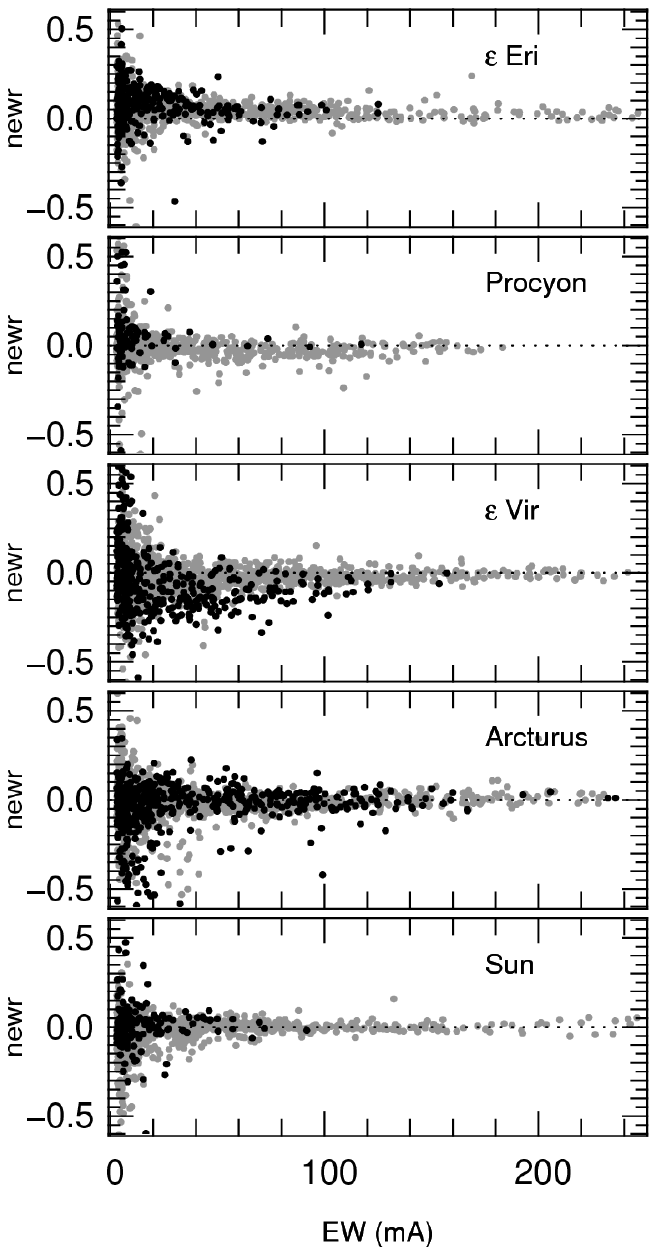}
\caption{As in Fig.~\ref{initial} but after 100 iterations of the \loggf\ calibration
routine.}
\label{first_gf_corr}
\end{minipage}
\hfill
\begin{minipage}[t]{6cm}
\includegraphics[width=6cm,viewport=30 28 232 390]{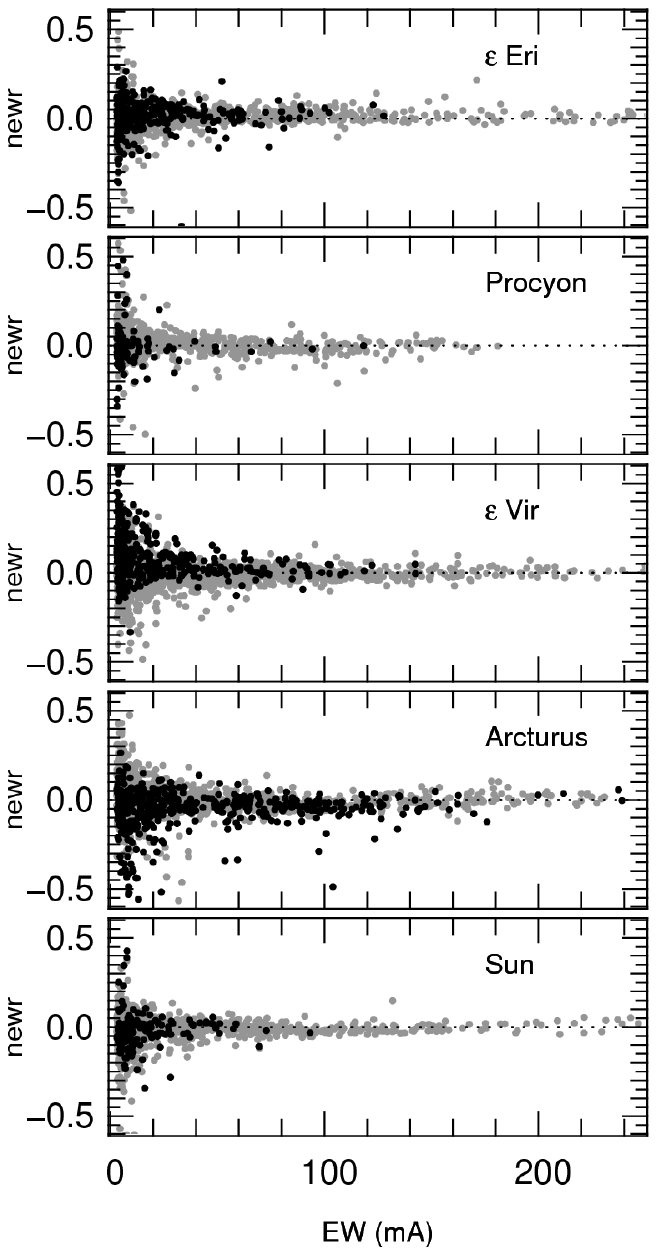}
\caption{As in Fig.~\ref{initial} but at the final stage, after many
iterations of the calibration routine and properly adjusted abundances.}
\label{final_corr}
\end{minipage}
\end{figure*}

\subsection{The \loggf s calibration routine and the abundances
correction}\label{sec_corr_routine}
The calibration routine consists of two parts: the $\log gf$s calibration and
the abundances correction. The first part is semi-automatic, the second part
is manual.
The procedure begins with the first synthesis of the five
spectra by using the code MOOG. At the beginning, we fix the abundances that
are known and these remain unchanged through the whole process with few
exceptions illustrated later on.
For the Sun the abundances are fixed at [El/H]=0~dex. For Arcturus
we fix the abundances of 16 elements as given Ram{\'{\i}}rez
\& Allende Prieto \cite{ramirez}. For the other elements we set
[El/H]=[M/H]. Assuming the elemental abundances of the other
stars to be unknown\footnote{Precise abundances of these stars have been derived by
Jofr{\'e} et al. \cite{jofre15}, whose results were not yet public at the time of this work.},
we assumed at the beginning [El/H]=[M/H].
Then the first \loggf\ calibration continues as follows:

\begin{enumerate}
\item The 5 spectra are synthesized by using the adopted  line list and
atmosphere models.
\item The observed spectra are re-normalized with the same routine used for
the \Space\ code (see Sec.\ref{sec_renorm}). In this case the
interval has a radius of 5\AA\ and only normalized fluxes larger than 0.98 are
considered.
\item With MOOG (driver {\it ewfind}) the equivalent widths $EW_i^k$ of the lines
for the 5 spectra are computed. These are the expected EWs of the lines
if they were isolated.
\item The isolation degree parameter $iso$ for the $i$-th line
and $k$-th star is computed as
$$
iso_i^k=\frac{EW_i^k}{\sum_{\lambda_i-0.3\AA}^{\lambda_i+0.3\AA} EW_i^k}
$$
where $\lambda_i$ is the central wavelength of the $i$-th line.
It approximates the fraction of the flux absorbed by the $i$-th line over the
total flux absorbed by any other line present in an interval 0.6\AA\ wide centered on the $i$-th line. 
\item The Normalized Equivalent Width Residual ($NEWR$) for the $i$-th line and the
$k$-th star is computed as follows
$$
NEWR_i^k=\frac{F_i^{k,synt}-F_i^{k,obs}}{1-F_i^{k,obs}}
$$
where $F_i$ is the flux integrated over an interval
centered on the $i$-th line. The width of the interval is 0.05\AA\ if the
$EW<70$m\AA, 0.10\AA\ if $70\leq EW<150$m\AA, 0.15\AA\ if $150\leq EW<200$m\AA,
0.20\AA\ if $200\leq EW<250$m\AA, and 0.30\AA\ if $EW\geq 250$m\AA.
The $NEWR$ represents the residual between the strengths of the synthetic and the
observed line. When the $NEWR$ is negative this means that the synthetic line is
stronger than the observed one (and vice versa).

\item The $\log gf$ calibration of the $i$-th line is performed by adding the quantity 
$$
\Delta \log gf_i=-\frac{\sum_k NEWR_i^k \cdot iso_i^k}{\sum_k iso_i^k}
$$
to the $\log gf_i$ .
If the line belongs to an atom the weighted sum considers all the five spectra,
otherwise (i.e., if it belongs to a molecule) the spectrum of Procyon is neglected (because no
molecular line is visible on its spectrum).
If $\Delta \log gf_i>0.05$, we confine it to this value to avoid
divergences. If $\Delta \log gf_i<0.01$ then we set it to zero.
If $\log gf_i<$-9.99 or $\log gf_i>$3.0 the line is
removed from the line list.
\item The EWs of the lines are computed with the driver {\it ewfind} of MOOG. The
lines with $EW\leq$3m\AA\ in all of the five spectra are removed from the
line list.
\item The routine is repeated from step~1.
\end{enumerate}

This routine is always followed with three exceptions: 
i) for very strong lines (some tens of lines) the
chosen interval (step~5 of the routine) is larger than 0.05\AA\ to
match the full line instead of the core alone;
the interval was chosen after eye inspection;
ii) many intense lines on the star $\epsilon$ Eri have particularly
wide wings, which cannot be well synthesized; therefore, for these lines
the calibration $\Delta \log gf$ was computed by neglecting
the $\epsilon$ Eri spectrum; iii) for strong lines like the \Hi\ lines,
the \Nai\ doublet at 5889 and 5895\AA, the \Caii\ triplet in the infrared
region, and few other \Fei\ intense lines the \loggf s were set by hand
after eye inspection.

\begin{figure*}[t]                                                            
\centering                                                                   
\includegraphics[width=16cm,clip,viewport=98 123 475 599]{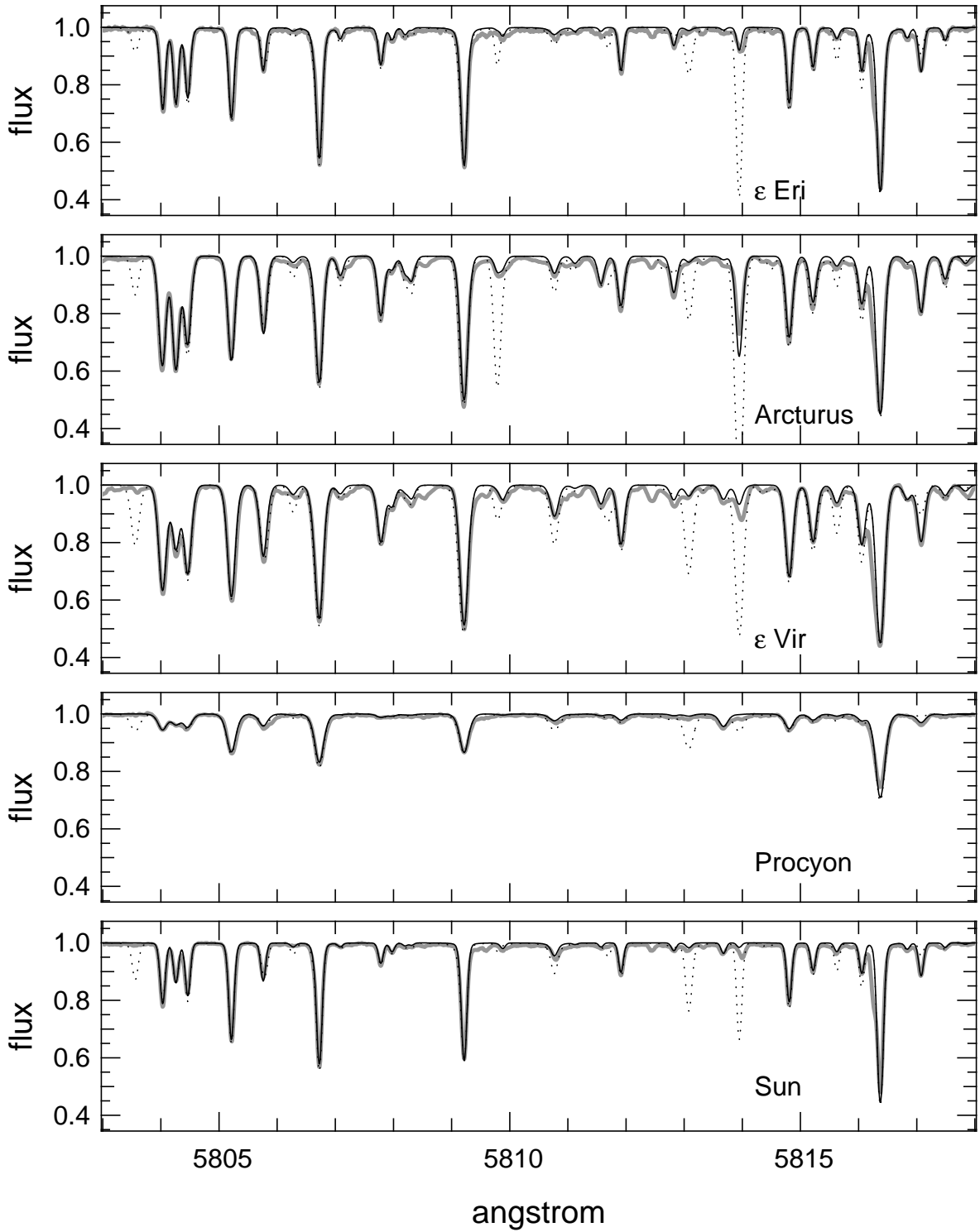}
\caption{Gray thick lines: observed spectra. Dotted lines: synthesized spectra
with VALD \loggf s but final abundances. Black lines: synthetic spectra
with calibrated \loggf s and final abundances.
}                                          
\label{final_abd_ini_llist}                
\end{figure*}

\begin{table*}
\caption{Final chemical abundances for the 5 stars at the end of the
\loggf\ calibration compared with the abundances by Ram{\'{\i}}rez \&
Allende Prieto and Allende Prieto et al. \cite{allende}. 
The values with an asterisk, ``*", are the abundances of the non-ionized 
element derived by Ram{\'{\i}}rez \& Allende Prieto
\cite{ramirez}.}\label{tab_5stars_abd}
\centering
\begin{tabular}{l|l|cc|cc|cc|c}
\hline
\noalign{\smallskip}
element & N &\multicolumn{2}{c}{Arcturus} & \multicolumn{2}{c}{Procyon} & 
\multicolumn{2}{c}{$\epsilon$ Eri}& $\epsilon$ Vir\\
\hline
        &   &   [El/H]       &  [El/H] &   [El/H]  & [El/H] &   [El/H]  & [El/H] & [El/H]\\
        &   & us & Ramirez  & us & Allende Prieto & us & Allende Prieto   &  us   \\
\hline
 Mg & 24   & -0.15   & -0.15   & -0.04  & -0.01   &  -0.09  & -0.03   &   0.15\\
 Si & 228  & -0.19   & -0.19   & -0.02  &  0.07   &   0.01  & -0.01   &   0.25\\
 Ca & 74   & -0.41   & -0.41   & -0.04  &  0.25   &  -0.09  & -0.01   &   0.15\\
 Sc & 76   & -0.37   & -0.37*  & -0.04  &  0.07   &  -0.09  &  0.02   &  -0.02\\
 Ti & 463  & -0.25   & -0.25*  &  0.02  &  0.13   &  -0.07  &  0.01   &   0.01\\
 V & 265  & -0.32   & -0.32   & -0.04  &         &  -0.02  &         &   0.02\\
 Cr & 225  & -0.57   & -0.57   & -0.04  &         &  -0.09  &         &   0.08\\
 Fe & 1436 & -0.52   & -0.52   & -0.06  &  0.03   &  -0.05  & -0.06   &   0.15\\
 Co & 186  & -0.23   & -0.43   & -0.04  &  0.05   &  -0.09  & -0.08   &   0.15\\
 Ni & 194  & -0.46   & -0.46   & -0.04  &  0.07   &  -0.09  & -0.06   &   0.13\\
\hline
\end{tabular}
\end{table*}

At the beginning of the calibration routine the  $NEWR$s are distributed
as shown in Fig.~\ref{initial}
(in this and in the two following figures we only show the $NEWR$s of the elements
Fe and Ti for the sake of clarity).
After 100 iterations the $NEWR$ distributions emerge as in
Fig.~\ref{first_gf_corr}.  Note that the dispersion of the points has decreased
and that the Fe lines (gray points) in Procyon, and the Ti lines (black
points) in $\epsilon$ Vir have an offset.  The
offsets are due to the assumption of the wrong Fe and Ti abundances for these stars at the
beginning of the procedure.  To continue with the \loggf s calibration we
must apply the second part of the calibration process, i.e., the abundance
correction, which is done manually. The negative offset of the Ti lines in $\epsilon$
Vir indicates that the Ti lines are too strong, therefore, to match the observed
spectrum, the Ti abundance must be decreased. 
Similarly, when the offset is positive, the abundance must be increased.
These evaluations and the consequent changes in abundance are done by observing the
distribution of the $NEWR$ for lines for which the isolation degree parameter $iso$
is larger than 0.99 (which implies the selection of isolated lines alone) in order to guess the right
abundances from isolated lines. If no isolated lines are present, the
abundances remain [El/H]=[M/H].
The abundance correction is performed after the \loggf s calibration and
both are carried out many times until no more lines 
are rejected by the calibration routine and the $NEWR$ distributions 
are centered on zero.\\

We want to spend a few more words on the abundance correction. The optimal
condition for the \loggf s calibration would be to have precise elemental
abundances for all the stars employed (as required by
equation~(\ref{eq_ew_manylines})) 
so that the abundance correction would not be necessary. 
Because at the time of this work precise abundances of these stars were not available,
in order fulfill the condition in equation~(\ref{eq_ew_manylines}) 
as much as possible, we adopt the known
abundances, i.e., the Sun and the Arcturus abundances. For the
other stars (or elements) for which we do not have chemical abundances, 
we adopted [El/H]=[M/H] and then followed the method described before 
(i.e., observing the distribution of the $NEWR$s of the isolated lines of 
an element and change its abundance to minimize the average of the absolute
$NEWR$ values) whenever the adopted initial abundance was not satisfactory for this element. 
In the case of the element \ion{Co}{} on Arcturus, we decided to follow this method
and we adjusted its abundance to [Co/H]=$-0.23$~dex because by using the
value [Co/H]=$-0.43$~dex derived by Ram{\'{\i}}rez \& Allende Prieto it was
not possible to minimize the absolute average $NEWR$ values for all the stars.\\
We want to stress that the solar abundances were never
changed during the whole process.  The Sun is synthesized with the
Grevesse \& Sauval \cite{grevesse} solar abundances and
these abundances must not be changed because this is the reference point
on which the whole calibration procedure is based.  Without this reference
point no calibration is possible, otherwise the equation system
(\ref{eq_ew_manylines}) would become underdetermined.\\

\subsection{Setting the microturbulence}\label{sec_microt}

\begin{figure}
\includegraphics[width=9cm,viewport=28 28 408 196]{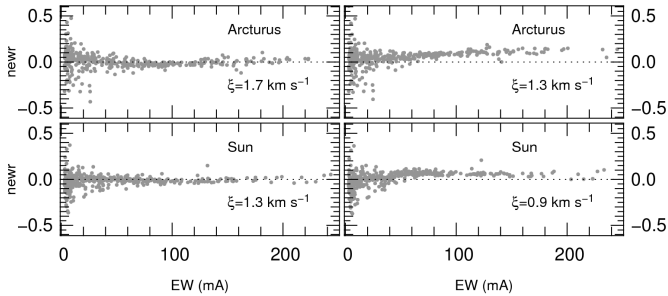}
\caption{$NEWR$ distributions for the Fe lines in Arcturus (top panels) and
the Sun (bottom panels) using microturbulences $\xi$ that differ by 0.4
\kmsec. Here we use Fe lines with isolation degree parameter $iso>0.99$
(i.e., these Fe lines are isolated).}
\label{newr_microt_variation}
\end{figure}

At the beginning of this work we tested several times the calibration
routine in order to find the best way to follow. In some of these tests
we adopted the microturbulence values reported in Jofr\'e et al.
\cite{jofre}, which are 1.2, 1.3, 1.8, 1.1, and 1.1 \kmsec\ for the Sun,
Arcturus, Procyon, $\epsilon$ Eri, and $\epsilon$ Vir, respectively. With
these values we could not find satisfactory results, which means that the
$NEWR$s of the isolated Fe lines did not align close to the $NEWR=0$ line for some
of the stars, no matter what Fe abundance was adopted. 
Therefore, we decided to change the $\xi$ values interactively
during the calibration process to minimize the absolute $NEWR$s values. The
$NEWR$s values are sensitive to microturbulence and in
Fig.~\ref{newr_microt_variation} we show the difference in the $NEWR$s
distributions observed by changing the microturbulence by 0.4~\kmsec\ (while
all other parameters are fixed) for the
Sun and Arcturus, with the best performing $\xi$ value in the lefthand panels. 
Our final best $\xi$ values are reported in Tab.~\ref{tab_st_params}.

\subsection{The final line list}\label{sec_final_ll}
The whole calibration
procedure described in the previous section is performed by 
applying alternatively the \loggf\
routine and the abundance correction iteratively until convergence of the
abundances and until no more lines are removed by the \loggf\ routine.  The
process began with a line list of 8\,947 lines.  After the convergence, the
final line list counts 4\,643 lines. The $NEWR$s distribution of the final
line list is shown in Fig.~\ref{final_corr}, while in Tab.~\ref{tab_5stars_abd} 
we report the abundances derived with the abundance correction procedure for
the elements that we consider reliable (see Sec.~\ref{sec_loggf_accuracy}
for further explanations). The final line list with the calibrated \loggf s
is released together with the GCOG library.
                                                                              
At the end of the calibration procedure we verified by eye inspection 
the good match between the synthetic and observed
spectra over the whole wavelength range considered. In some
cases there are unidentified absorption lines for which none of the lines
given in the VALD database seems to match. These lines are neglected 
during the analysis performed by \Space.
We removed by hand some lines of the line list because
were clearly erroneous (but not removed by the calibration routine
because they lie under incorrectly fitted lines, like under the \Can\
triplet lines, for instance) or because their \loggf s were badly affected
by unidentified lines. In Fig.~\ref{final_abd_ini_llist} we compare
part of the synthetic and the observed spectra before and after the \loggf s
calibration. For this figure the synthesis was performed using the 
final abundances reported in Tab.~\ref{tab_5stars_abd}, so that the
differences between the spectra synthesized with the VALD \loggf s and the
calibrated \loggf s are only due to the difference in \loggf s.
Fig.~\ref{final_abd_ini_llist} 
shows that the spectra synthesized with the new calibrated \loggf s 
match the observed spectra better than the ones synthesized with the
VALD \loggf s.

\subsection{Validation of the calibrated \loggf
s}\label{sec_loggf_validation}
\begin{figure}
\includegraphics[width=9cm,bb=14 30 390 224]{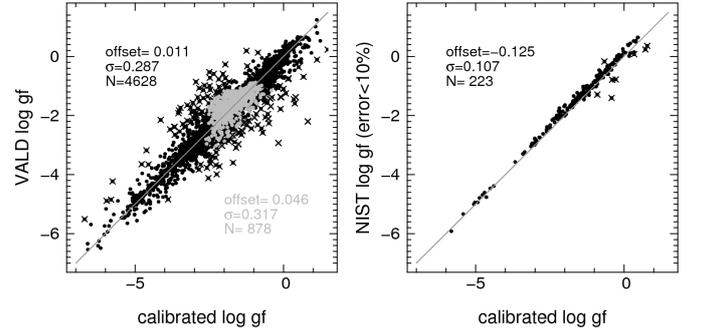}
\caption{{\bf Right}: comparison between the VALD and our calibrated \loggf s. 
Black and gray points represent atomic and molecular CN lines, respectively. {\bf Left}: comparison
between the NIST and our \loggf s values for those lines with NIST \loggf\ precision
better than 10\%. Only lines having $EW>5$m\AA\ in the solar spectrum are
reported here. 
The offsets and standard deviations are computed as ``calibrated minus
reference" after rejecting the outliers (crossed points) with a 3$\sigma$
clipping.} 
\label{comp_loggf}
\end{figure}

In Fig.~\ref{comp_loggf} (left panel) we compare the original \loggf s
of the VALD database with our final calibrated \loggf s for those lines that
have $EW>5$m\AA\ in the solar spectrum. The residuals have an average offset
of $\sim+0.01$ with a dispersion of $\sim0.29$ (statistic computed excluding
strong lines for which \loggf\ was calibrated by hand and after
rejection of outliers with a 3$\sigma$ clipping). 
Because VALD is a database that collects data from several sources with
different degrees of precision, we want to verify the robustness of our calibrated 
\loggf s comparing them with precise values.
For this purpose, we accessed the NIST database \cite[Kramida et al.][]{nist}
and select lines that have a $gf$ precision better than 10\% ($\sim$0.04
in \loggf) and $EW>5$m\AA\ on the solar spectrum. With these 
criteria we found 328 lines in common with our line list. After removing the
lines with $EW<5.0$m\AA\ (for which we expect the largest calibrated \loggf\ errors) we
are left with 223 lines belonging to
the elemental species \Ci, \Ni, \Oi, \Nai, \Mgi, \Siii, \Sci, \Scii, \Tii,
\Vi, \Cri, \Mni, \Fei, and \Coi. The comparison between the NIST and the
calibrated \loggf s is shown in right panel of Fig.~\ref{comp_loggf}.
The comparison with high precision \loggf s shows that our calibrated \loggf
s have an offset of $\sim-0.12$~dex with a dispersion of $\sim$0.1~dex. 
The negative offsets say that our \loggf s are more negative than the
corresponding NIST \loggf s. This can be due to several causes, 
which may be
i) an inappropriate line profile of the synthetic spectra (the line
profiles can vary for lines with different strengths and lines broadening
parameters)
ii) an inappropriate continuum normalization of the observed spectra,
iii) neglected Non Local Thermodynamic Equilibrium (NLTE) and 3D effects, 
iv) errors in the stellar parameters adopted to synthesize the spectra.
As a last remark, one may regard at the solar abundances adopted (Grevesse \&
Sauval \citealp{grevesse}) as 
the cause of the offset of our calibrated \loggf s.
If these abundances (and in particular the Fe
abundance, which has the higher number of lines represented in
Fig.~\ref{comp_loggf}) were too high, the calibration would render
\loggf s lower than expected. Some works, based on laboratory \loggf s,
derived a solar iron abundance\footnote{Here we define
$\log[\epsilon(Fe)]=\log\frac{N(Fe)}{N(H)}+12$.} of 
$\log[\epsilon(Fe)]=7.44$ (\citealp[Ruffoni et al.][]{ruffoni_ges};
\citealp[Bergemann et al.][]{bergemann2012}). Since we adopted
$\log[\epsilon(Fe)]=7.50$, this would explain part of the offset
observed for our \loggf s. 

\subsection{On the accuracy of the calibrated \loggf
s}\label{sec_loggf_accuracy}
Although we verified by eye the good match between the spectra synthesized
with our final line list and the observed spectra, this does not ensure the
good accuracy of the astrophysically calibrated \loggf s for all the lines of
the line list.
In fact, the spectra exhibit many blended features composed of many lines
that cannot be fully resolved, because for such blends the equation system
(\ref{eq_ew_manylines}) is underdetermined. On the other hand, weak lines
($EW\lesssim10$m\AA) are the ones more affected by imprecision of the continuum
placement or by blends. A difference in 0.5\% of the normalized flux between
the synthetic and the observed spectra can look like a ``good match" to
an eye inspection, but it leads to a very poor accuracy of the \loggf\ of a 
line having an EW of a few m\AA. Another source of uncertainty comes from the abundance
correction procedure when the lines of one element are all weak in the Sun' s
spectrum. With the Sun as reference point, when the lines are weak the
match with the synthetic spectrum is subject to the uncertainties discussed above,
so that the reference point becomes uncertain. For this reason, 
the derived elemental abundances output by the code \Space\ (in its
present version), are for those 
species for which the number and strength of
lines are big enough for a good abundance estimation of the
five stars during the abundance correction process outlined in
Sec.~\ref{sec_corr_routine}. These elemental abundances are the ones
reported Tab.~\ref{tab_5stars_abd}. The abundances of other elements are also
internally derived by \Space\ but are used as ``dummy" elements and 
rejected at the end of the analysis.\\
There are further reasons why the calibrated \loggf s of some lines 
may be not physically meaningful. We employed stellar atmosphere models
that are one-dimensional and the physical processes are assumed to take
place in Local Thermodynamic Equilibrium (LTE). This is an approximation that, in
some cases, is too rough to describe real stellar atmospheres. Some absorption lines
suffer of non-LTE effects, which can affect the observed EW.
Therefore, if we perform an astrophysical calibration of the \loggf s of
one of these lines under LTE assumptions, the calibrated \loggf\ value 
can be significantly different from the real value (which expresses the
probability of the electronic transition) and the difference accounts for 
the neglected non-LTE effect. This is not the right way to correct for non-LTE
effects and it may lead to systematic errors when stellar parameters 
and the chemical abundances are derived.\\
During the \loggf s calibration and abundance correction procedure, we identified several
strong lines that cannot be correctly synthesized in our five standard
stars. The profiles of these synthetic lines have too strong (or too weak) wings
with respect to the observed lines in the spectra of the standard stars. 
Some of these lines are reported in Tab.~\ref{tab_lines_reject} with a
qualitative goodness of fit of the wings (and strength) between the synthetic and observed
lines. For some lines (such as most of the \Hi\ lines and the \Nai\ doublet at
$\sim$5890) we changed the \loggf s (and also the damping constants for
the Paschen \Hi\ lines) by hand in order to match the strength
of these lines in the solar spectrum. However, the match is often not satisfactory. 
Most of the \Mni\ lines show a line width too narrow in synthetic spectra
with respect to the observed ones, and in the Sun synthetic spectrum these lines
are too strong at the core, although their EWs seem to be close to the observed ones. 
The \Mni\ abundance is therefore rejected from the \Space\ results.
All these discrepancies can be due to non-LTE effects, 3D effects, and
hyperfine splitting of the lines that we do not take into account in the present
work.
\begin{table}[t]
\caption[]{Qualitative match of the wings of some intense lines between the synthetic
and the observed ones. The symbols ``+" and ``-" mean that the synthetic
line is too strong or too weak (respectively) with respect to the observed
ones. ``Ok" means that the match is satisfactory. These lines may suffer of
non-LTE and/or 3D effects.}
\label{tab_lines_reject}
\vskip 0.3cm
\centering
\begin{tabular}{l|ccccc}
\hline
\noalign{\smallskip}
wavelength & Sun & Arcturus & Procyon & $\epsilon$ Eri& $\epsilon$ Vir\\
\hline
\noalign{\smallskip}
5269.537 \Fei & ok & + & ok & ok & + \\
5328.039 \Fei & ok & + & ok & ok & + \\
5371.489 \Fei & - & + & ok & -- & ok \\
5405.775 \Fei & ok & + & ok & ok & + \\
5889.9510 \Nai & ok & + & - & ok & - \\
5895.9240 \Nai & ok & + & + & ok & - \\
8498.023 \Caii & ok & - & ok & ok & - \\
8542.091 \Caii & ok & - & ok &  & - \\ 
8662.141 \Caii & ok & - & ok &  & - \\ 
8806.756 \Mgi & ok & - & ok & ok & - \\ 
\noalign{\smallskip}
\hline
\end{tabular}
\end{table}
Because these ``poorly matching" lines can negatively affect the stellar parameter estimations, they
are rejected from the analysis performed by \Space.\\
However, the fact that the spectra synthesized with our line list with
calibrated \loggf s match reasonably well\footnote{The residuals between the
synthetic and the observed spectra have a standard deviation of
$\sim1-2$\% of the normalized flux.} the great majority of the spectral
range of our standard stars (which span a wide range in temperature and
gravity) and that most of the abundances derived
during the abundance correction process are close to the ones reported in
high-resolution studies (see Tab.~\ref{tab_5stars_abd}), suggests that our line
list under LTE assumption can be employed to derive reliable stellar parameters and
chemical abundances in the \temp\ and \logg\ ranges covered by the
five calibration stars adopted in this work.

\begin{figure}
\includegraphics[width=9cm,bb=97 286 493 499]{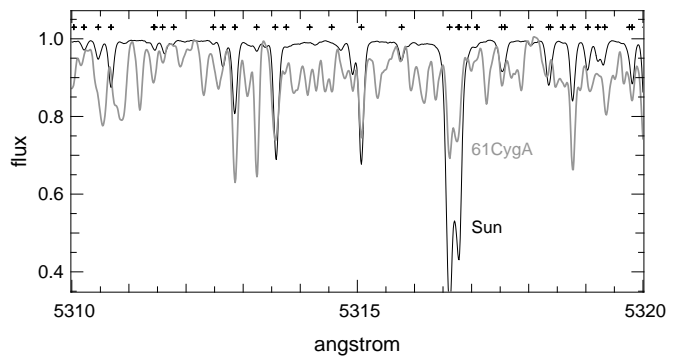}
\caption{Normalized spectra of the Sun (black line) and 61CygA (gray line).
The ``plus" symbols indicate the positions of the atomic lines.}
\label{comp_sun_61CygA}
\end{figure}

\subsection{The molecular lines}\label{sec_CN}
In the previous sections we discussed mainly the atomic lines, although
molecular lines of several species are present in the wavelength ranges
considered. During the preparation of this work we did several tests
to verify whether a \loggf s calibration of atomic and molecular lines together
was possible. We found that i) the calibration is not always 
possible, and ii) when it is possible, the calibrated \loggf s are
physically meaningless and can be only used as dummy values.
The first point applies to the wavelength range 5212-6860\AA\ where the
very high number of molecular lines of the species CN, CH, MgH, and TiO
generate a forest of weak lines in cool star spectra that makes 
the identification of the lines impossible and the equation system (\ref{eq_ew_manylines}) 
becomes underdetermined.
In Fig.~\ref{comp_sun_61CygA} we compare the spectrum of the Sun (normalized by Hinkle et al.
\citealp{hinkle}, \temp=5777~K, \logg=4.44, \met=0.0~dex) and 61CygA (normalized by
Blanco-Cuaresma et
al. \citealp{blanco-cuaresma}, \temp=4374~K, \logg=4.63, \met=$-0.33$~dex). 
The forest of weak molecular lines in 61CygA is 
so dense that it creates a ``pseudo-continuum" that hides the real continuum
and prevents the correct estimation of the EWs of the atomic lines.
This convinced us that, at present, our method cannot calibrate \loggf s of
molecular lines in the interval 5212-6860\AA. 
Besides, to calibrate \loggf s of atomic lines we need spectra ``free" of
molecular lines. Therefore we verified that the standard stars employed for the
\loggf\ calibration are not significantly affected by molecular lines.
We verified that this is true for dwarf stars having \temp$\gtrsim5000$~K 
and for a giant star like Arcturus.\\

In the interval 8400-8924\AA\ the second answer above applies: 
here we can identify the CN lines and calibrate their \loggf s, but we strongly doubt 
the accuracy of the calibration. When the original \loggf\ by Kurucz
are applied, the synthetic CN lines of Arcturus are far too strong 
with respect to the observed ones. Molecular lines are known to be
prone to NLTE effects
(Hinkle \& Lambert,\citealp{hinkle_molecule}; Schweitzer et al.
\citealp{schweitzer}; Plez, \citealp{plez}) and 3D effects \cite[Ivanauskas
et al.][]{ivanauskas}, and their strengths may not be correctly
reproduced under 1D LTE assumption. 
In order to match the strenghts of the CN lines observed on the Sun and on
Arcturus at the same time we needed to set the Arcturus \Cn\ and \Nn\ abundances 
to [C/H]=[N/H]=$-0.34$~dex, which lie between the atomic
abundances by Ram{\'{\i}}rez \& Allende Prieto
\cite{ramirez} who found [C/H]=$-0.09$~dex and [N/H]=$-0.42$~dex
and the ones of Smith et al. \cite{smith} who found [C/H]=$-0.56$~dex and
[N/H]=$-0.28$~dex.\\
We believe that Arcturus' low \Cn\ and \Nn\ abundances found
by us merely counterbalance the 3D NLTE effects that we could not take in account. 
Thus, the CN lines
in the wavelength interval 8400-8924\AA\ are employed by \Space\ as ``dummy"
lines and the results are rejected after the estimation process.


\section{The Equivalent Widths (EW) library}\label{sec_ewlibrary}
We built the EW library using the driver {\it ewfind} of the code MOOG,
which computes the expected EW of the absorption lines for a given stellar
atmosphere model. We employed the atmosphere models grid ATLAS9 
by Castelli \& Kurucz \cite{castelli} updated to the
2012 version. The Castelli \& Kurucz grid has steps
in stellar parameters (500~K in \temp, 0.5 in \logg, and 0.5 in \met) that 
are too wide for our needs. We linearly interpolated the models to obtain a 
finer grid with steps of 200K in \temp, 0.4 in \logg, and 0.2~dex in \met\ and
covering the ranges 3600-7400~K in \temp, 0.2 to 5.4 in \logg\footnote{The
Castelli \& Kurucz grid has a \logg\ upper limit of 5.0. To explore the
$\chi^2$ space around this limit (necessary for a cool dwarf stars that can
have \logg$\sim4.8$, for instance) \Space\ needs to construct spectral models
with \logg$>$5.0. Thus, we extended the EW library to \logg=5.4 by
computing the EW of the lines with a linear extrapolation.}, 
and $-2.4$ to $+0.4$~dex  in \met. In the following we always refers to this grid.
The microturbulence $\xi$ assigned to each atmosphere model is computed
as a function of \temp\ and \logg. This function is described in
Appendix~\ref{appx_microt}.\\ 
Note that in Sec.~\ref{sec_method} we defined
the GCOG as a function of
the three variables \temp, \logg, and [El/H] (and not \met). 
However, to construct the EW library we need the metallicity \met\ of the atmosphere
model. In fact, besides the \temp, \logg, and the abundance [El/H], the EW of a line 
also depends on the opacity of the stellar atmosphere in which the line
forms, which is driven by atmospheric metallicity \met. This means that to compute the GCOG
of a line we must also define the metallicity of the atmosphere model,
making (in this specific case) the GCOG a function of four variables. 
Therefore, we define the stellar parameter grid in the three dimensions \temp,
\logg, and \met\ plus a fourth dimension that accounts for the relative abundance
[El/M].
To construct the EW library, for every point of the grid and every line of our line list we computed
the EW of the lines at 6 different abundance enhancements with respect to
the nominal metallicity of the atmosphere model, that means (for the generic element
El) [El/M]$=-0.4$, $-0.2$, $0.0$, $+0.2$, $+0.4$, and $+0.6$. These 6 points
belong to the COG of the lines for every grid point.
The EW library so constructed contains the EWs of the lines synthesized as they were
isolated. Because \Space\ constructs the spectrum model by summing up the absorption lines
with given EWs, the spectrum model is realistic if the lines are
isolated or, in case of blends, if the EWs of the involved lines are small 
(i.e., weak line approximation). 
Because these conditions are not always satisfied in a real spectrum,
in the following we discuss how to remove the weak line approximation.

\subsection{The weak line approximation problem}
\begin{figure}[t]
\centering
\includegraphics[width=9cm,bb=83 484 277 669]{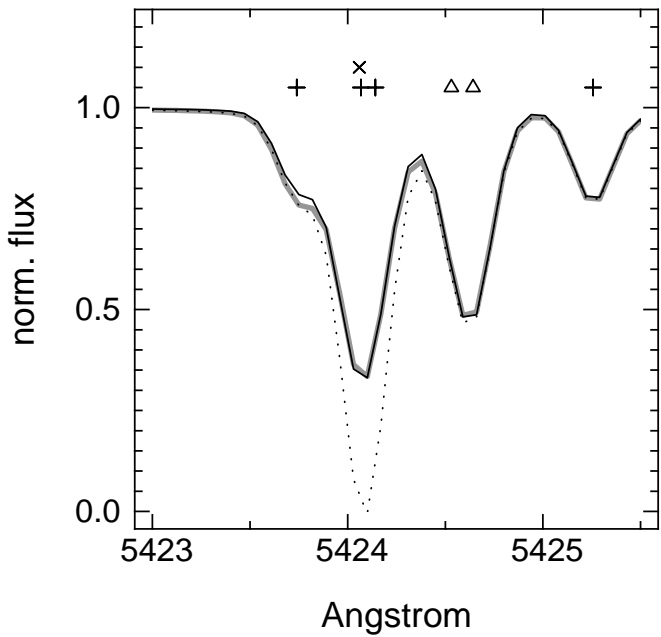}
\caption{Comparison between the synthetic spectrum with stellar parameters 
\temp=4200~K, \logg=1.4, and \met=0.0~dex (the black solid line)
and the corrispondent spectrum model 
(the gray solid line) constructed by \Space\ using the EWs corrected 
for the opacity of the neighbor lines as described in Sec.~\ref{sec_corr_opac}.
The dotted line is the spectrum model constructed using the EW of the lines 
computed as if they were isolated (i.e., no correction for the opacity of
the neighbor lines). Plus, cross, and triangle symbols indicate the 
position of the \Fen, \Vn, and \Nin\ lines, respectively.} 
\label{EW_corr_example}   
\end{figure}

Consider the case of two or more lines that are instrumentally blended but 
physically isolated in a spectrum. We can write\\
\begin{equation}\label{eq_blend_instr}
EW_{tot}=\sum^n_i EW_i,
\end{equation}
where $EW_{tot}$ is the total EW of the blended feature, $EW_i$ are the EWs of
the lines computed as isolated, and $n$ the number of
lines considered.\\ 
Consider now the same lines as before, but now they are physically blended.
If the lines have small EWs, then equation~(\ref{eq_blend_instr}) is still 
(approximately) valid because their line opacity is
small and does not affect the local opacity significantly.
This is what we call weak line approximation. Under these conditions
we can use the $EW_i$ of the lines contained in
the EW library and, 
assuming a line profile, subtract the lines from a normalized continuum to obtain a 
spectrum model which approximates the synthetic spectrum well.
Unfortunately, the weak line approximation can rarely be applied because
strong and broad absorption lines are common in real spectra.
In case of strong lines, equation~(\ref{eq_blend_instr}) is not true anymore, 
because the opacities of the lines diminish reciprocally the flux absorbed by them,
and equation~(\ref{eq_blend_instr}) becomes the inequality\\
\begin{equation}\label{eq_blend_physic}
EW_{blend}<\sum^n_i EW_i
\end{equation}
where $EW_{blend}$ indicates the total equivalent width of the blend and
$EW_i$ is as in equation~(\ref{eq_blend_instr}).
In this case, by summing up the $EW_i$ of the lines contained in the EW
library would render a spectrum model where the blends are too strong with 
respect to the synthetic ones.
This is shown in Fig.~\ref{EW_corr_example} where a blended
feature constructed using equation~(\ref{eq_blend_instr}) (dotted line)
with EWs from the EW library turns out to be much stronger than the
$EW_{blend}$ of the feature synthesized by MOOG (black line).
To correctly
reproduce the blend, the $EW_i$ of the EW library
must be corrected for the opacity of the neighboring lines, so that the
EWs employed to construct the spectrum model are smaller than the
corresponding isolated lines.
These corrected quantities that we call ``equivalent widths corrected 
for the opacity of the neigbouring lines" ($EW_i^c$) are smaller than $EW_i$ 
and satisfy the equation\\
\begin{equation}\label{eq_blend_corr}
EW_{blend}=\sum^n_i EW_i^c.
\end{equation}
The quantity $EW_i^c$ cannot be computed with MOOG. In fact, to know the
quantity $EW^c$ we need to compute the fraction of the contribution
function due to each absorber present in the stellar atmosphere
(continuum, atoms, molecule) that form the blend. These fractions of the contribution
function are not usually computed by spectral synthesis codes. This information is
lost when the spectral synthesis code computes the 
total opacity $\kappa_{\lambda}$ at wavelength $\lambda$ 
by summing up the opacities of all the absorbers to obtain the optical depth
$\tau_{\lambda}$.
The way to compute the fraction of the contribution function is discussed
in Sec.~\ref{sec_contrib_lines} and, although a rigorous solution was found it cannot be
used to correct the EWs of the library. An approximate solution must be
adopted and this is outlined in Sec.~\ref{sec_corr_opac}.

\subsection{Separating the contributions of each 
absorber}\label{sec_contrib_lines}
\begin{figure}[t]
\centering
\includegraphics[width=9cm,bb=89 484 277 669]{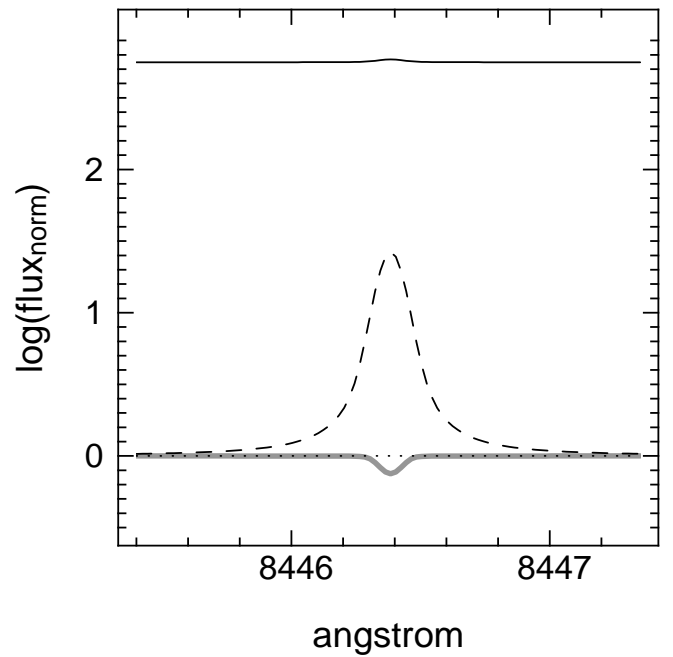}
\caption{Emerging (synthetic) fluxes obtained when opacities of the continuum and the
line are accounted separately.
The gray solid line is the synthetic spectrum of the line \Fei\ at
8446.388\AA, i.e. the emerging flux when the line and continuum opacities are
accounted for. The black solid line represents the emerging flux 
when the emissions alone are accounted for.
The dotted line is the level of the continuum when the continuum opacity
alone is accounted for. The dashed line is the level of the continuum across the
line when both line and continuum opacities are accounted for. 
The y-axis is expressed as logarithm (base
10) of the normalized flux.}
\label{flux_level_8446.388}   
\end{figure}

In the attempt to obtain the quantity $EW^c$,
we tackled and solved the problem to compute the
fractions of the contribution function due to each absorber individually. 
Unfortunately, the result turned out to be inapplicable for our purpose: we can 
compute the flux absorbed by each absorber but this cannot be written in terms of EW. 
To explain this apparent paradox, we here outline the general result
and point the reader to Appendix~\ref{appendix_cf} for the full detailed solution.
Although the problem concerns blends, the simple case of one isolated line is
also illustrative for multiple absorbers like in blends. In fact, in the case of one line
the opacity is due to two absorbers: the continuum and the line.
In Fig.~\ref{flux_level_8446.388} we show an absorption line 
(gray solid line) and the continuum in absence of the absorption line (dotted line).
The EW of this line, as commonly defined, is represented by the area
between the gray and the dotted line. In this way, the EW does not
represent the total flux absorbed by the line, because the continuum (dotted
line) has been computed in absence of the line, i.e. the opacity of the
line has been neglected. When the line opacity is taken in account, 
the continuum level is higher (the dashed line of
Fig.~\ref{flux_level_8446.388}) because it absorbs less radiation. 
In fact, when the line is present, its
opacity diminishes the intensity of the radiation and the continuum absorber is left
with less radiation to absorb. Therefore, the real flux absorbed by the line
is represented by the area included between the gray and the dashed lines of
Fig.~\ref{flux_level_8446.388}, which is much bigger than the EW as
usually defined. This proves that, although we can
precisely determine the real quantity of flux absorbed by a line (in a
synthetic spectrum), we still miss the solution of our problem.
In fact, to reconstruct a spectrum model by summing up
the absorbed fluxes we need to consider the continuum level at any
wavelength. At the stage of development of our work, the variation of the
continuum level as function of the strength of the lines looks too
complicated to be implemented. Therefore we must follow another method to 
approximate the $EW^c$ quantities and apply
equation~(\ref{eq_blend_corr}) to construct the spectrum models.

\begin{figure}[t]
\centering
\includegraphics[width=6cm,angle=0,bb=150 73 630 551]{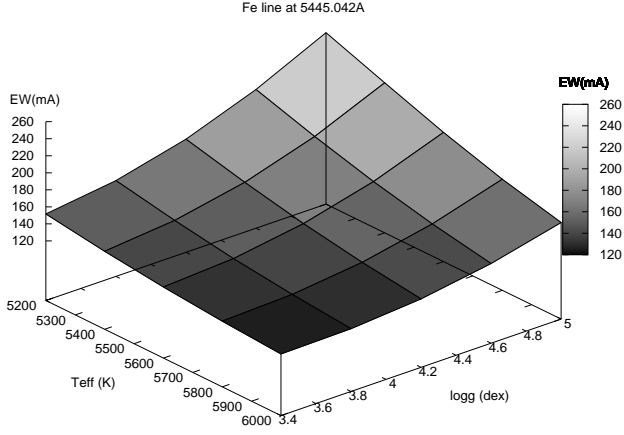}
\caption{Two dimensional section of the General COG of the \Fei\ line at 5445.042\AA\ as a function of
\temp\ and \logg. The \met\ and Fe abundance has been fixed at 0.0~dex.}
\label{ew_5445}   
\end{figure}

\subsection{Approximated correction for the opacity of the neighbor lines}\label{sec_corr_opac}
The method to approximate the quantity $EW^c$ is based on the idea that when
the first derivative of the COG (expressed as EW as function of abundance)
inside a blend is small this means that its contribution to the absorbed flux
(i.e., the EW) is small too, similarly to what happens to the isolated
line.  The method, outlined in the following example, makes use of
EWs of synthesized isolated lines and blended features. 
The EWs of isolated lines are computed with the MOOG driver {\it ewfind}
which numerically intergrates the depression of the synthetic line with
respect to the continuum.  Because this driver does not handle more than one
line per time, to compute the EW of blends we need to synthesize the blend
and numerically integrate the depression\footnote{For a faster procedure we
modified the driver {\it ewfind} to make it handle more than one line per
time.}.\\

Consider one line in a blend composed of two (or more) lines indexed with $i$. The lines belong to
different elements $El_i$. By using MOOG we compute the equivalent widths EWs of 
the lines as isolated for 6 different abundances
[$El_i$/M]=$-0.4$, $-0.2$, 0.0, +0.2, +0.4, +0.6~dex
so that we have six points of the COG of the line (we call it COG$^i_{iso}$). 
Similarly, for every line we synthesize the whole blend and we measure the
total equivalent width of the blend $EW^i_{blend}$.
This is done by synthesizing the blend in which all the lines have constant
[$El_i$/M]=0.0 but for the $i$-th line which assumes six different abundances
[$El_i$/M]. The six EWs of the blend so obtained represent the 
COG of the $i$-th line in the blend (we call them
COG$^i_{blend}$). If the opacity of one line is not affected by the other line,
then COG$^i_{iso}$=COG$^i_{blend}$, otherwise COG$^i_{iso}\ne$COG$^i_{blend}$. In
particular, if the first derivative of COG$^i_{blend}$ is smaller than the
one of COG$^i_{iso}$, it means that the contribution of the
$i$-th line to the absorbed flux of the blend (this is the $EW^c$ quantity we look for) is 
smaller than the one absorbed when the $i$-th line is isolated. 
Thus, the quantity COG$^i_{blend}$ can be used to approximate $EW^{c}$ as
follows:
\begin{enumerate}
\item Compute the first derivatives of the curves-of-growth $\delta COG^i_{iso}$ and
$\delta COG^i_{blend}$.
\item Perform a first correction of $EW^i_{iso}$ as follows\\
$$
EW^{i,c}_{iso}=EW^i_{iso}\cdot\frac{\delta COG^i_{blend}}{\delta COG^i_{iso}}
$$

\item Under the assumption that the ratio of $EW^{i,c}_{iso}$ between the lines is
conserved in the blend, the contribution to the absorbed flux of the  
$i$-th line in the blend is approximated as\\
$$
EW^{i,c}=EW^{i,c}_{iso}\cdot\frac{EW^i_{blend}}{\sum_i EW^{i,c}_{iso}}.
$$
\end{enumerate}

In the general case of a blend of $n$ lines and $m$ elements with $m<n$, there
are two or more lines that belong to the same element $El$. In this case, the
strengths of the $El$ lines would change together when we change the
abundance [$El$/M] during the synthesis of the blend, and this must be avoided
in order to evaluate the COG$^i_{blend}$ of the target line. 
This problem is solved by changing
the \loggf\ (and not the abundance) of the line, so that the target line is
the only line the strength of which changes in the blend.\\
The corrected values $EW^{i,c}$ are computed for all the EWs contained in
the EW library considering any line closer than $\Delta \lambda=0.5$\AA\ 
to the target line. Fig.~\ref{EW_corr_example} shows the improvement
obtained for a blend  when the corrected values $EW^c$s are used to construct the spectrum model
(gray solid line) with respect to the model constructed with EWs (dotted
line).\\
The limit of 0.5\AA\ is satisfactory for most of the lines. For a few intense
and broad lines (for instance, the \Tiii\ at 5226.538\AA), which can affect lines
farther than 0.5\AA, a larger limit would be necessary. At this stage of
development, \Space\ neglects these lines during the analysis.

\begin{figure*}[t]
\centering
\includegraphics[width=18cm,bb=91 342 545 614]{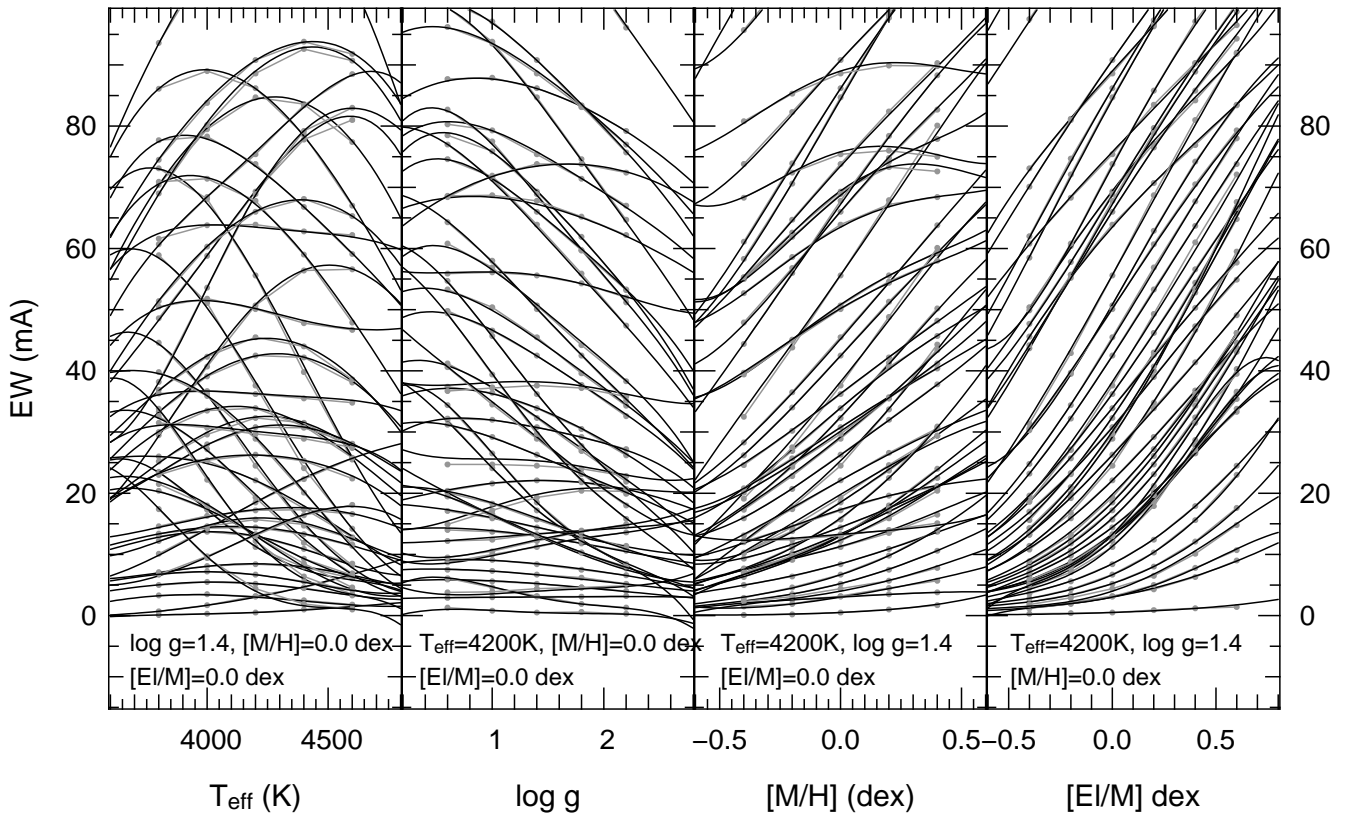}
\caption{One dimensional sections of the GCOGs of several absorption lines in the range 
5212-5222\AA\ as a function of \temp, \logg, \met, and abundance [El/M].
The three (out of four) fixed stellar parameters are reported in the
panels. The gray points connected with gray lines represent the $EW^{i,c}$s of the
lines. The black lines are one dimensional sections of the polynomials GCOGs of the same
absorption lines.} 
\label{GCOG_TGME_sections}   
\end{figure*}

\section{The General Curve-Of-Growth (GCOG) library}
The COG of a line is a function that gives the  
EW of a line as a function of the abundance 
of the element the line belongs to. 
This function can be recovered from the EW 
library where, for each absorption line, we stored six points 
of the COG between the [El/M] values $-0.4$ and $+0.6$~dex in steps of
0.2~dex.
The EW library also contains the EWs that the lines 
assume over a grid spanning a wide range in the stellar parameters \temp,
\logg, and \met. 
Because the EW of a line changes not only as a function of the abundance
but also as a function of \temp\ and \logg, we extend the concept of COG.
We call {\it General Curve-of-Growth} (GCOG) the function that
describes the EW of a line as function of the variables \temp, \logg, and
[El/H], where [El/H] represents the abundance of the generic element El the
line belongs to (see Fig.~\ref{ew_5445}).
Unfortunately the GCOG has no analytical form, therefore, to obtain the EW
of a line we must approximate the GCOG with a polynomial function in the
parameter space.
In principle the GCOG has a three dimensional domain (\temp, \logg,
[El/H]). As already reported in Sec.\ref{sec_ewlibrary}, because we rely on a grid of stellar 
atmosphere models the opacity of which depends on \met, we must construct the polynomials
in a four-dimensional space (\temp, \logg, \met, [El/M]). We refer to these
functions as ``polynomial GCOGs", and they are constructed to approximate the
GCOG of the lines. Thanks to the polynomial GCOGs, \Space\ can compute the
expected EW of any line at any point of the parameter space (\temp, \logg,
\met, [El/H]) removing in this way the discontinuity of the grid in the EW library.

\subsection{The polynomial GCOGs}\label{sec_poly_GCOG}
\begin{figure}[t]
\centering
\includegraphics[width=9cm,bb=24 25 403 239]{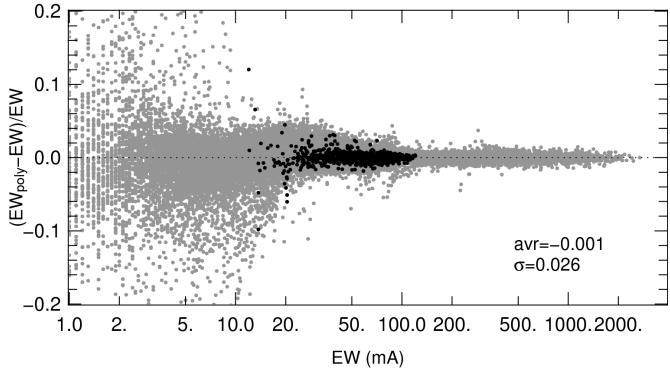}
\caption{{\bf Grey points}: residuals between the EW given by the polynomial GCOG and the
EW of the library as a function of EW for 100 absorption lines (in the range
5212-5235\AA) at the grid point \temp=4200~K, \logg=1.4 and \met=0.0~dex.
{\bf Black points}: as before but for the line \Fei\ at 5231.395\AA\ alone.
The residuals are normalized for the
EW so that the values in the y-axis and the statistic in the panel 
express the errors of the polynomial GCOG normalized to the expected EW.
}
\label{GCOG_residuals}   
\end{figure}

We fit a polynomial GCOG for every absorption line in the parameter space. 
Because of the difficulties in fitting a
function over points covering the whole parameter space, the polynomial
GCOGs fit the EWs that the line assumes over a limited stellar parameter
interval surrounding the points of the grid. The width of this interval is 800~K in \temp, 1.6 in \logg,
0.8~dex in \met, and 1.0~dex in [El/M], which includes five grid points for the
first three dimensions and six for the last dimension. 
For instance, for the grid
point \temp=4200~K, \logg=1.4, and \met=0.0 the polynomial GCOG fits the EWs
that the line has at \temp=3800, 4000, 4200, 4400, and 4600~K, \logg=0.6, 1.0,
1.4, 1.8, and 2.2, \met=$-0.4$, $-0.2$, 0.0, $+0.2$, and $+0.4$~dex, and the six
abundance points [El/M]=$-0.4$, $-0.2$, 0.0, $+0.2$, $+0.4$, and $+0.6$~dex 
(Fig.~\ref{GCOG_TGME_sections}). In total, every polynomial GCOG
fits 750 EWs. The polynomial GCOG function has the form\\
\begin{equation}\label{poly_GCOG}
EW_{poly}=\sum_{i,j}^{i+j \leq 4} a_{ij} (\mbox{\temp})^i (\mbox{\logg})^j.
\end{equation}
So defined, the polynomial GCOG has 70 coefficients $a_{ij}$ that are computed by using a 
minimization routine that minimizes the $\chi^2$ between the polynomial and the
given EWs. The residuals between the $EW_{poly}$ (given by the polynomial
GCOG)  and the EWs of the library are shown in Fig.~\ref{GCOG_residuals}.
The residuals are on average 2.6\% of the expected EW, which is 
equivalent to an error of $\sim 0.01$~dex in chemical abundance.
Fig.~\ref{GCOG_TGME_sections} shows the polynomial GCOG of 50 absorption
lines compared with the expected EW plotted as a function of \temp,
\logg,\ \met, and abundance [El/M]. 

\section{The \Space\ code}\label{sec_space_code}
In the following we outline the main structure of the code. This is not intended
to be a user manual. A detailed tutorial on how to use \Space\ and the full
description of the available functionalities of the code is provided
together with the code.\\
The \Space\ code is written in FORTRAN95. It processes one spectrum per
run. The observed spectrum must be wavelength calibrated, continuum
normalized and radial velocity corrected.
When launched, \Space\ reads the parameter file that must include the name
of the spectrum to process, the address of the GCOG library, a first guess
of the FWHM, and other optional settings. 
In the following we
outline the algorithm that summarized the \Space\ analysis procedure
specifying the most important routines that we explain later.
This algorithm carries out the following steps:\\
\begin{enumerate}
\item Upload the observed spectrum.
\item Make a first rough estimation of the stellar parameters \temp, \logg,
and \met. (This is performed by the ``starting
point routine".)
\item Find the closest grid point to the estimated \temp, \logg, and \met\
and upload the corresponding polynomial GCOG.
\item Derive \temp, \logg, and \met. (This is performed by the TGM routine.) 
\item Find the closest grid point to the derived \temp, \logg, and \met.
If it is different from the previous grid point, then upload the polynomial 
GCOG of the new grid point and go to step~4, otherwise continue.
\item Re-normalize the observed spectrum. (This is performed by the
re-normalization routine.)
\item Derive \temp, \logg, \met\ from the re-normalized spectrum. (This is performed by the TGM routine.) 
\item Find the closest grid point to the estimated \temp, \logg, and \met.
If it is different from the previous grid point, then upload the polynomial 
GCOG of the new grid point and go to step~6, otherwise continue.
\item Derive the chemical abundances [El/M]. (This is performed by the ABD routine.) 
\item Go to step~6 and repeat until convergence.
\item Derive the confidence limits for \temp, \logg, \met, and [El/M]
(optional)
\item End the process and write out the results.
\end{enumerate}
Every step is composed of routines and sub-routines. The
most important ones are described in the following.\\
These algorithm can be executed with or without step~10. 
This is controlled by the keyword {\it ABD\_loop} (abundance loop) that can be
used by \Space. When {\it ABD\_loop} is switched on, step~10 is executed,
otherwise it is skipped. 
The two settings show significant differences when run on real 
and synthetic spectra. This is discussed in Sec.\ref{sec_validation}.

\subsection{The ``make model" routine}\label{sec_makemodel}
To derive the stellar parameters and the chemical abundances, 
\Space\ constructs several spectrum models and compares them to the observed spectrum,
looking for the model that renders the minimum $\chi^2$.
The routine that constructs the model (called the ``make model" routine) is
therefore particularly important and it follows this algorithm:\\
\begin{enumerate}
\item Set the initial spectrum model with the same number of
pixels and wavelengths of the observed spectrum and initial flux normalized
to one. We call it the ``working model".
\item Consider the stellar parameter with which the model must be constructed.
\item Consider the first absorption line of the line list.
\item Compute the $EW^c$ of the absorption line by using
its polynomial GCOG.
\item Compute the strength of the line profile at every pixel around the center of the
line and subtract it from the working model. The result is the
new working model.
\item Consider the next line and go to step~4 until the last
line has been reproduced.
\end{enumerate}
The line profile adopted is a Voigt function approximated with the
implementation by McLean et al. \cite{mclean}. We modified this
implementation so that the line profile becomes broader as a
function of \logg\ and EW with a law that can be different for some special lines 
(for instance, lines with large damping constants). 
For a detailed description of the line profile adopted we refer the reader to
Appendix~\ref{app_voigt}.

\subsection{The ``starting point" routine}
This routine finds the first rough estimation of the stellar parameters.
It uses the ``TGM routine" outlined in the next section, with the
difference that the polynomial GCOG employed has been computed not over a
small volume of the parameter space (as explained in Sec.~\ref{sec_poly_GCOG})
but over the whole parameter space. This polynomial GCOG has larger errors
with respect to the other polynomials contained in the GCOG library, but it
permits a rough and fast estimation of the parameters, which is used as
starting point by the next TGM routine. 

\subsection{The ``TGM routine"}
This part of the code is responsible for deriving the stellar parameters.
It employs the Levenberg-Marquadt method to minimize the $\chi^2$ between
the models and the observed spectrum in the parameter space (\temp, 
\logg, \met). At the fourth 
step of the main algorithm, the TGM routine uses the observed spectrum as 
provided by the user, while at the seventh step it uses the observed spectrum 
after the re-normalization (explained in section~\ref{sec_renorm}). 
Because the polynomial GCOGs were computed over an
interval of stellar parameters that covers 800~K in \temp, 1.6 in \logg, and
0.8~dex in \met\ (see Sec.\ref{sec_poly_GCOG}) centered on a grid point (we
call it the ``central point"), its reliability decreases with the distance
from the central point.
When the stellar parameters given by the TGM routine are close to a grid
point that is not the central point, the TGM routine stops,
uploads the polynomial GCOG for this new grid point, and repeats. 
In this way the routine always finds the minimum $\chi^2$ close to the
central point, where the EWs provided by the polynomial GCOG are the most
reliable.
For some spectra (e.g., spectra with very low S/N, spectra of very cool/very 
hot stars, or spectra of stars with high rotational velocity \Vrot)
the TGM routine may try to move beyond the extension of the GCOG library. In
this case \Space\ writes a warning message and exits with no results.\\
Apart from the stellar parameters, the TGM routine also estimates two other
parameters:  the radial velocity and the FWHM of the instrumental profile. 
Because \Space\ only processes radial-velocity-corrected spectra, the radial
velocity estimation of \Space\ is not a real measurement, but it is an internal setting
to improve the match between the model spectrum and the observed spectrum.
Usually this quantity amounts to a small fraction of FWHM.
Similarly, \Space\ searches for the FWHM that matches best the instrumental
profile.
The optimization of the FWHM gives to \Space\ some flexibility in estimating
the stellar parameters for stars with a rotational velocity \Vrot\ different
from zero. However, the Voigt profile adopted by
\Space\ cannot properly fit the shape of the lines of stars with high \Vrot,
and the \Vrot\ limit beyond which the line profile becomes inadequate depends
on the spectral resolution. This limit is higher for low-resolution spectra
in which the instrumental line profile dominates over the physical profile of
the line. 

\subsection{The re-normalization routine}\label{sec_renorm}
\begin{figure}[t]                                                            
\centering                                                                   
{\includegraphics[width=9cm,bb=66 287 503 562]{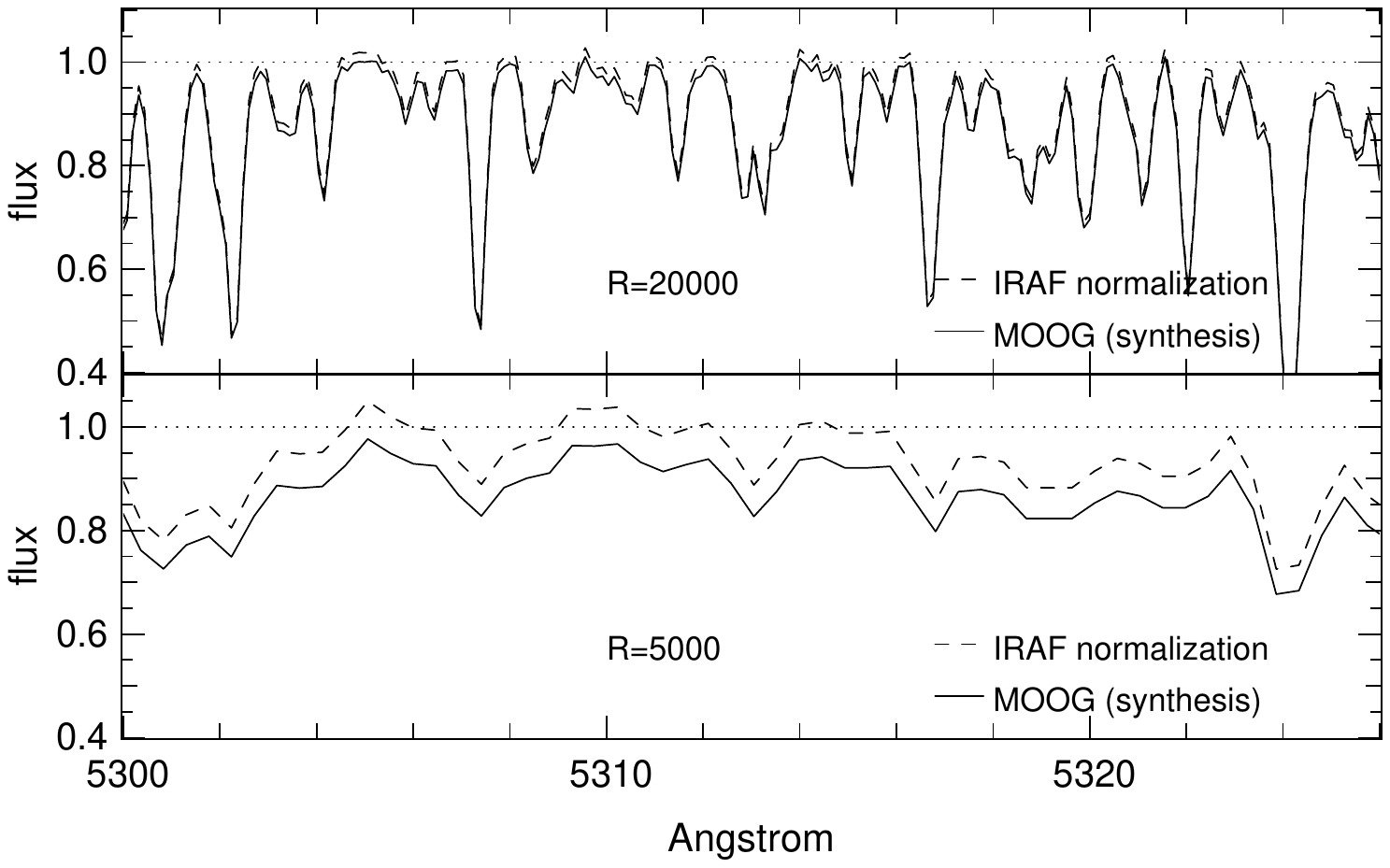}}
\caption{Synthetic spectrum with the stellar parameters \temp=4212~K, logg=1.8,
\met=0.0~dex, and S/N=100 at resolution R=20\,000 (top) and R=5\,000
(bottom). The black line is the correctly normalized spectrum (synthesized by
MOOG and noise added),
while the dashed line is the same spectrum after normalization with the task
{\it continuum} of IRAF.}
\label{comp_spectra_SN100}
{\includegraphics[width=9cm,bb=69 286 503 562]{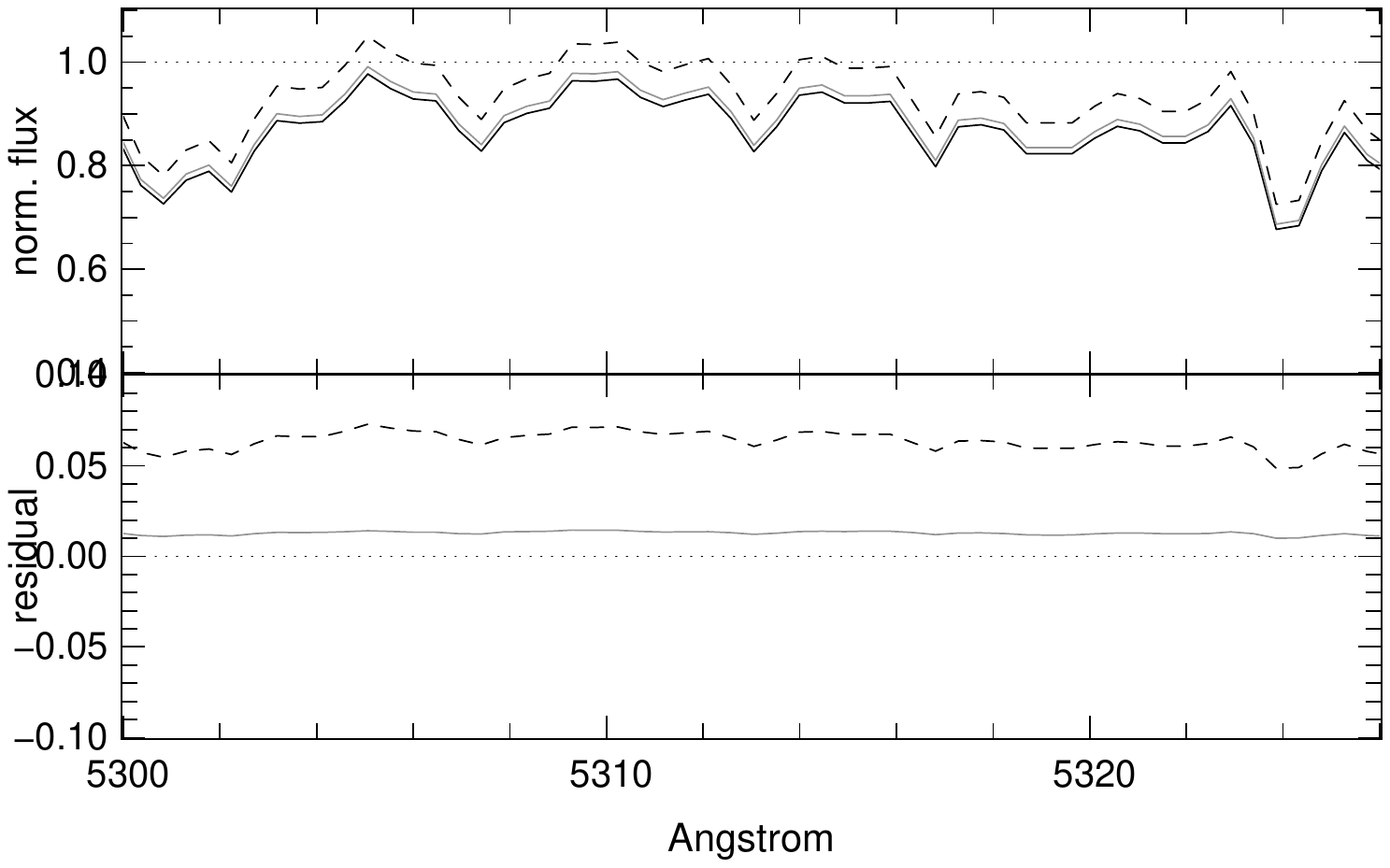}}
\caption{Comparison between the IRAF and \Space\ normalization. 
{\bf Top}: Solid and dashed black lines are as in bottom panel of
Fig.~\ref{comp_spectra_SN100}. The gray line was obtained by processing the
dashed line (IRAF normalized spectrum) with the \Space\ re-normalization 
routine. {\bf Bottom}: residuals between the IRAF
normalized (dashed line) and \Space\ normalized (gray line) spectra with
respect to the correctly normalized spectrum.}
\label{plot_normaliz}
\end{figure}

As stressed before, \Space\ can only handle flux-normalized spectra.
However, it can perform a re-normalization to adjust the continuum
level. This operation may be unnecessary 
for high-resolution and high-S/N spectra, where the continuum is 
clearly detectable and a normalization done with the commonly used IRAF task 
{\it continuum} is usually satisfactory.
For low-resolution, low-S/N spectra, and in particular for spectra crowded 
of lines, the continuum cannot be clearly identified. At low resolution, the
lines are instrumentally blended and they create a pseudo-continuum that can lie
under the real continuum. In this case, the IRAF task {\it continuum} cannot
correctly estimate the continuum and renders a too low continuum level, leading to an
underestimation of the metallicity (as well as the other stellar parameters, 
since they are correlated). As an example, in Fig.~\ref{comp_spectra_SN100} we show
a low- and a high-resolution spectrum (synthetic) of a high-metallicity star, and the result of the
continuum normalization performed with the IRAF task {\it continuum}, using
a spline function and {\it low\_rej=1} and {\it high\_rej=4}, settings that
take the presence of absorption lines in the spectra in account. 

Because of the instrumental blend of the lines, the low-resolution spectrum
suffers of a too low continuum estimation and its normalized flux is too high.
When the same {\it continuum} settings are used for very low S/N spectra
the normalized spectra suffer the opposite problem: 
the noise dominates the spectrum and the flux distribution becomes nearly
symmetric with respect to the continuum. 
(This commonly happens in spectroscopic surveys when the number of spectra 
to normalize is big and the parameters of the task {\it continuum} 
cannot be set by hand for every spectrum).
In this case, the settings {\it low\_rej=1} 
and {\it high\_rej=4} are not appropriate and cause a too high estimated continuum
(and an overestimation of the metallicity).
To fix this problem, \Space\ re-normalizes the observed spectrum.
In the following, $f_{obs}$ indicates the normalized flux of the
observed spectrum and $f_{model}$ indicates the normalized flux of the spectrum
model. The re-normalization routine works as follows:\\
\begin{enumerate}
\item Consider the $i$-th pixel at wavelength $\lambda(i)$ and the
$n$ pixels for which $\lambda-\lambda(i)<30\AA$
\item Compute the average of the observed flux $\overline{f_{obs}}$, the average
of the residuals $\overline{(f_{obs}-f_{model})}$  and their standard deviation
$\sigma_{res}$ of the $n$ pixels defined before.
\item From the set of $n$ pixels defined in step~1, reject the pixels with
$\overline{f_{obs}}\leq(\overline{f_{obs}}-2\cdot\sigma_{res})$. The new set of
pixels has now $m\leq n$ number of pixels.
\item With the new set of $m$ pixels, compute the new average of the observed 
flux $\overline{f_{obs}}$, the average of the model
$\overline{f_{model}}$.
\item Compute the continuum level at the $i$-th pixel as
$$
cont(i)=1.+(\overline{flux_{obs}}-\overline{flux_{model}})
$$ 
\item Re-normalize the $i$-th pixel of the observed spectrum as
$$
f_{renorm}(i)=\frac{f_{obs}(i)}{cont(i)}
$$
\item Move to the next $i$-th pixel and go to step~1 until all the pixels
have been processed.
\end{enumerate}

In Fig.~\ref{plot_normaliz} we show the synthetic spectrum at R=5\,000 (also seen
in the bottom panel of Fig.~\ref{comp_spectra_SN100}) together with
the IRAF normalized spectrum and the spectrum after the
re-normalization. The re-normalization routine employed by \Space\ can greatly
decrease the offset caused by the IRAF normalization.

\subsection{The ABD routine}
The routine to derive the chemical abundances (called ``the ABD routine")
works similarly as the TGM routine. The ABD routine is run after the TGM
routine. The abundances [El/M] are varied by a minimization routine until
the $\chi^2$ between the model and the observed spectrum is minimized.

\subsection{Internal errors estimation}\label{sec_error_est}
\Space\ can estimate the expected errors for the parameters \temp, \logg, and
the chemical abundances [El/M]. The routine dedicated to this task finds the
confidence limits of the stellar parameters intended as the region of the parameter
space {\it that contains a certain percentage of the probability distribution
function} \cite[Press et al.][]{nr}. If we want to determine the extension of the region
that has a 68\% of probability to include the resulting parameter (say
\temp$^{best}$) with the lowest $\chi^2$ ($\chi^2_{best}$), this region is an interval 
that has an upper and a lower limit \temp$^{up}$
and \temp$^{low}$ with $\chi^2=\chi^2_{best}+1$. Because
the stellar parameters are correlated, the confidence limits of one
parameter are a function of the others, so that the upper and lower confidence limits of
\temp\ as a function of \logg\ correspond to the largest and smallest values
of \temp\ with $\chi^2=\chi^2_{best}+\Delta\chi^2$ where $\Delta\chi^2$
depends on the number of degrees of freedom \cite[Press et al.][]{nr}. 
The higher the number of degrees of freedom, the larger the confidence limits. 
Because the determination of
these limits is computationally expensive, we limited this determination
to three variables, namely \temp, \logg, and \met\ for these three stellar
parameters, and \temp, \logg, and [El/M] for the
chemical abundance of the generic element El. This means that we determine the upper and
lower limits of any parameter at $\chi^2=\chi^2_{best}\pm3.53$. 
These confidence limits must be considered as internal errors (therefore smaller than the real
errors) because they do not take external errors into account like 
the mismatch between the atmosphere model and the real stellar atmosphere, 
uncertainties in the atomic transition probability, in the continuum
placements and other uncertainties in the spectrum model construction.\\
The error estimation is an option left to the user because it is computationally
expensive: when done, it can easily double (or more) the time required to process 
the same spectrum without error estimation.

\subsection{Output results}
At the end of the process, \Space\ writes four output files. One file called
``space\_TGM\_ABD.dat" contains the resulting stellar parameters \temp, \logg, and chemical
abundances with their confidence intervals, the number
of lines measured, and a few other parameters like the $\chi^2$ of the best 
matching spectrum model, internal RV and FWHM. The second output file
called ``space\_model.dat" contains a table the columns of which correspond to i)
the pixel wavelength of the observed spectrum, ii) the flux of the observed
spectrum,
iii) the flux of the observed spectrum after re-normalization, iv) the flux of the 
best matching model,  v) the continuum level adopted for the
re-normalization, and vi) the
weights of the pixels (rejected pixels have weight=0).  The third output file
``space\_ew\_meas.txt" contains the EW employed by \Space\ to construct
the model. We want to stress that {\em these are not the EWs of the
absorption lines} but merely the $EW^{c}$ (EW corrected for the opacity of the
neighboring lines) computed from the polynomial GCOG during
the construction of the best matching model.
The fourth output file
``space\_msg.txt" contains the warning messages generated when something goes wrong 
during the analysis.

\begin{table*}[th]
\caption[]{Ages, iron abundance ranges, and coefficients $m$ and $q$ of the
linear law $[El/Fe]=m\cdot[Fe/H]+q$ that expresses the chemical abundances 
used to synthesize the spectra of the three mock stellar populations.}
\label{tab_three_pop}
\vskip 0.3cm
\centering
\begin{tabular}{ll|c|cc|cc|cc|cc}
\hline
\noalign{\smallskip}
mock & age & [Fe/H] range & \multicolumn{2}{c|}{\Cn, \Nn, \On} &
\multicolumn{2}{c|}{\Mgn} & \multicolumn{2}{c|}{\Aln, \Sin, \Can, \Tin} &
\multicolumn{2}{c}{other elements}\\
population   &(Gyr)& (dex)        & $m$  &   $q$    & $m$ & $q$ & $m$  & $q$ &
$m$ & $q$\\

\noalign{\smallskip}
\hline
\noalign{\smallskip}
thin disk stars& 5 & $0.0\leq$[Fe/H]$\leq+0.2$& 0.0 & 0.0 & 0.0 & 0.0 & 0.0 & 0.0 & 0.0
& 0.0\\
         &   & $-0.8<$[Fe/H]$<0.0$& $-0.4$ & 0.0 & $-0.3$ & 0.0 & $-0.15$
& 0.0 & 0.0 & 0.0\\
\hline
\noalign{\smallskip}
halo/thick disk stars& 10 & $-1.0<$[Fe/H]$<-0.2$& $-0.5$ & $+0.1$ & $-0.3$ &
 $+0.1$ & $-0.2$ & $+0.1$ & 0.0 & 0.0\\
               &      & $-2.2\leq$[Fe/H]$\leq-1.0$& $0.0$ &$+0.6$ & $0.0$ & $+0.4$ &
$0.0$ & $+0.3$ & 0.0 & 0.0\\
\hline
\noalign{\smallskip}
accreted stars & 10 & $-2.2\leq$[Fe/H]$\leq-1.0$& 0.0 & 0.0 & 0.0 & 0.0 & 0.0
& 0.0 & 0.0 & 0.0\\
\noalign{\smallskip}
\hline
\end{tabular}
\end{table*}

\section{Validation}\label{sec_validation}
To establish the precision and accuracy of the stellar parameters and
chemical abundances derived by \Space, we run the code on several sets of 
synthetic and real spectra with well known parameters and compare them with
the parameters derived with \Space. We test \Space\ on the wavelength ranges
5212-6270\AA\ and 6310-6900\AA, which avoids the range 6270-6310\AA\ where
the presence of telluric lines can affect the analysis. The tests are
performed on spectra with spectral resolution between 2\,000 and
20\,000\footnote{At R$\lesssim$20\,000 the line profile implemented in \Space\ 
is expected to work best. See Appendix~\ref{app_voigt} for a short discussion
on the accuracy of the line profile as a function of the spectral
resolution.}.\\
Before the presentation of these tests, we illustrate and discuss the
accuracy with which \Space\ constructs the spectrum models (i.e. how close these
models match the synthetic spectra from which they are derived).

\subsection{Spectrum models accuracy}
\begin{figure}[t]
\centering
{\includegraphics[width=9cm,bb=64 290 505 686]{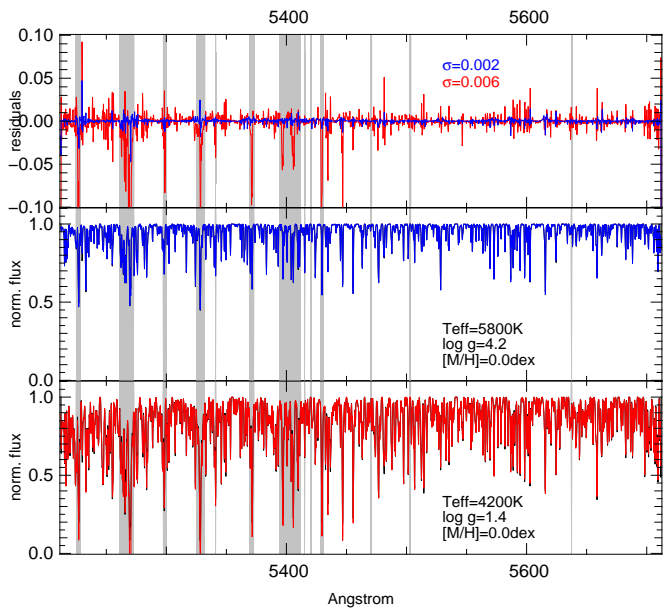}}
\caption{Comparison between spectra models and the corrispondent synthetic
spectra. The black lines are synthetic spectra of a cool giants stars
(bottom panel) and a warm dwarf stars (middle panel) with stellar parameters
as reported in the panels. The overplotted colored lines represent the
spectrum models constructed by \Space\ for the corrispondent giant (red
line) and dwarf (blue line) star. The shaded areas indicate the wavelength
ranges rejected during the \Space\ analysis (see text for more details). 
The residuals (model minus synthetic spectra) are reported in the top panel, together with the standard
deviation of the residuals for the dwarf and giants spectra (in blue and red
colors, respectively) after the exclusion of the gray shaded areas. The
color version of this plot is available in the electronic edition.}  
\label{test_reconstr}
\end{figure}

In Sec.~\ref{sec_makemodel} we outlined the algorithm that constructs the
spectrum model which must be compared with the observed spectrum. Our goal
was to make a spectrum model that looks as close as possible to a synthetic spectrum.
To evaluate the accuracy with which the spectrum models constructed by
\Space\ match the corresponding synthetic spectra, we synthesized two
spectra  and
compared them with the corresponding spectrum models constructed with the
same stellar parameters. The goodness of the match illustrates the precision with
which the strength of the lines (encoded in the EW library first, and then
in the GCOG library) and the line profile adopted, can reproduce a realistic
spectrum model. We chose to synthesize the spectra of a dwarf star 
(\temp=5800~K, \logg=4.2, \met=0.0~dex) and of a giant star (\temp=4200~K, \logg=1.4, and
\met=0.0~dex) degraded to a spectral resolution of R=12\,000\footnote{The
accuracy with which the model spectra match the synthetic ones can change
with spectral resolution and stellar parameters. For the sake of brevity, we
only present here two spectra as exemplary cases.}. The
relatively high metallicity adopted generates spectra rich in lines, allowing 
us to verify how well \Space\ can reproduce blended features (more numerous in
spectra of giants) and
how good it fits the profile of strong lines (usually broader in dwarf stars).
In the case of blended features we test the correction for the opacity of
the neighboring lines applied to the EW library (Sec.~\ref{sec_corr_opac}),
for isolated strong and weak lines we verify the goodness of the Voigt profile adopted (described in
Appendix~\ref{app_voigt}). In general, any line is affected by the precision 
with which the polynomial GCOGs represent the expected EWs.
The comparison between models and synthetic spectra is shown in
Fig.~\ref{test_reconstr}. The top panel of the figure shows that the
normalized flux of the models differs by no more than 1\% for most of the
wavelengths, with a general standard deviation $\sigma$ of 0.2\% and 0.6\%
for the dwarf and the giant, respectively. This statistic was computed after
the rejection of the gray shaded areas, which were rejected during the
analysis because in the case of real spectra they are affected by unidentified lines, lines with NLTE effects, or
lines for which the correction for the opacity of the neighboring lines is not
satisfactory (see last paragraph of Sec.~\ref{sec_corr_opac}).\\
Although not perfect, the spectral models match the corresponding synthetic 
spectra with a satisfactory degree of accuracy .

\subsection{Tests on synthetic spectra}
To verify the ability of \Space\ to distinguish the stellar parameters and chemical abundances of
different Milky Way stellar populations of different ages, metallicity, and
evolutionary
stages, we synthesized the spectra of three mock stellar populations with
characteristics that mimic the thin disk stars, the halo/thick disk stars, and
accreted stars with non-enhanced $\alpha$ abundances (a dwarf galaxy accreted by the Milky
Way, for instance). All the synthetic spectra were synthesized with
MOOG, adopting the final line list described in Sec.~\ref{sec_final_ll} and
atmosphere models from the grid ATLAS9 by Castelli \& Kurucz \cite{castelli}
(updated to the 2012 version) linearly interpolated to the wanted stellar parameters.

\subsubsection{Construction of the synthetic mock populations}
We construct a mock sample for a total number of 1200 spectra for the three
populations (300, 600, and 300 spectra for
the thin, halo/thick, and accreted populations, respectively), 
randomly chosen from the PARSEC isochrones (Bressan et al.
\citealp{bressan} complemented by Chen et al. \citealp{chen_bressan}) 
to cover the stellar parameter range 3600 to 7500K in \temp, 0.2 to 5.0 in
\logg, and $-2.0$ to $+0.3$~dex in [Fe/H].
In this way, the mock sample covers uniformly the chosen isochrones.\\
Their chemical abundances were chosen with the following characteristics:\\
\begin{itemize}
\item {\bf mock thin disk stars}: they cover the iron abundance ranges
$-0.8\leq[Fe/H]{ \mbox(dex)}<+0.3$ and their \temp\ and \logg\ were taken
from
isochrones with an age of 5Gyr. Their $\alpha$-elements enhancement [El/Fe]
becomes progressively higher for lower [Fe/H]. (In the next plots this
population is represented by blue points.)
\item {\bf mock halo/thick disk stars}: they cover the iron abundance
ranges $-2.0\leq[Fe/H]{\mbox(dex)}<-0.2$ and their \temp\ and \logg\ were
taken from the
isochrones with an age of 10Gyr. Their $\alpha$-element enhancement [El/Fe]
becomes progressively higher for lower [Fe/H] down to $[Fe/H]=-1.0$~dex, and
stays constantly high ($\sim+0.4$) for $[Fe/H]<-1.0$~dex. 
(In the next plots this
population is represented by red crosses.)
\item {\bf mock accreted stars}: they cover the iron abundance
ranges $-2.0\leq[Fe/H]{\mbox(dex)}<-1.0$ and their \temp\ and \logg\ were
adopted from
isochrones of an age of 10Gyr. Their $\alpha$-element enhancements [El/Fe]
are equal to zero. (In the next plots this
population is represented by dark green triangles.)
\end{itemize}

\begin{figure}[t]
\centering
{\includegraphics[width=9cm,bb=101 288 359 455]{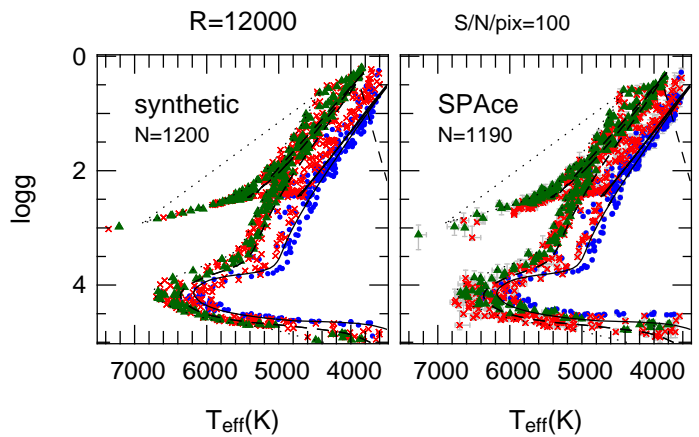}}
\caption{Distribution of the three mock populations on the \temp\ and 
\logg\ plane as synthesized (left panels) and derived by \Space\ (right
panel). The blue points, red crosses, and green triangles represent the
thin, halo/thick disc, and accreted stars, respectively.
The solid, dashed, and dotted black lines show isochrones at
[M/H]=0.0~dex and 5~Gyr, [M/H]=$-1.0$~dex and 10~Gyr, and [M/H]=$-2.0$~dex and
10~Gyr, respectively. The light gray errorbars represent the confidence
intervals of the individual measurements. A missing
errorbar indicates that the error is larger than the parameter grid. The
color version of this plot is available in the electronic edition.}
\label{TG_R12_SN100_synt_ABDloop}
\end{figure}
\begin{figure*}[t]
\begin{minipage}{18cm}
\centering
{\includegraphics[width=14cm,bb=85 286 531 524]{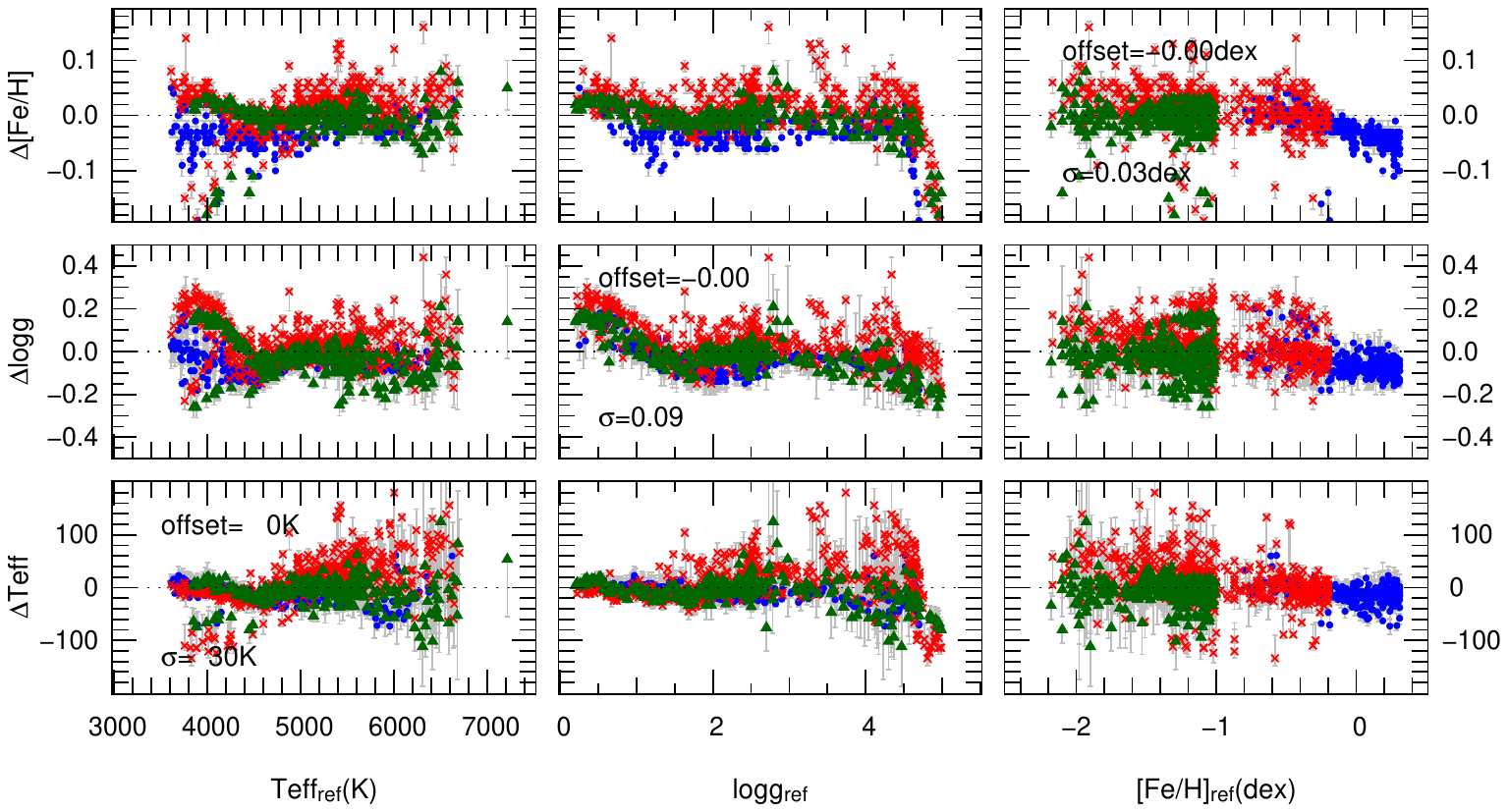}}
\caption{Residuals between derived and reference parameters (y-axis) as a
function of the reference parameters (x-axis). Symbols and colors are as in
Fig.\ref{TG_R12_SN100_synt_ABDloop}. The color version of this plot is
available in the electronic edition.}
\label{correl_R12_SN100_synt_ABDloop}
\vskip 0.3cm
\centering
{\includegraphics[width=14cm,bb=85 286 531 524]{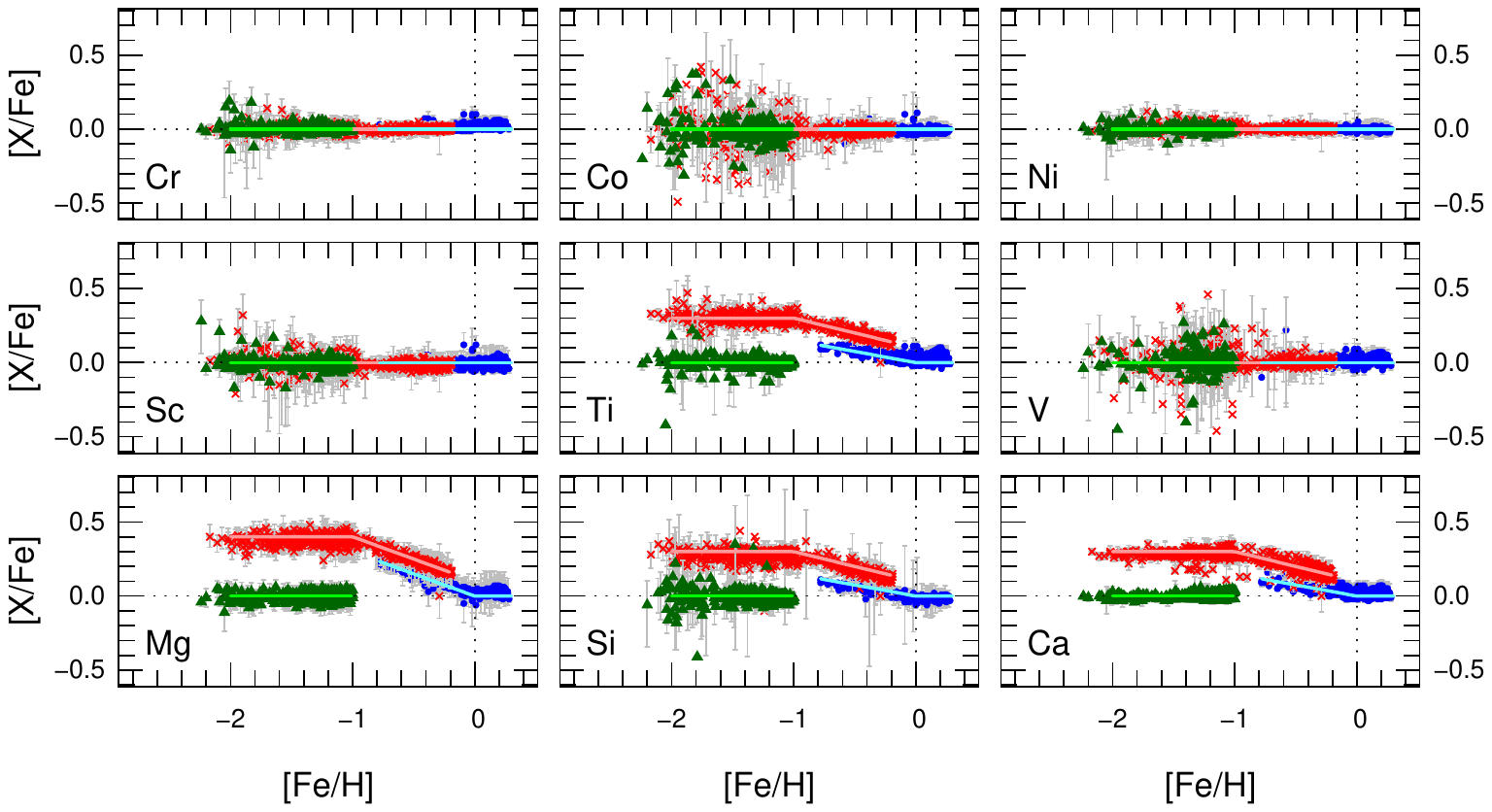}}
\caption{Chemical abundances derived by \Space\ for the three mock
populations. Symbols and colors are as in Fig.\ref{TG_R12_SN100_synt_ABDloop}. The
reference abundances of these three populations are represented by the 
light blue, light red, and light green solid lines for the thin disc, the 
halo/thick disc, and accreted populations, respectively. The color version
of this plot is available in the electronic edition.}
\label{XFe_R12_SN100_synt_ABDloop}
\end{minipage}
\end{figure*}

More precisely, the abundance of the generic element $El$ follows 
a linear law which can be expressed as $[El/Fe]=m\cdot[Fe/H]+q$, 
where $m$ and $q$ have different values for different [Fe/H] intervals 
and elements. The exact $m$ and $q$ values for each element
are listed in Tab.~\ref{tab_three_pop}. The distributions of these
populations in \temp\ and \logg\ are shown in the left panel of 
Fig.~\ref{TG_R12_SN100_synt_ABDloop}.
The samples were degraded to resolutions of R=20\,000, 12\,000, 5\,000, and
2\,000 and to signal-to-noise ratios of S/N=100, 50, 30, and 20 (by
adding Poissonian noise) for a total amount of
19\,200 spectra. The stellar parameters and abundances of these spectra were
derived with \Space\ and compared with the expected ones. 
The measurements were performed with the keyword {\it ABD\_loop} on.
We switched off the internal
re-normalization to evaluate the goodness of the GCOG library to provide the
right EW of the lines and the ability of \Space\ to reproduce the
correct line profile of the absorption lines.

\begin{figure}[t]
\centering
\includegraphics[width=9cm,bb=62 286 508 691]{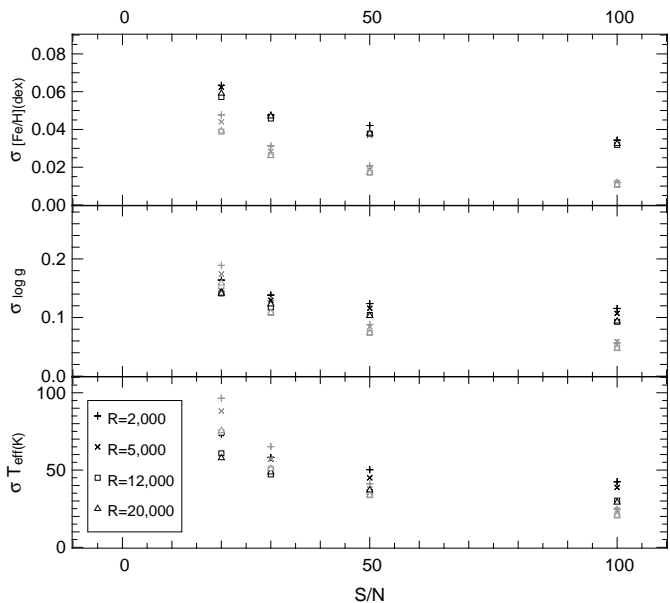}
\caption{{\bf Black symbols}: standard deviations of the residuals of the stellar parameters
(expressed as estimated minus reference values) as a function of S/N
for the four different resolutions considered here. {\bf Grey symbols}: as the black
symbols but the y-axis expresses the average semi-width of the estimated confidence
interval. The black and gray symbols would match if the error distributions
were Gaussian and there were no systematic errors.}
\label{internal_errors_TGM_R}
\end{figure}
\begin{figure}[t]
\centering
\includegraphics[width=8cm,bb=66 191 470 641]{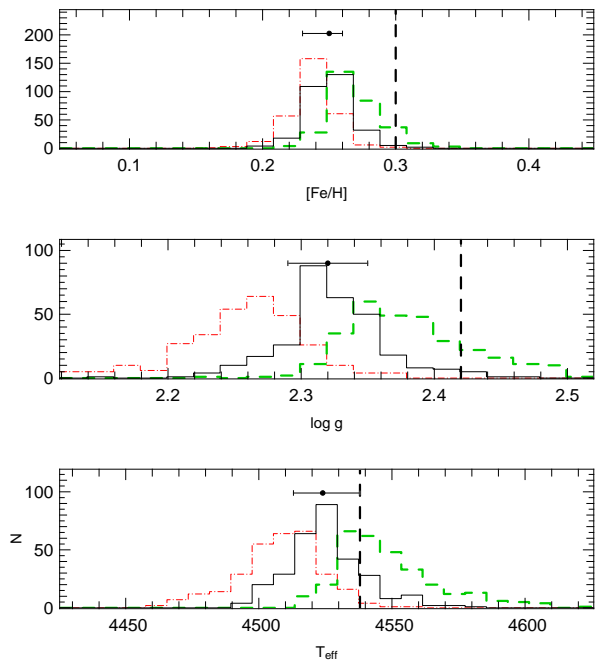}
\caption{Distributions of the derived stellar parameters (black solid histogram),
and the lower and upper limits of the confidence intervals (dotted-dashed red 
histogram and thick dashed green histogram, respectively) as computed in
Sec.~\ref{sec_error_est} for 100 Monte Carlo realizations 
of the synthetic spectrum with \temp=4538~K, 
\logg=2.42, [Fe/H]=0.30~dex, S/N=100, and R=12\,000. The black
point represents the median of the derived stellar parameters
while the errorbars show the lower and upper limit of the interval that
holds the 68\% of the measurements. The vertical dashed lines represent the 
reference values. The color version of this plot is available in the
electronic edition.}
\label{plot_bressan_MC_TGM}
\end{figure}
\begin{figure}[ht]
\centering
{\includegraphics[width=8cm,bb=70 287 505 689]{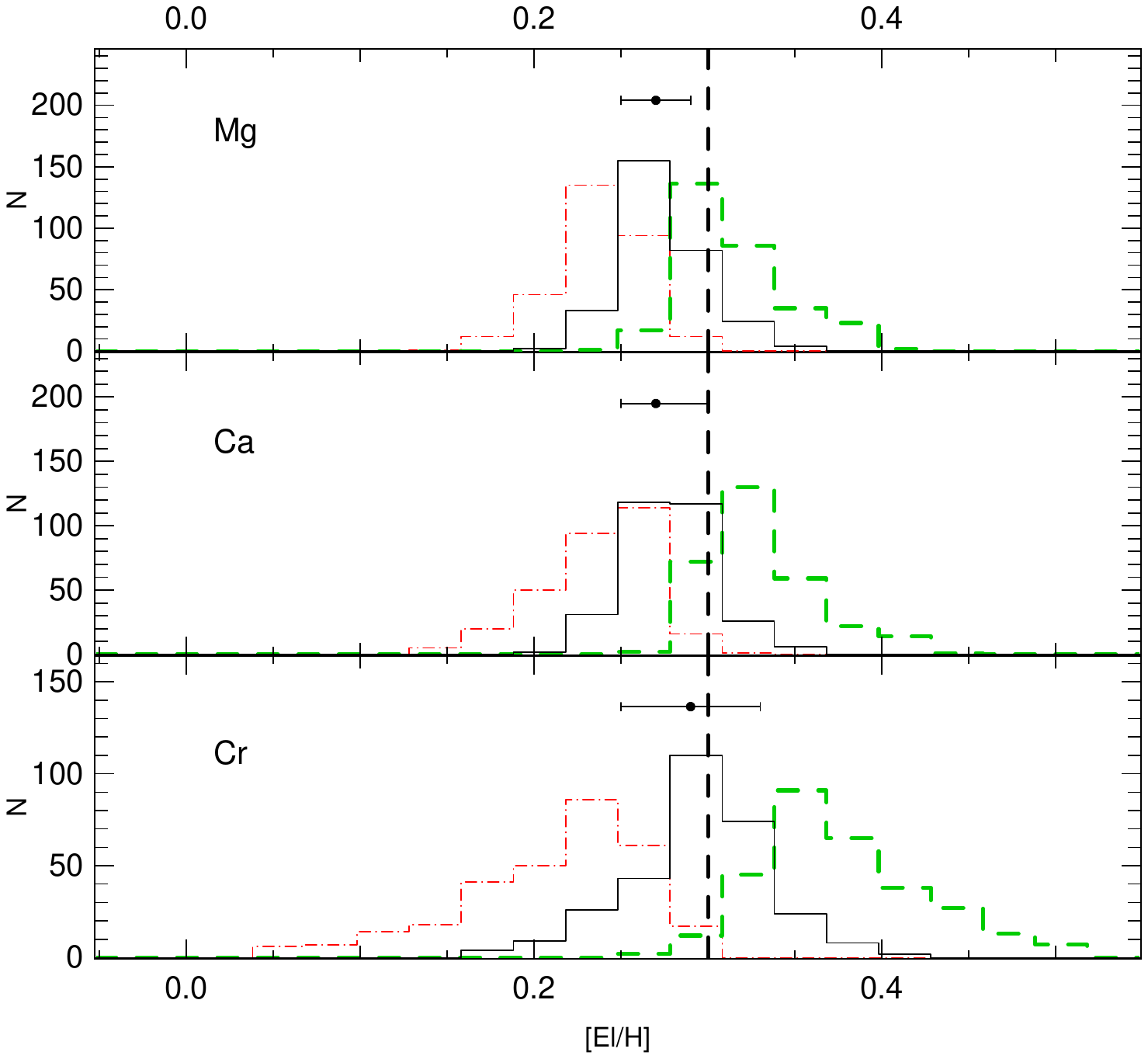}}
\caption{As in Fig.~\ref{plot_bressan_MC_TGM} but for the elemental abundances
of \Mgn, \Can, and \Crn. The color version of this plot is available in the
electronic edition.}
\label{plot_bressan_MC_ElH}
\end{figure}

\subsubsection{Results}
We run \Space\ on the sets of synthetic spectra just outlined. 
For the sake of brevity, we only report 
here the representative case of R=12\,000 and S/N=100.
In Fig.~\ref{TG_R12_SN100_synt_ABDloop} we show the
distribution in \temp\ and \logg\ of the synthetic spectra (left panel)
in comparison with the same parameters derived by 
\Space\ (right panel). The \temp\ and \logg\ derived by \Space\ follow
closely the isochrones for all three synthetic populations.
In Fig.~\ref{correl_R12_SN100_synt_ABDloop} we show
the residuals between the derived and reference values as a function 
of the reference stellar parameters while in 
Fig.~\ref{XFe_R12_SN100_synt_ABDloop} we report the distribution of
the chemical abundances in the chemical plane for nine elements 
derived from the same spectra.

\subsubsection{Errors estimation in synthetic spectra}
In the panels of Fig.~\ref{correl_R12_SN100_synt_ABDloop} we report the dispersions of the
measurements around the expected values of \temp, \logg, and \met, while the confidence
interval of the single measurements (computed as explained in
Sec.~\ref{sec_error_est}) are shown with the light gray errorbars.
The errorbars in Fig.~\ref{correl_R12_SN100_synt_ABDloop} are often smaller 
than the dispersion of the residuals, which suggests an underestimation of the errors. 
This is summarized in
Fig.~\ref{internal_errors_TGM_R} where the overall dispersion of the
residuals for the stellar parameters (derived minus reference, black symbols) are compared with the
half-width of the confidence intervals (gray symbols) for different resolutions and
S/Ns. The black and gray symbols are closer where the stochastic noise 
dominates, i.e., at low S/N, while for high S/N the confidence intervals always
underestimate the stellar parameter dispersions. The reason can be guessed from
Fig.~\ref{correl_R12_SN100_synt_ABDloop}: the overall dispersion $\sigma$
is inflated by the presence of systematic errors
in \temp, \logg, and [Fe/H] for which the computed confidence intervals
cannot account. The latter
can only account for the stochastic errors. We proved
the last statement by generating 100 Monte Carlo realizations of a few synthetic
spectra, derived their stellar parameters and chemical abundances, and
compared them with the confidence intervals computed by \Space. The
distributions of the parameters and the confidence intervals obtained with 
this test show that the confidence intervals only account for the stochastic noise
and fail to recognize the systematic errors when present (the shift between the average
of the black histogram and the expected value represented with a black dashed
line in Fig.~\ref{plot_bressan_MC_TGM}).  The chemical abundances
recovered by \Space\ for the three mock samples (see Fig.~\ref{XFe_R12_SN100_synt_ABDloop})
are accurate and follow the expected sequences traced by the colored solid
lines. No particular systematic error is visible and the errorbars appear to
be a good representation of the dispersion around the expected value.
This is supported by
the statistic obtained from the 100 Monte Carlo realizations cited before
and illustrated in Fig.~\ref{plot_bressan_MC_ElH}. A further discussion of
the systematics errors seen in this section can be found in
Sec.~\ref{sec_discussion}.

\subsection{Tests on real spectra}\label{sec_test_real}
We employed sets of publicly available spectra like the ELODIE spectral library
\cite[Prugniel et al.][]{prugniel}, the spectra of the benchmark stars \cite[Jofr\'e et
al.][]{jofre}, and the spectra of the S4N catalogue \cite[Allende Prieto et
al.][]{allende}.
For the ELODIE spectra we selected those spectra for which 
the authors report literature stellar parameters flagged as being of good and
excellent quality (quality flags ``3" and ``4") to be compared with the stellar 
parameters derived by \Space. For the benchmark and S4N stars we compare our
results with the high quality stellar parameters provided by Jofr\'e et al.
and Allende Prieto et al., respectively.
All these spectra have high spectral resolution
($>$60\,000) and high S/N. To test \Space\ on spectra of lower resolution 
and S/N, we degraded the spectra to resolutions of R=20\,000, 12\,000, 5\,000, and 2\,000 and to
S/N=100, 50, 30, and 20 by adding artificial Poissonian noise\footnote{Many of these
spectra have S/N that are not high (S/N$\sim$60-100) so that by adding
artificial noise the final S/N is actually lower than the nominal one.}. 
Although the original spectra were already normalized, we re-normalized them 
after degrading them with the IRAF task {\it continuum} in order to simulate the continuum
obtained when these spectra are normalized at their nominal 
spectral resolution. Then, the spectra were processed with \Space\ and the
derived stellar parameters compared with the reference values.
For the sake of brevity, we only report 
here the representative case of R=12\,000 and S/N=100.
We measured the spectra with the {\it ABD\_loop} keyword
in the wavelength range 5212-6270\AA\ and
6310-6900\AA\ to avoid the telluric lines in the range 6270-6310\AA\ that may
degrade the quality of the measurements. In this case we switched on the
internal re-normalization as usually done for spectra for which the continuum level
must be refined.

\begin{figure}[t]
\centering
{\includegraphics[width=9cm,bb=101 288 359 505]{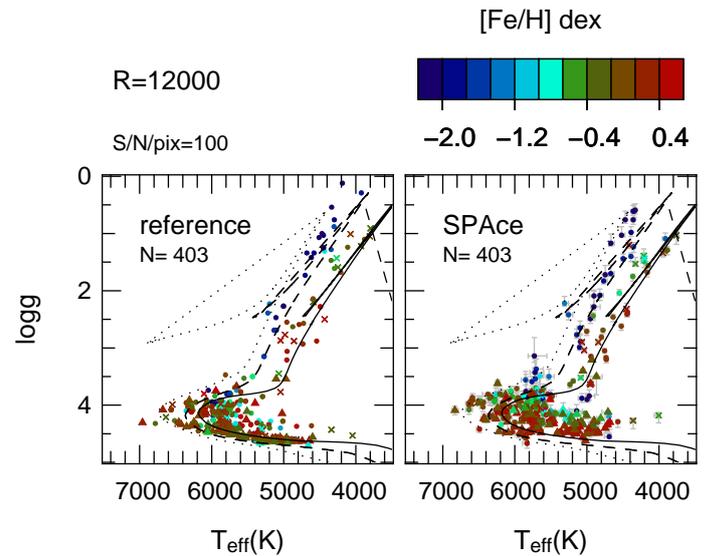}}
\caption{{\bf Left panel}: Distribution of the reference stellar parameters
in the \temp\ and 
\logg\ plane for the ELODIE library (solid points), the benchmark stars
(crosses), and the S4N stars (triangles). {\bf Right panel}: As before but
with the stellar parameters derived by \Space. The colors code the [Fe/H].
Errorbars are reported in light gray. The solid, dashed, and dotted black lines trace the isochrones for
[M/H]=0.0~dex and 5~Gyr, [M/H]=$-1.0$~dex and 10~Gyr, and [M/H]=$-2.0$~dex and
10~Gyr, respectively. The color version of this plot is available in the
electronic edition.}
\label{TG_R12_SN100_real_noABDloop}
\end{figure}

\subsubsection{Results}
The distributions on the \temp\ and \logg\ plane of the reference parameters
and the derived parameters with \Space\ are shown in the left and right
panels of Fig.\ref{TG_R12_SN100_real_noABDloop}, respectively.
The derived parameters appear to follow fairly well the isochrones.
In Fig.~\ref{correl_R12_SN100_real_ABDloop}
the residuals between the derived and expected values as a function 
of the reference stellar parameters show small residuals in all the panels,
except for the middle right panel, which reveals a systematically low \logg\
with lower [Fe/H]. This feature is discussed in
Sec.\ref{sec_logg_systematic}.

\subsubsection{Error estimation in real spectra}
\begin{figure}[t]
\centering
\includegraphics[width=9cm,bb=63 287 509 690]{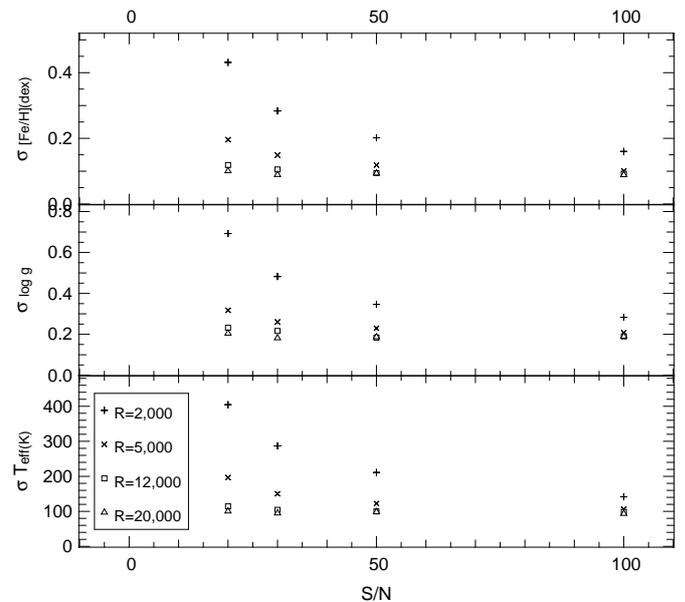}
\caption{Standard deviations of the residuals of the stellar parameters
derived from real spectra 
as a function of S/N for the fourth different resolutions considered.}
\label{errors_real}
\end{figure}

Because for real spectra we do not have exact stellar parameters
but estimations from high resolution spectra, our evaluation of the 
errors relies on the dispersion of the
residuals between the parameters derived by \Space\ and the high resolution
parameters that we use as reference. This is
summarized in Fig.~\ref{errors_real} for different resolutions and S/N
ratios. Because
the reference parameters also suffer from errors, the 
dispersions reported in Fig.~\ref{errors_real} are actually
an overestimation of the \Space\ errors because they result from the
quadratic sum of the reference errors plus the \Space\ errors\footnote{
The reference values of the S4N and the benchmark stars' gravity was derived from parallaxes.
Therefore we expect that, being such estimation usually much smaller than
the ones derived from spectra, for these stars the \Space\ errors are the
major contributors to the dispersion of the redisuals.}.
For the individual elements, Fig.~\ref{XFe_R12_SN100_real_ABDloop} 
shows nine relative abundances derived from the same spectra. For these quantities
reference values can be only found for the S4N spectra, for which we have
seven elements in common. The comparison between derived and reference
abundances is reported in Fig.~\ref{plot_s4n}.

\subsection{Measure of the whole ELODIE spectral library}
We derived stellar parameters and chemical abundances for the whole ELODIE
spectral library at a resolution of R=12\,000 and S/N=100. 
\Space\ provided results for 1386 spectra (out of 1959 spectra of the
library) while it did not converge for those spectra which stellar parameters
are beyond the stellar parameter volume covered by the GCOG library. 
The high number of stars provided by the full ELODIE library gives a more 
robust statistic with respect to the previous test. The derived
parameters are reported in appendix~\ref{appendix_tests_real}, 
Tabs.~\ref{ELODIE_table}, \ref{benchmark_table}, and
\ref{S4N_table}, and are plotted in Fig.~\ref{R12_SN100_elodie_whole_noABD}.

\subsection{Discussion}\label{sec_discussion}
\begin{figure*}[t]
\begin{minipage}{18cm}
\centering
{\includegraphics[width=14cm,bb=85 286 531 524]{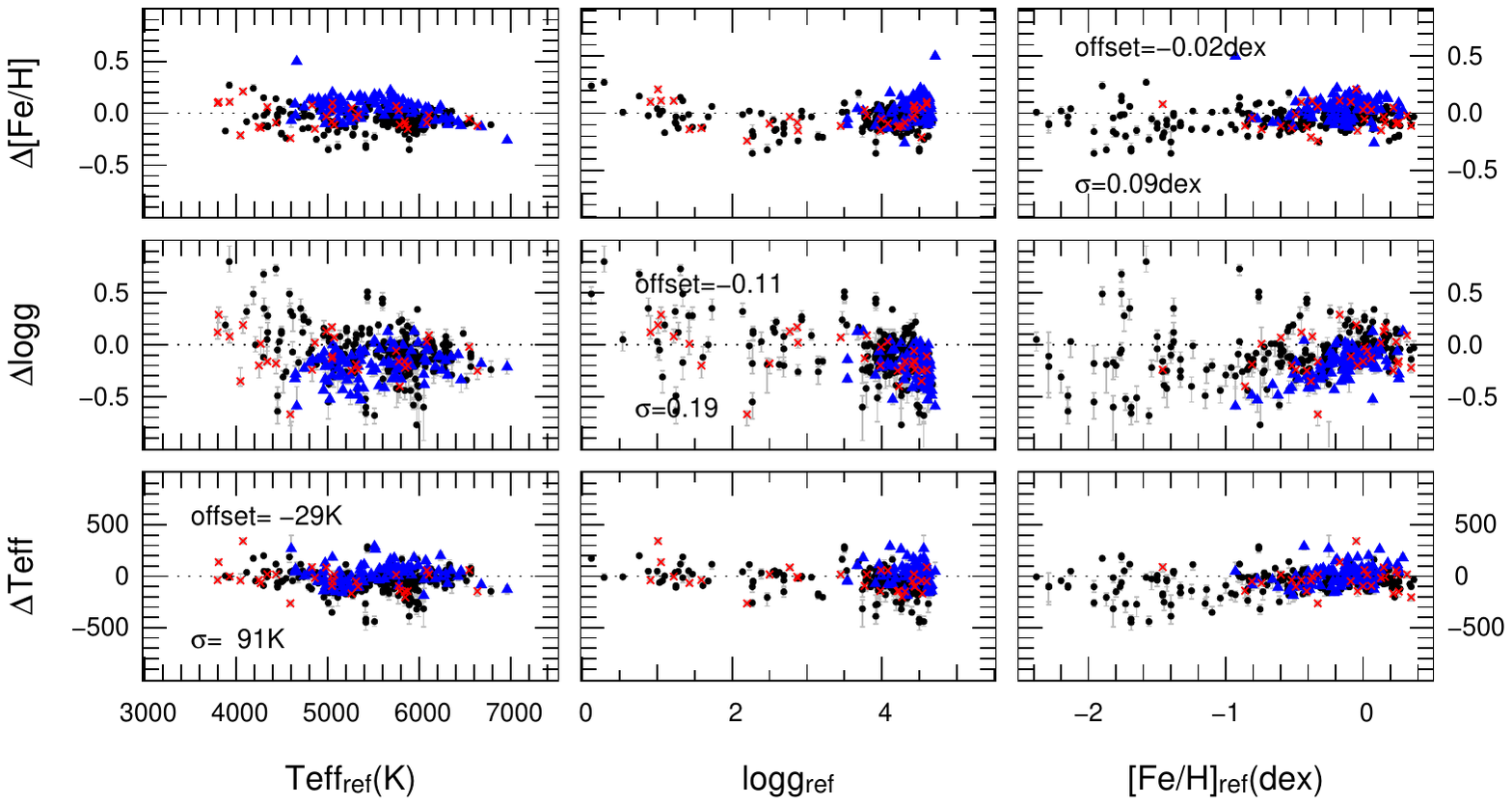}}
\caption{Residuals between derived and reference parameters (y-axis) as a
function of the reference parameters (x-axis). ELODIE, benchmark, and S4N
stars are indicated with black points, red crosses, and blue triangles,
respectively. Errorbars are reported in light gray. A missing
errorbar indicate that the error is larger than the parameters grid.
The color version of this plot is available in the electronic edition.}
\label{correl_R12_SN100_real_ABDloop}
\vskip 0.3cm
\centering
{\includegraphics[width=14cm,bb=85 286 531 524]{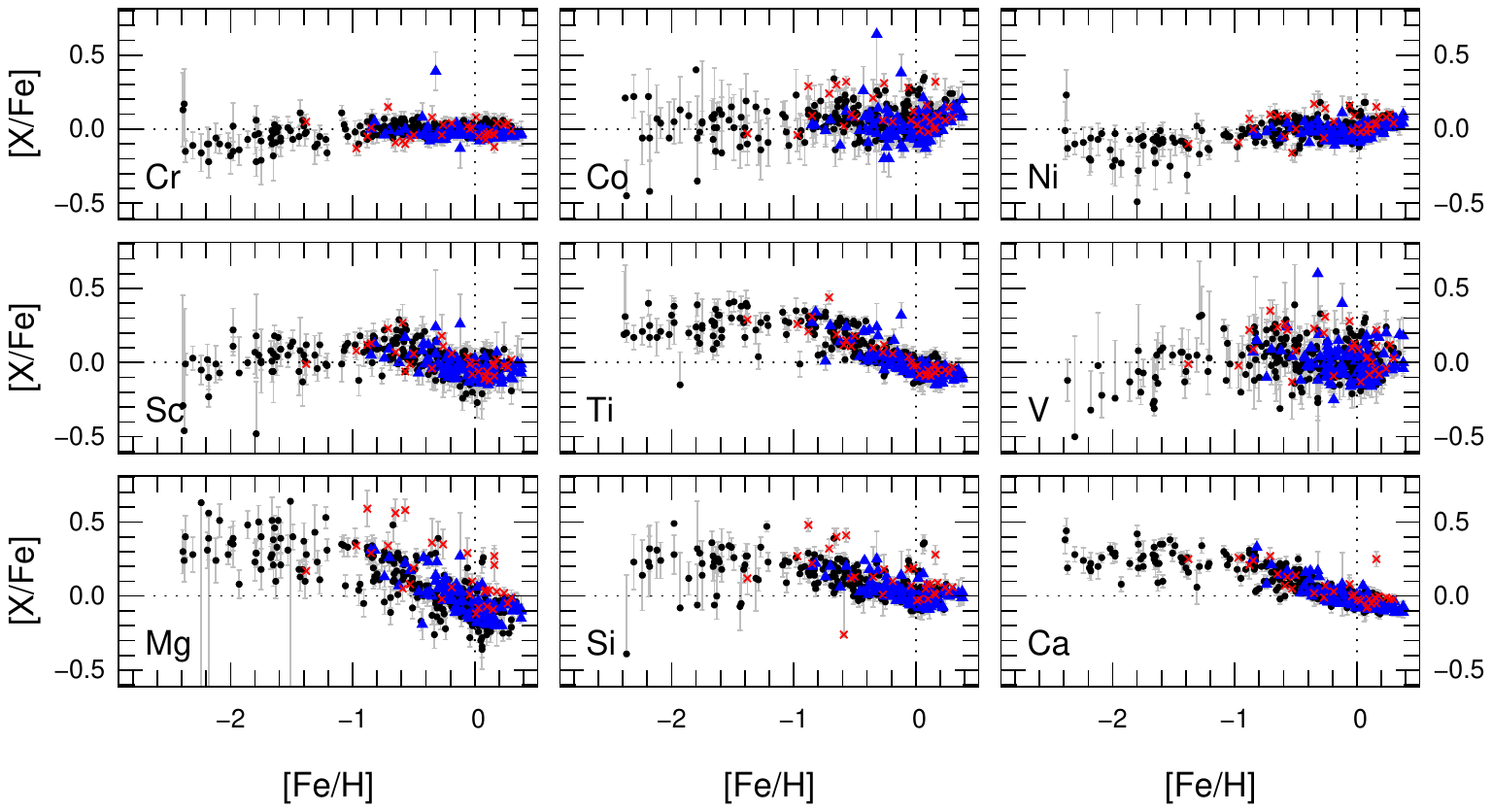}}
\caption{Chemical abundances derived by \Space\ for the ELODIE, benchmark,
and S4N stars. Symbols are as in Fig.\ref{correl_R12_SN100_real_ABDloop}.
The color version of this plot is available in the electronic edition.}
\label{XFe_R12_SN100_real_ABDloop}
\end{minipage}
\end{figure*}

\begin{figure*}[t]
\centering
{\includegraphics[width=16cm,bb=58 113 570 375]{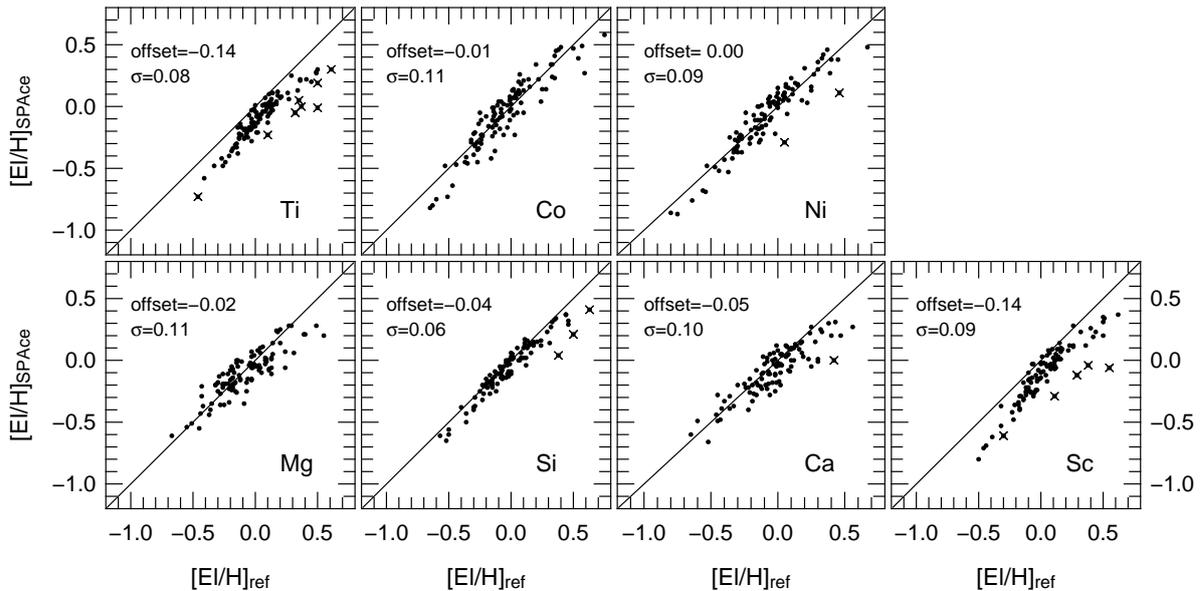}}
\caption{Comparison between derived and reference chemical abundances of
seven elements for the S4N
spectra. Points marked with crosses have been excluded by the statistic reported in the
panels.}
\label{plot_s4n}
\end{figure*}

Tests on synthetic and real spectra showed that the derived stellar
parameters and chemical abundances are reliable and have a good precision. 
The accuracy suffers from systematic errors (in particular for \temp, \logg, and
[Fe/H]) highlighted in the test on
synthetic spectra. This particularly affects spectra that have high density of
strong lines, for which the correction for the opacity of the neighbor
lines (seen in Sec.~\ref{sec_corr_opac}) applied to a wavelength interval
0.5\AA\ wide becomes insufficient.  In this case, the expected EW of the
lines stored in the EW library (and encoded in the GCOG library) is too
large and leads to misestimations of the stellar parameters with the
systematic errors seen in Fig.~\ref{correl_R12_SN100_synt_ABDloop}.
However, these errors are relatively small (up
to 100~K in \temp, 0.2 in \logg, and 0.1~dex in [Fe/H]). While these errors 
affect mostly metal rich cool dwarfs (\temp$<$4500~K, [Fe/H]$>$0~dex) and, 
to some extent, cool giants (\logg$<$0.5) in synthetic spectra, in the test
with real spectra they do not seem seem to play a significant role
(Fig.~\ref{correl_R12_SN100_real_ABDloop}) perhaps because they are smaller
than the stochastic errors. On the other hand, the measurements done with the
whole ELODIE sample (Fig.~\ref{R12_SN100_elodie_whole_noABD}) show an
underestimation of the gravity for dwarfs stars cooler than \temp$<$4800K
(they do not follow the isochrones, as expected) and an apparent
gravity overestimation
of the red clump stars of $\sim+0.25$ in a
general picture that confirms the goodness of the results in every other
respect.\\
Another source of systematic errors is the adopted line profile
(Appendix~\ref{app_voigt}). The \Space\ line profile is an empirical function of the EW, broadening
constants, and \logg\ of the star, and it proved to fit reasonably well the
lines for most of the stellar parameters. However, there are some
discrepancies that causes the systematic deviations from the expected stellar
parameters just discussed. An improved function for the line profile can reduce
the systematic errors and this is one of the many possible improvements that 
are left for the next version of \Space.\\
Despite the errors in stellar parameters, the resulting chemical abundances
are reliable, as shown in Fig.~\ref{XFe_R12_SN100_synt_ABDloop} and
Fig.~\ref{XFe_R12_SN100_real_ABDloop} for synthetic and real spectra,
respectively. For the synthetic spectra the distribution of the derived chemical
abundances on the chemical plane follows closely the expected values (light colored
solid lines in Fig.~\ref{XFe_R12_SN100_synt_ABDloop}), while for real
spectra the chemical abundance distributions follow fairly well the pattern
expected for the
Milky Way stars. A one-to-one comparison of the derived chemical abundances
with the reference abundances of the S4N spectra (Fig.~\ref{plot_s4n})
reveals that some of the elements may suffer from systematic errors. In
particular \Scn\ and \Tin\ seem to be underestimated by $\sim$0.1~dex with
respect to the S4N estimations.  For the
other elements, the abundances agree fairy well.  
It is not clear what may cause the underestimation of the \Scn\ and \Tin.
The absorption lines of these two elements
are weak in the Sun, which makes the calibration
of the \loggf s and the determination of the abundances of the other
standard stars used for the calibration (Sec.~\ref{sec_gf_calibration})
difficult. 
This can lead to a systematic offset of the calibrated \loggf s of the lines
of these elements, and therefore to an offset in the derived abundances.
On the other hand, in
Sec.~\ref{sec_loggf_validation} we showed how the calibrated \loggf s seems
to be smaller than the good quality \loggf s we took as reference. This would
lead to an overestimation of the chemical abundances, which is the opposite
of the underestimation seen. Moreover, it would affect all the elements and
not \Scn\ and \Tin\ alone.\\
Most of the systematic errors seen in synthetic spectra become
indistinguishable in the tests with real spectra, where the stochastic errors are larger.
However, there is at least one systematic error highlighted by the
tests on real spectra that must be discussed. It affects the \logg\ and it is
discussed in the next section.

\subsubsection{On the systematic error in \logg\ in real
spectra}\label{sec_logg_systematic}
\begin{figure}[t]
\centering
\resizebox{\hsize}{!}
{\includegraphics[width=9cm,bb=75 286 378 455]{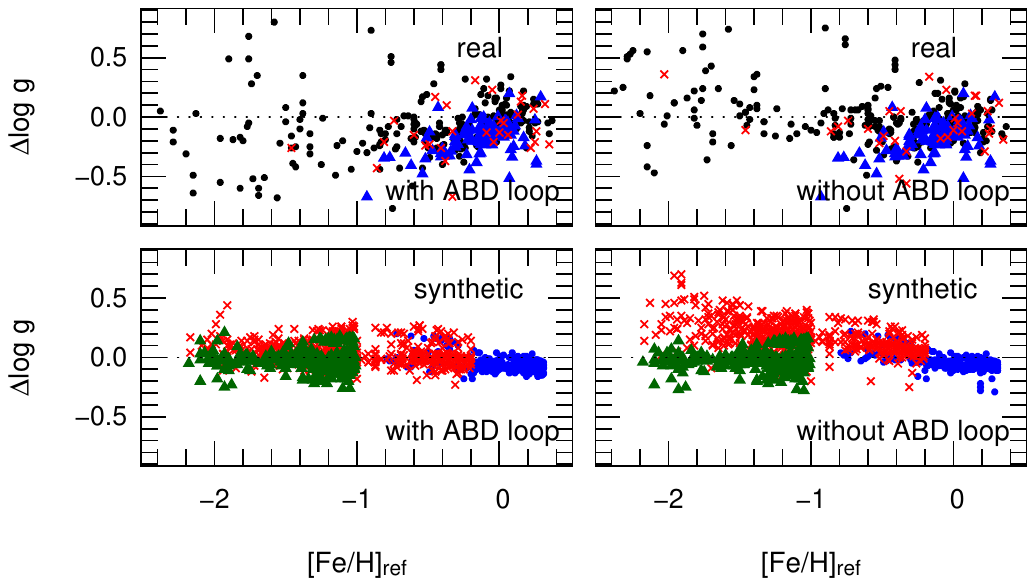}}
\caption{Residuals between derived and reference \logg\ for real spectra
(top) and synthetic spectra (bottom) when the long (left panels) and short
(right panels) versions of \Space\ are used to derive the stellar
parameters. The symbols are as in Fig.~\ref{correl_R12_SN100_real_ABDloop} and
Fig.\ref{TG_R12_SN100_synt_ABDloop} for real and synthetic spectra,
respectively. The color version of this plot is available in the electronic
edition.} 
\label{plot_logg_comp}
\end{figure}

The results obtained with synthetic and real spectra prove to be reliable
and in fair agreement for all the stellar parameters but for \logg, for
which \Space\ derives a too low gravity for metal poor spectra. The
absence of this systematic error in the tests on synthetic spectra
excludes that the error may originate in the way in which \Space\ constructs the
spectrum models. In the attempt to shed light on this, we tested \Space\
with different settings, and we found that running \Space\ with and without the 
keyword {\it ABD\_loop} (which executes or skips the
step~10 of the algorithm outlined in Sec.\ref{sec_space_code})
leads to results that are in agreement for all the 
stellar parameters but for \logg. In Fig.~\ref{plot_logg_comp} the residuals in 
\logg\ are shown as a function of [Fe/H] for both settings for comparison
purposes. With {\it ABD\_loop}, real spectra show the systematic error just
cited, whereas this is absent in synthetic spectra. Conversely, without {\it
ABD\_loop} the systematic error in \logg\ for real spectra is greatly
reduced, but the same systematic with opposite sign appears in synthetic
spectra for the $\alpha$-enhanced stars (red crosses in
Fig.~\ref{plot_logg_comp}) and not for the non-enhanced ones (green triangles). 
This seems to be in agreement with
the real spectra, because for the stars here considered, the low metallicity
stars are $\alpha$-enhanced too.  This suggests that the only stars affected by 
this systematic are the $\alpha$-enhanced stars. 
This indicates that the discrepancy observed between synthetic and real
spectra does not depend on the method employed, but that it may originate from 
i) incorrect microturbulence adopted for the EW library or from
ii) the adopted 1D atmosphere models and LTE assumption, for which the discrepancy
to the physical conditions of real stars becomes larger for lower metallicity (Asplund
\citealp{asplund2005}; \citealp[Bergemann et al.][]{bergemann2012}). We are inclined to exclude that
the atomic parameters like damping constants or oscillator strengths may
play a significant role in this systematic error because otherwise it should equally 
affect metal-rich and metal-poor stars.
The future development of a new GCOG library that accounts for NLTE effects and 3D
atmosphere models should shed light on the origin of this systematic and, hopefully,
remove it. Because this problem cannot be solved in the present work, we
choose to leave the option to the user whether to use the {\it ABD\_loop}
keyword. In the appendix of this work, the
results of the tests on real spectra run without the {\it ABD\_loop}
keyword are presented.\\
A further systematic error is the underestimation ($\sim-0.2$) of \logg\
for dwarf (\logg$\gtrsim4.2$)stars. This effect is smaller in the
test with synthetic spectra than in real spectra.
The fact that for synthetic spectra this is small seems to
suggest that the origin of the problem may, as before, lie in the basic assumption made,
such as the LTE assumption and the stellar atmosphere models adopted.\\

\subsection{Remark}
In this paper we mostly aimed at validating the method proposed. For the sake of
brevity, we do not discuss the tests done by using other functions available in
\Space\ and we just briefly mention two of them here\footnote{The full list of
functions available is provided with the tutorial that accompany the
code \Space.}. \Space\ accepts keywords like {\it T\_force} and {\it G\_force}
that force \Space\ to look for solutions with fixed \temp\ and/or \logg\
given by the user. This is particularly useful for low S/N spectra
for which \Space\ cannot converge to precise stellar parameters. For
instance, a robust photometric temperature passed to \Space\ with the
keyword {\it T\_force} helps to improve
the \logg\ and chemical abundances estimations. Another useful keyword is {\it
alpha} which forces \Space\ to derive the abundance of the $\alpha$-elements (\Mgn, \Sin,
\Can\ and \Tin) as if they were one single element while any other elements
(excluding
\Cn, \Nn, and \On) are considered to be a separate single element called ``metals". As
before, this is useful to get abundances from low-quality spectra that carry little
information.

\section{Publication}
The source code of \Space\ will be publicly available soon together with the
line list and the GCOG library. The code will be released under a GPL license.
In addition, a VO-integrated service allowing operation of \Space\
without installation is available \cite[Boeche et al.][]{boeche2015}.  
As simple Web front end to this service can be found at
http://dc.g-vo.org/SP\_ACE.

\section{Future work}
In this work we outlined the method the code \Space\ relies upon and
the solutions chosen up to now, which prove to work but are far from perfect.
Many improvements and further developments are possible. Among the most
important we cite the following ones:
\begin{itemize}
\item {\bf Extension of the stellar parameter grid}: it is possible to
extend the coverage of the GCOG library to hotter temperature than the
actual covered ones (\temp$<$7400~K) and to higher gravities. The latter
have been
extended to \logg=5.4 with an extrapolation because stellar atmosphere
models with \logg$>$5.0 are not provided by the grid ATLAS9. An extension of
the stellar atmosphere grid (and subsequent extension of the GCOG library)
up to \logg$\sim$6 is planned for the near future.
\item {\bf Extension of the line list}: currently the wavelength range
covered by the GCOG library is 5212-6860\AA\ and 8400-8920\AA. We plan to
extend the wavelength range, in particular toward bluer wavelengths. With
the extension of the grid to hotter stars, the line list will be augmented
by ionized/high excitation potential lines only visible at high temperatures. 
\item {\bf Molecular lines}: at the present time the molecular lines we
take in account are the CN lines in the range 8400-8920\AA. An extension to
the optical region would improve the derivation of stellar parameters for
cool stars. How this problem can be solved in the framework of the method used
by \Space\ is still unclear.
\item {\bf Opacity correction}: an improved method to correct the EW for
the opacity of the neighboring lines has to be found. A further investigation
of the rigorous solution proposed in Appendix~\ref{appendix_cf}, or
new techniques of line deconvolution (like the one proposed by Sennhauser et
al. \citealp{sennhauser}) may lead to a solution.
\item{\bf Improved line profile}: the present line profile adopted in
\Space\ is an empirical function that represents fairly well the shape of the
lines over a wide range of parameters, but still is not good enough at
the borders of the parameter grid. It is possible and desirable to find a
new improved line profile function that would permit the removal of some of the
systematic errors seen in synthetic spectra. 
\item {\bf Extension to other stellar atmosphere models}: the present EW library has
been constructed with the 2012 version of the ATLAS9 atmosphere grid by Castelli \& Kurucz
\cite{castelli}, but it can be done with any other atmosphere models. The creation of
EW and GCOG libraries based on MARCS \cite[Gustafsson et al.][]{gustafsson} or PHOENIX
\cite[Husser et al.][]{husser} models is desirable and we plan to do it in the
near future.
\item {\bf Extension to 3D models and NLTE assumptions}: although the
construction of a whole EW library with 3D atmosphere model and NLTE
assumptions seems still prohibitive in terms of computing costs, the
integration of the present EW library with few important absorption lines
the EWs of which have been computed under NLTE assumptions and/or a 3D atmosphere model is
doable. For instance, computing the EWs of H$\alpha$, 
the \Fei\ at 5269.537\AA, and one line of the \Caii\ triplet with 3D
atmosphere models and/or under NLTE assumptions and integrating these EWs into
the present EW library would greatly increase the ability of \Space\ to constrain
the stellar parameters, in particular for low metallicity or low S/N
spectra where only strong lines can be seen.
\end{itemize}

While for some of the above points the amount of work may be considerable,
for other points the necessary work is small and it would bring significant
improvements in a short time.


\section{Conclusions}
In this work we proposed and described a new method to derive stellar
parameters and chemical elemental abundances from stellar spectra. Based on calibrated
oscillator strengths of a complete line list and on 1D
atmosphere models and under LTE assumptions, this method relies on 
polynomial functions (stored in the GCOG library) that describe the 
EWs of the lines as a function of
the stellar parameters and chemical abundance. The method is implemented
in the code \Space, which constructs on the fly spectral models and
minimizes the $\chi^2$ computed between the models and the observed spectrum.
The method has a full-spectrum-fitting approach, which means i) it assures the reliability
of the spectrum models by calibrating the oscillator strengths of the line
list adopted in high-resolution spectra of stars with well-known stellar
parameters, and ii) it employs all the possible absorption lines (thousands)
in a wide wavelength range to derive the stellar parameters and abundances, 
exploiting the information carried by lines that are usually rejected in the classical
analysis because they are blended or have unreliable theoretical atomic parameters. This
approach proved to be successful, obtaining reliable stellar parameters and 
chemical abundances even in spectra carrying little information,
such as low-resolution or low S/N spectra, for which the classical analysis
based on EW measurements cannot be applied.\\
The method is far from perfect, but we believe it shows considerable
promise already
at the present stage of development. It is highly automated, so
that it is suitable for the analysis of large spectroscopic surveys. It is
flexible, in the sense that its internal re-normalization and internal 
re-setting of the radial velocity of the spectrum make \Space\ independent
from the initial quality of the normalization and RV correction  performed by 
previous users or reduction pipelines. It is independent from the
stellar atmosphere models used to create the GCOG library on which
\Space\ relies. In fact, the GCOG library can be constructed by using
any stellar atmosphere models available in the literature, or under LTE or NLTE
assumptions, with no need to change the code \Space.\\

An on-line version of the code \Space\
is available on the German Astrophysical Virtual Observatory
web server at the address http://dc.g-vo.org/SP\_ACE.
The source code will be made publicly available soon.

\begin{acknowledgements}
B.C. thanks: H.-G.~Ludwig for the numerous useful discussions on
atomic parameters and stellar atmosphere models; M. Demleitner and H. Heinl
for their support in preparing the web front-end of \Space\ and useful
discussions.
We acknowledge advice and assistance in publishing the web service
provided by the German Astrophysical Virtual Observatory (GAVO).
We acknowledge funding from Sonderforschungsbereich SFB 881 ``The Milky Way
System" (subproject A5) of the German Research Foundation (DFG).
\end{acknowledgements}
\newpage

\appendix

\section{Microturbulence}\label{appx_microt}
\begin{figure}[t]                                                            
\centering                                                                   
{\includegraphics[width=9cm,bb=99 484 512 685]{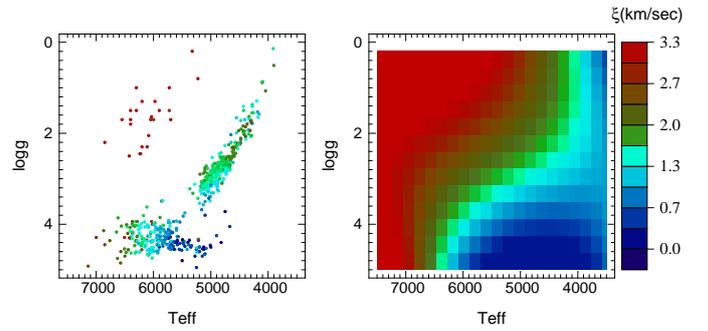}}
\caption{{\bf Left}: distribution on the (\temp,\logg) plane of the stars
employed to determine the microturbulence function. {\bf Right}: polynomial
function adopted to represent the microturbulence $\xi$. For both panels the
value of $\xi$ is given by the color bar on the right. The color version of
this plot is available in the electronic edition.}
\label{microt_Teff_logg}
\end{figure}

\begin{figure}[t]                                                            
\centering                                                                   
{\includegraphics[width=9cm,bb=70 289 566 557]{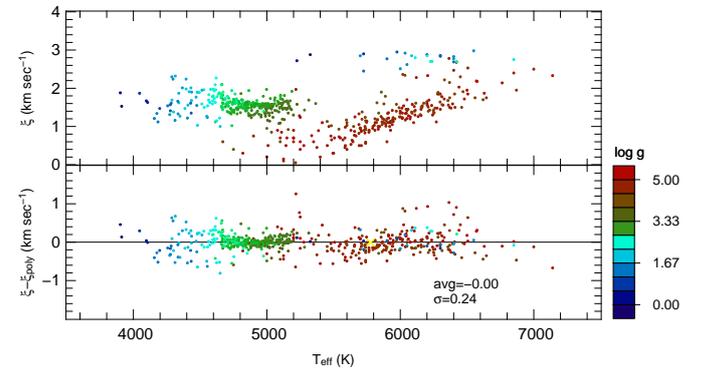}}
\caption{Reference microturbulence (top) and residuals (bottom) between
reference ($\xi$) and polynomial microturbulence ($\xi_{poly}$) as a
function of \temp. The colors code the gravity \logg\ as given by the
color bar on the right. The color version of this plot is available in the
electronic edition.}
\label{microt_poly_residuals}
\end{figure}

Because \Space\ cannot determine the microturbulence, the EW library
must be constructed by assuming the microturbulence value $\xi$ at each
point of the grid before the EW computation.  
As done in a previous work (Boeche et al., \citealp{boeche11})
we can set the microturbulence as a function of the stellar parameters
(see the work done by M. Bergemann and V. Hill in the Gaia-ESO
collaboration cited in Jofr{\'e} et al. \citealp[][]{jofre}; Allende Prieto
at al. \citealp{allende}). To determine
such a function, we investigated how the microturbulence $\xi$ varies on the
(\temp, \logg) plane for some hundreds of stars studied in high-resolution
spectroscopy works found in the literature
(Fuhrmann \citealp{fuhrmann}; Allende Prieto et al. \citealp{allende};
Bensby et al. \citealp{bensby}; Fulbright et al. \citealp{fulbright}; Luck
et al. \citealp{luck1}, \citealp{luck2}; Luck \& Heiter \citealp{luck3}; Luck et
al. \citealp{luck4}). From such studies we used 620 derived microturbulence
$\xi$ values for dwarfs, giants, and Cepheid stars to determine a polynomial
function that approximates the microturbulence on the (\temp, \logg)
plane\footnote{The total number of stars considered in these studies
altogether is 921. Because the particularly high $\xi$ of some stars
(mostly Cepheids) the fitting polynomial function was not satisfactory for
normal stars. Because \Space\ is designed manly for normal stars, we excluded
stars with $\xi>3$~\kmsec.}. Such stars are shown in the left panel of
Fig.~\ref{microt_Teff_logg}.
To fit $\xi$ we used a fourth-degree polynomial function. Higher
degrees do not improve the fit. 
The polynomial function is written as follows:
\begin{equation}\label{microt_eq}
\xi_{poly}(\mbox{km s$^{-1}$})=\sum_{i,j=0}^{2} a_{ij}
(\mbox{\temp})^i(\mbox{\logg})^j
\end{equation}
where the coefficients $a_{ij}$ are
\begin{eqnarray}
a_{00}&=&-5.11308\\\nonumber
a_{01}&=&0.58507\\\nonumber
a_{02}&=&0.471885\\\nonumber
a_{10}&=&0.00207105\\\nonumber
a_{11}&=&1.70456\cdot10^{-5}\\\nonumber
a_{12}&=&-0.000257162\\\nonumber
a_{20}&=&-1.0543\cdot10^{-7}\\\nonumber
a_{21}&=&-3.21628\cdot10^{-8}\\\nonumber
a_{22}&=&2.94647\cdot10^{-8}\\\nonumber
\end{eqnarray}
Because this polynomial function diverges to large positive values for high \temp\ and
low \logg\, and to negative values at low \temp\ and high \logg, we imposed
the following conditions: $\xi=3$~\kmsec where $\xi>3$~\kmsec and
$\xi=0.5$~\kmsec where $\xi<0.5$~\kmsec. The values of the polynomial
function and the just cited conditions on the (\temp, \logg) plane are
shown in the right panel of Fig.~\ref{microt_Teff_logg}.
In Fig.~\ref{microt_poly_residuals} the residuals between the computed
$\xi$ and the literature works are shown together with the literature $\xi$
values as a function of \temp. 
This polynomial function has been employed to set the microturbulence during
the construction of the EW library.

\section{Separating the contributions of different absorption lines to the
total absorbed flux}\label{appendix_cf}

In Sec.~\ref{sec_contrib_lines} we shortly discussed how to separate the 
contribution of blended lines to the total $EW_{blend}$. Here we give a
detailed explanation on how this can be done in the case of two blended lines.
The extension to many lines is straightforward. Consider two blended
absorption lines and the equation
\begin{equation}\label{eq_sep_ew}
EW_{blend}=EW_A^c+EW_B^c,
\end{equation} 
where $EW_A^c$ and $EW_B^c$ are the individual contributions to the EW of
the blends of each absorption line $A$ and $B$. We want to verify that
equation~(\ref{eq_sep_ew}) is satisfied, or, in other words, that 
$EW_A^c$ and $EW_B^c$ can be inferred from the radiative transfer
equation.\\
Because the EW of a line
is the measure of the absorbed flux over the line with respect 
to the continuum, the previous problem is solved by verifying that
the flux (intensity of radiation $I_{out}$) emerging from a layer of a
stellar atmosphere over a small wavelength interval $d\lambda$ can be written as\\
\begin{equation}\label{eq_sep_contr}
I_{out}=I_{in}-I_{abs}^A-I_{abs}^B
\end{equation}
where $I_{abs}^A$ and $I_{abs}^B$ are the individual absorbed intensities
due to the absorbers $A$ and $B$, respectively. We are going to prove that
equation~(\ref{eq_sep_contr}) is true for a layer that is homogeneous
in its physical conditions and chemical composition\footnote{This can be
approximated by taking a very thin layer of a stellar atmosphere. The
1D atmosphere models are composed of $N$ layers that satisfy these
conditions.}. Then we reach the answer by extending the result to the 
whole atmosphere.\\

\subsection{Pure absorption case}
Consider one optically thick layer with two species of absorbers $A$ and $B$
in it, and the variation of the radiation that goes through it. 
The optical depth of the layer is $\tau=\tau^A+\tau^B$. In the case of pure
absorption the transfer equation is\\
$$
\frac{dI}{d\tau}=-I,
$$
and the solution of this equation is\\
\begin{equation}\label{RTE_absorption1}
I_{out}=I_{in}e^{-(\tau^A+\tau^B)}
\end{equation}
where $I_{in}$ is the intensity of the entering radiation, 
$I_{out}$ is the intensity of the emerging radiation. We rearrange
equation~(\ref{eq_sep_contr}) and using equation~(\ref{RTE_absorption1}) 
we write the absorbed intensity\\
\begin{equation}\label{eq_abs0}
I_{abs}=I_{in}-I_{out}=I_{in}-(I_{in}e^{-(\tau^A+\tau^B)})=I_{in}(1-e^{-(\tau^A+\tau^B)})
\end{equation} 
We can ideally divide this optically thick layer (with optical deph $\tau$) in many
optically thin layers, each of them with optical depth $d\tau^i$. Thus, for
the $i$-th thin layer we can write the equation~(\ref{eq_abs0}) as follows:
$$
I_{abs}^i=I_{in}^i(1-e^{-(\tau^{i,A}+\tau^{i,B})})
$$
where $I_{in}^i$ is the incoming intensity in the $i$-th layer and
$\tau^{i,A}$ and $\tau^{i,B}$ are the optical depths of the $i$-th layer due
to the $A$ and $B$ absorbers. By adopting the Taylor expansion of the exponential
function $e^{-(\tau^{i,A}+\tau^{i,B})}$ valid for optically thin layers, and
truncating it to the first order (the residual goes to zero when $\tau\rightarrow0$), it
becomes\\
\begin{eqnarray}\label{eq_abs_}\nonumber
I_{abs}^i & = & I_{in}^i(1-(1-\tau^A-\tau^B))\\\nonumber
          & = & I_{in}^i(\tau^{i,A}+\tau^{i,B})\\\nonumber
\end{eqnarray}
By writing
$I_{abs}^{i,A}=I_{in}^i\tau^{i,A}$ and $I_{abs}^{i,B}=I_{in}^i\tau^{i,B}$
we separate the contributions of the absorbers $A$ and
$B$ to the absorbed intensities\\
\begin{eqnarray}\label{eq_abs1}
I_{abs}^i & = & I_{abs}^{i,A}+I_{abs}^{i,B}.
\end{eqnarray}
Now we consider the ratio between the absorbed intensities\\
\begin{equation}\label{eq_abs2}
\frac{I_{abs}^{i,A}}{I_{abs}^{i,B}}=\frac{\tau^{i,A}}{\tau^{i,B}}.
\end{equation}
From equation~(\ref{eq_abs2}) we isolate the term $I_{abs}^{i,B}$ and
put it in equation~(\ref{eq_abs1}) to obtain\\
\begin{equation}
I_{abs}^i=I_{abs}^{i,A}\Big(1+\frac{\tau^{i,B}}{\tau^{i,A}} \Big).
\end{equation}
We can write the contributions of the two absorbers $A$ and $B$
as follows:\\
\begin{eqnarray}\label{eq_abs3}
I_{abs}^{i,A} & = & I_{abs}^i\cdot C^{i,A}\\\label{eq_abs4}
I_{abs}^{i,B} & = & I_{abs}^i\cdot C^{i,B}
\end{eqnarray}
where\\
$$
C^{i,A}=\frac{\tau^{i,A}}{\tau^{i,A}+\tau^{i,B}}
\mbox{\hspace{1cm}and\hspace{1cm}}
C^{i,B}=\frac{\tau^{i,B}}{\tau^{i,A}+\tau^{i,B}}\\
$$
Since the thick layer is homogeneous, what holds for the $i$-th thin layer
(i.e., equations~(\ref{eq_abs3}) and (\ref{eq_abs4})), must hold for the whole
thick layer.
(This can actually be proved by using the limit $\tau^i\to 0$.
The residual of the Taylor expansion goes to zero faster that $\tau^i$, and
the equations~(\ref{eq_abs3}) and (\ref{eq_abs4}) are therefore valid.) 
Therefore, for the whole thick layer we can write\\
\begin{equation}\label{eq_abs5}
I_{abs}=\sum_i I_{abs}^i\cdot C^{i,A}+ \sum_i I_{abs}^i\cdot C^{i,B}
\end{equation}
Because $C^{i,A}$ and $C^{i,B}$ are constant through the whole thick layer
($\tau^{i,A}$ and $\tau^{i,B}$ do not change with $i$), we can write\\
\begin{eqnarray}\nonumber
I_{abs} & = & C^A \sum_i I_{abs}^i + C^B \sum_i I_{abs}^i\\\nonumber
   & = & C^A\cdot I_{abs} + C^B\cdot I_{abs}\\\label{eq_abs6}
   & = & I_{abs}^A + I_{abs}^B.
\end{eqnarray} 
The terms\\
\begin{eqnarray*}
I_{abs}^A & = & C^A\cdot I_{abs}\\
I_{abs}^B & = & C^B\cdot I_{abs}\\
\end{eqnarray*}
are the individual contributions to the
absorbed flux of the absorbers $A$ and $B$, respectively.\\
By using equation~(\ref{eq_abs0}),
$I_{abs}^A$ and $I_{abs}^B$ can be written as\\
\begin{eqnarray}\label{eq_abs7}\nonumber
I_{abs}^A & = & C^A\cdot I_{in} (1-e^{-(\tau^A+\tau^B)})\\\nonumber
I_{abs}^B & = & C^B\cdot I_{in} (1-e^{-(\tau^A+\tau^B)})\\
\end{eqnarray}
and equation~(\ref{eq_sep_contr}) is validated.

\subsection{Atmosphere model case}
Here we want to apply the previous result to an atmosphere model, which is
composed of $N$ number of layers (for instance, the Castelli \& Kurucz atmosphere models
have $N=72$). Such
layers have different physical conditions, but inside every layer, the
physical conditions and the composition of the gas are
constants and homogeneous. Under such conditions,
the equations~(\ref{eq_abs7}) are valid for every $i$-th layer.\\

Consider the transfer equation
$$
\frac{1}{\rho}\cdot\frac{dI}{dz}=-(\kappa^a+\kappa^b)(I-S)
$$
where $a$ and $b$ refer to two different absorbers and $S$ is the source
function. The equation can also be written as
$$
\frac{dI}{d\tau}=-(I-S)
$$
where $\tau^a=\rho\kappa^a dz$, $\tau^b=\rho\kappa^b dz$ and
$\tau=\tau^a+\tau^b$. $\tau$ grows from the bottom toward the
surface of the atmosphere, so that the optical depth is 0 at the bottom and
$\tau_{surf}\ne0$ at the surface. $I$ is the monochromatic
intensity of the radiation in a plane-parallel atmosphere. The solution of this
equation is:\\
\begin{equation}
I(\tau_2)=\int^{\tau_2}_{\tau_1}S(t)e^{-(\tau_2-t)}dt+I(\tau_1)e^{-(\tau_2-\tau_1)}
\end{equation}
which gives the intensity of the flux at the optical depth $\tau_2$.
We can follow the intensity $I(\tau)$ as function of the optical depth
$\tau$ by subdividing the atmosphere into N layers, precisely at optical depths
$\tau_1$,$\tau_2$,...,$\tau_N$ and write
\begin{equation}\label{rad_solution1}
I(\tau_{i})=\int_{\tau_{i-1}}^{\tau_{i}}S(t)e^{-(\tau_{i}-t)}dt+I(\tau_{i-1})
e^{-(\tau_{i}-\tau_{i-1})}
\end{equation} 
where $\tau_{i-1}$ and $\tau_i$ are the optical depths at the bottom and the top
of the $i$-th layer, respectively, and $I(\tau_{i-1})$ is the incoming intensity at
the bottom of the $i$-th layer (and the emerging intensity from the
$i-1$-th layer). Because every layer has constant physical conditions
and constant optical thickness,
we can set the optical depths $\tau_i$ to match the optical depths
of the top of the $i$-th layer of the atmosphere model. Because the constancy of the
physical conditions in each layer, the $S(t)$ function of the integrand in
equation~(\ref{rad_solution1}) is constant, and by solving the integral the equation
becomes:\\
\begin{equation}\label{sol_serie}
I(\tau_i)=S^i(1-e^{-\Delta\tau_i})+I(\tau_{i-1}) e^{-\Delta\tau_{i}}
\end{equation}
where $\Delta\tau_i=\tau_i-\tau_{i-1}$ is the optical thickness of the $i$-th layer, $S^i$ 
is the source function and $I(\tau_{i-1})$ the incoming intensity in the $i$-th
layer. Equation~\ref{sol_serie} is a
recurrence relation. The line of reasoning followed for
the case of pure absorption can be applied
to the source function as well. In fact, the product
$S^i(1-e^{-\Delta\tau_i})$ is equivalent to the absorbed flux expressed in
equation~(\ref{eq_abs0}).
Then, equations~(\ref{eq_sep_contr}) and (\ref{eq_abs5}) can be applied to the two terms of
the sum in equation~(\ref{sol_serie}), and it becomes\\
\begin{equation}\label{eq_sep_contr1}
I(\tau_i) = 
(S^{i,A}+S^{i,B})
        + (I(\tau_{i-1})-I_{abs}^{i,A}-I_{abs}^{i,B})
\end{equation}
where\\
\begin{eqnarray}
S^{i,A} & = & C^{i,A}\cdot S^i (1-e^{-\Delta\tau_i})\\
S^{i,B} & = & C^{i,B}\cdot S^i (1-e^{-\Delta\tau_i})\\
\label{frac_I_A}
I_{abs}^{i,A} & = & C^{i,A}\cdot I(\tau_{i-1})(1-e^{-\Delta\tau_i})\\
\label{frac_I_B}
I_{abs}^{i,B} & = & C^{i,B}\cdot I(\tau_{i-1})(1-e^{-\Delta\tau_i})\\
\label{frac_A}
C^{i,A} & = & \frac{\tau_i^a}{\tau_i^a+\tau_i^b}\\
\label{frac_B}
C^{i,B} & = & \frac{\tau_i^b}{\tau_i^a+\tau_i^b}\\
\end{eqnarray}
Rearranging the terms, equation~(\ref{eq_sep_contr1}) can be written as\\
$$
I(\tau_i) =
I(\tau_{i-1})-(I_{abs}^{i,A}-S^{i,A})-(I_{abs}^{i,B}-S^{i,B})
$$
where the values in the last two terms represent the contributed intensities of
the lines A and B to the intensity after the $i$-th layer.

\subsection{Application to two blended lines}
We applied the formula to the two lines OI at 8446.359\AA\
(indicated with the letter $a$) 
and FeI at 8446.388\AA\ (indicated with letter $b$) 
and solar atmosphere model. We also need to take
in account the continuum absorption as third absorber. The continuum is
indicated with the letter $c$. For the $i$-th layer the source function is
$$
S_i(\theta)=\frac{2.0\tau^i_{ref}\cdot(\kappa_{\lambda}^i+\kappa_{\nu}^i)}{0.4343
\kappa_{ref}^i}\cdot \mathcal{B}_i \sin(\theta)*d\theta
$$
where $\tau^i_{ref}$ and $\kappa_{ref}$ are the optical depth and opacity at
the reference optical depth, $\kappa_{\lambda}$ and $\kappa_{\nu}$ opacities
of the continuum and the lines, $\mathcal{B}_i$ is the Planck 
function, and $\theta$ the angle of view of
the stellar disc, from the center ($\theta=0$) to the limb ($\theta=\pi/2$).
To follow how the radiation $S_i(\theta)$ gets absorbed by the upper layers, we 
introduce a running index $k$ which runs from $i$ to $1$ and indicates
the layers crossed by the radiation along its path to the surface.
We start from the layer $k=i$. The flux $S_k(\theta)$ generated in the $k$-th 
layer gets extincted in the same layer. The corresponding outcoming flux is
then
$$
I_{k,out}^i(\theta)=S_k(\theta)
\exp{\Big(-\frac{\tau_{\lambda}^k+\tau_{\nu}^k}{\cos(\theta)}\Big)}.
$$
where $I_{k,out}^i$ is the  emerging radiation from the $k$-th layer 
that was generated in the $i$-th layer (in this case the layer is the
same, $k=i$), and $\tau_{\lambda}^k$ and $\tau_{\nu}^k$
are the optical depths of the $k$-th layer due to the continuum and the
two lines, respectively. 
The following layer $k-1$ absorbs the flux $I_{k,out}^i(\theta)$ emerging
from the $k$-th layer,
and the emerging flux from the $k-1$-th layer is
\begin{equation}\label{I_k_abs}
I_{k-1,out}^i(\theta)=I_{k,out}^i(\theta)
\exp{\Big(-\frac{\tau_{\lambda}^{k-1}+\tau_{\nu}^{k-1}}{\cos(\theta)}\Big)}.
\end{equation}
and so on for the following layers. This is a recurrence formula, which
continues until the layer $k=1$ (that is at the
surface) is reached. For this surface layer the emerging radiation is
$$
I_{1,out}^i(\theta)=I_{2,out}^i(\theta)
\exp{\Big(-\frac{\tau_{\lambda}^1+\tau_{\nu}^1}{\cos(\theta)}\Big)}.
$$
It represents the {\em contribution function of the $i$-th layer}.
At every $k$-th layer we can compute the fraction of the radiation absorbed
by the lines $a$ and $b$ by using the equations~(\ref{frac_I_A}),
(\ref{frac_I_B}),
(\ref{frac_A}), and (\ref{frac_B}). Similarly, this can be done for the continuum $c$.
The quantity $I(\tau_{k-1})(1-e^{-\Delta\tau_k})$ in
equations~(\ref{frac_I_A}) and
(\ref{frac_I_B}) indicates the variation of intensity before and after the $k$-th
layer. This can be easily computed from the equation~(\ref{I_k_abs}) by
computing $I^k-I^{k-1}$. Thus, 
the fractions of intensity absorbed by the two lines
$a$, $b$, and the continuum $c$ in the $k$-th layer are
\begin{eqnarray}
\label{line_a_abs}
I^{i,a}_{k,abs}(\theta) & = &
\frac{\kappa^a_k}{\kappa^a_k+\kappa^b_k+\kappa^c_k}\big(I^k(\theta)-I^{k-1}(\theta)\big)\\
\label{line_b_abs}
I^{i,b}_{k,abs}(\theta) & = &
\frac{\kappa^b_k}{\kappa^a_k+\kappa^c_k+\kappa^c_k}\big(I^k(\theta)-I^{k-1}(\theta)\big)\\
\label{line_c_abs}
I^{i,c}_{k,abs}(\theta) & = &
\frac{\kappa^c_k}{\kappa^a_k+\kappa^b_k+\kappa^c_k}\big(I^k(\theta)-I^{k-1}(\theta)\big).\\\nonumber
\end{eqnarray}

By summing over the index $k$ in the equations~(\ref{line_a_abs}),
(\ref{line_b_abs}), and (\ref{line_c_abs}) we obtain the flux absorbed by the lines $a$,
$b$, and $c$. The contribution
function of the $i$-th layer can be therefore written as
\begin{equation}\label{contrib_formula_i}
I^i=S^i-I^{i,a}_{abs}-I^{i,b}_{abs}-I^{i,c}_{abs}
\end{equation}
where
\begin{eqnarray}\nonumber
I^{i,a}_{abs} & = & \sum^{k=i}_{1} I^{i,a}_{k,abs}(\theta)\\\nonumber
I^{i,b}_{abs} & = & \sum^{k=i}_{1} I^{i,b}_{k,abs}(\theta)\\\nonumber
I^{i,c}_{abs} & = & \sum^{k=i}_{1} I^{i,c}_{k,abs}(\theta)\\\nonumber
\end{eqnarray}

\begin{figure}[t]
\centering
\includegraphics[width=9cm,bb=24 24 285 302]{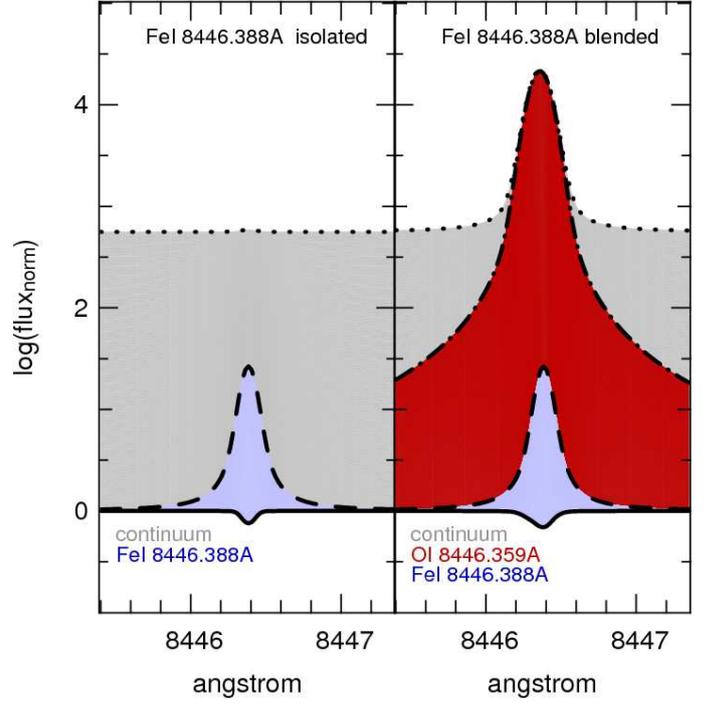}
\caption{{\bf Left panel}: the black solid line represents the \Fei\ line at 8446.338\AA\
synthesized as isolated line, i.e., the emerging flux when the line and
continuum absorptions are accounted for.  The dashed line represent the flux in
presence of the line when the continuum absorption alone is accounted for. 
The dotted line represent the emerging flux when the emissions alone are accounted
for. The gray area represents the flux absorbed 
by the continuum, while the light blue area is the flux absorbed by the
line. {\bf Right panel}: as in the left panel, but for the two blended lines 
OI and FeI lines at 8446.359\AA\ and 8446.338\AA, respectively.
The flux absorbed by the continuum and the Fe line are still 
represented by the gray and light blue areas, respectively, 
while for the OI line the absorbed flux is the red area.
The y-axis represents the logarithm (base 10) of the normalized flux.
The color version of this plot is available in the electronic edition.}
\label{flux_sep_8446.388}
\end{figure}

By summing over the $i$ index in equation~(\ref{contrib_formula_i}) we 
obtain the total flux at angle $\theta$. 
To obtain the observed flux we must integrate from $\theta=0$ to $\theta=\pi/2$. 
The quantities of
the absorbed flux by the lines $a$, $b$, and the continuum $c$ are
illustrated in Fig.~\ref{flux_sep_8446.388}, where the logarithm of the
normalized flux is shown. The gray, red, and light blue areas show the flux absorbed by
the continuum, the OI line, and the FeI line, respectively.\\
This figure shows that by using the outlined way to
compute the flux we can estimate how much flux has been individually 
absorbed by the lines in a blend. It is now clear why this
solution is not helpful in constructing the EW library we need:
while the classical EW refers to the flux
absorbed by a line with respect to the level of the continuum measured in
absence of the line, the absorbed flux obtained in
equation~(\ref{contrib_formula_i}) is absolute and is much larger than the
EW because the continuum level and the emitted intensity change across the line. 

\section{The \Space\ line profile}\label{app_voigt}
The shape of an absorption line is a function of many
variables (such as stellar parameters, atomic
parameters, chemical abundances, radial velocity of the atmosphere layers in
which the line forms) which cannot be analytically expressed
but only numerically computed through spectral synthesis.
However, experience says that spectral lines can be fairly 
well described by a Voigt function, which 
is the convolution of a Lorentzian function with a Gaussian function.  
The Lorentzian FWHM ($\gamma_L$) rules the width of the wings, while the 
Gaussian FWHM ($\gamma_G$) rules the width of the core of the Voigt
function.
Lines with larger damping constants have broader wings than the ones with
smaller damping constants, and the wings' widths also vary with the gravity
\logg\ of the star. Besides, the wings of the line
profile become broader (i.e.,  the ratios $\gamma_L/\gamma_G$ become bigger)
for larger EW\footnote{For growing EW the core of the line grows slower
than the wings, therefore the line become broader.}. 
Thus, the profile of an absorption line can be approximated with 
a Voigt function in which $\gamma_L$ and $\gamma_G$ depend on the instrumental 
FWHM, EW, and \logg.
To allow \Space\ to handle the shape of the absorption
lines, we adopted the Voigt function implementation given in 
McLean et al.  \cite{mclean}, modified to have
the $\gamma_G$ equal to the instrumental FWHM, while $\gamma_L$ follows the
empirical law\\
$$
\gamma_L=dl\cdot EW\cdot dp\cdot \Big(1.-\exp\Big[-\big(\frac{EW\cdot
dp}{\sigma}\big)^2\Big]\Big)
$$
where EW is the equivalent width of the line, $dp$ is the
``broadening" parameter (with $dp>0$), and $\sigma$ and $dl$ are functions 
of \logg\ as described as follows\\
$$
\left\{
\begin{array}{ll}
\sigma=0.14, dl=0.8 & \textrm{if \logg$>$4.5}\\
\sigma=0.16, dl=(0.7+(\log g-3.5)\cdot0.1) & \textrm{if 3.5$<$\logg$\leq$4.5}\\
\sigma=0.20, dl=(0.6+(\log g-2.5)\cdot0.1) & \textrm{if 2.5$<$\logg$\leq$3.5}\\
\sigma=0.20, dl=0.6 & \textrm{if 1.5$<$\logg$\leq$2.5}\\
\sigma=0.20, dl=(0.6+(1.5-\log g)\cdot0.1) & \textrm{if \logg$<$1.5}\\
\end{array} \right.
$$

The $dp$ parameter has been empirically derived by hand for each line 
with a simple manual "trial-and-error" method\footnote{Out of
thousands of lines of the \Space\ line list, the lines that
need $d\ne1$ are of the order of one hundred. 
All the other lines have $dp=1$. 
This allows us to find the $dp$ parameter manually.}:
the $dp$ parameter is varied until the matches 
between the line profile in the spectra models constructed by
\Space\ and the line profile in five synthetic spectra of 
different stellar parameters was satisfactory.\\
This empirical implementation of the line profile is satisfactory in most of
the stellar parameters here considered. However, progressive deviation from 
the correct line profile is observed at \temp$<5000$K,
which causes systematic errors
(this is shown in tests with synthetic and real spectra in
Sec.~\ref{sec_validation}). 
The mismatch between the \Space\ and the synthetic/real line profiles can be
reduced with a new improved line profile law that will be implemented in one
of the future releases of \Space.

We here want to shortly discuss how the 
adopted line profile affects the performance of \Space\ as a function of the
spectral resolution. At low resolution the observed line profile is
dominated by the instrumental profile, which is symmetric and constant for
all the absorption lines. This characteristic makes the observed line profile
easy to model. The higher the resolution, the more the physical
profile of the lines dominates over the instrumental profile, making the
observed profile difficult to model because, depending on the absorption line
considered, the line can be asymmetric or deviate from the adopted Voigt function.
Therefore, the line profile adopted by us matches at best the observed line
profile at low resolution (where the instrumental profile dominates) and
progressively loses accuracy at higher resolution. We verified that the
accuracy of \Space\ in reproducing the observed line profile is satisfactory
up to R$\sim$20\,000. For higher resolutions the \Space\ line profile may still be
satisfactory for most of the absorption lines, but the robustness of the
results has not been proved with extensive tests yet.

\section{Tests on real spectra}\label{appendix_tests_real}
We report here the tables and the figures of the stellar parameters and chemical abundances
derived with \Space\ of the stellar spectra ELODIE, benchmark, and S4N
degraded to R=12\,000 and S/N=100.

\begin{table*}[t]
\caption[]{Stellar parameters and chemical abundances obtained with \Space\
from the ELODIE stars. Here we report part of the data for two stars. The full
table is only available in electronic form
at the CDS via anonymous ftp to cdsarc.u-strasbg.fr (130.79.128.5)
or via http://cdsweb.u-strasbg.fr/cgi-bin/qcat?J/A+A/.}
\label{ELODIE_table}
\vskip 0.3cm
\centering
\fontsize{6}{6}\selectfont
\begin{tabular}{l|cccccccccccccccc}
\hline
\noalign{\smallskip}
star & \temp & \temp$_{inf}$ & \temp$_{sup}$ & \logg & \logg$_{inf}$& \logg$_{sup}$ & 
[M/H] & [M/H]$_{inf}$& [M/H]$_{sup}$ & 
[Fe/H] & [Fe/H]$_{inf}$& [Fe/H]$_{sup}$ & 
[Mg/H] & [Mg/H]$_{inf}$& [Mg/H]$_{sup}$ & 
... \\
\hline
\noalign{\smallskip}
HD000245 & 5842  & 5663& 5850 & 4.16 & 3.95 & 4.19 & -0.55 &-0.60 &-0.54 & -0.59
& -0.63 & -0.58 & -0.37 & -0.43 & -0.31 & ...\\
HD000358 & 6182 & 6123 & 6215 & 4.03 & 3.97 & 4.11 & -0.30 &-0.33 &-0.28 & -0.29
& -0.31 & -0.28 & -0.32 & -0.39 & -0.26 & ...\\
\noalign{\smallskip}
\hline
\end{tabular}
\normalfont
\end{table*}

\begin{table*}[t]
\caption[]{Stellar parameters and chemical abundances obtained with \Space\
from the benchmark stars. Here we report part of the data for two stars. 
The full table is only available in electronic form
at the CDS via anonymous ftp to cdsarc.u-strasbg.fr (130.79.128.5)
or via http://cdsweb.u-strasbg.fr/cgi-bin/qcat?J/A+A/.}
\label{benchmark_table}
\vskip 0.3cm
\centering
\fontsize{6}{6}\selectfont
\begin{tabular}{l|cccccccccccccccc}
\hline
\noalign{\smallskip}
star & \temp & \temp$_{inf}$ & \temp$_{sup}$ & \logg & \logg$_{inf}$& \logg$_{sup}$ & 
[M/H] & [M/H]$_{inf}$& [M/H]$_{sup}$ & 
[Fe/H] & [Fe/H]$_{inf}$& [Fe/H]$_{sup}$ & 
[Mg/H] & [Mg/H]$_{inf}$& [Mg/H]$_{sup}$ & 
... \\
\hline
\noalign{\smallskip}
HD 49933 & 6570 & 6434 & 6582 &  4.03 & 3.87 & 4.10 & -0.49 & -0.53 & -0.48 &
-0.50 & -0.53 & -0.49 & -0.42 & -0.51 & -0.36 & ...\\
$\xi$ Hya & 5063 & 5014 & 5072 &  3.05 & 3.01 & 3.11 &  0.09 & 0.04 & 0.10 & 0.09 & 0.08
& 0.11 & 0.12 & 0.04 & 0.18 & ...\\
\noalign{\smallskip}
\hline
\end{tabular}
\normalfont
\end{table*}

\begin{table*}[t]
\caption[]{Stellar parameters and chemical abundances obtained with \Space\
from the S4N stars. Here we report part of the data for two stars.
The full table is only available in electronic form
at the CDS via anonymous ftp to cdsarc.u-strasbg.fr (130.79.128.5)
or via http://cdsweb.u-strasbg.fr/cgi-bin/qcat?J/A+A/.}
\label{S4N_table}
\vskip 0.3cm
\centering
\fontsize{6}{6}\selectfont
\begin{tabular}{l|cccccccccccccccc}
\hline
\noalign{\smallskip}
star & \temp & \temp$_{inf}$ & \temp$_{sup}$ & \logg & \logg$_{inf}$& \logg$_{sup}$ & 
[M/H] & [M/H]$_{inf}$& [M/H]$_{sup}$ & 
[Fe/H] & [Fe/H]$_{inf}$& [Fe/H]$_{sup}$ & 
[Mg/H] & [Mg/H]$_{inf}$& [Mg/H]$_{sup}$ & 
... \\
\hline
\noalign{\smallskip}
Sun & 5685 & 5649 & 5705 &  4.28 & 4.25 & 4.31 & 0.04 & 0.02 & 0.05 & 0.04
& 0.03 & 0.06 & -0.04 & -0.8 & 0.00 & ...\\
HIP 171 & 5330 & 5246 & 5354 &  4.08&  3.96 & 4.09 & -0.82 & -0.88 & -0.80 &
-0.82 & -0.85 & -0.80 & -0.51 & -0.56 & -0.43 & ...\\
\noalign{\smallskip}
\hline
\end{tabular}
\normalfont
\end{table*}

\clearpage

\begin{figure*}[t]
\centering
\includegraphics[width=8cm,bb=95 287 363 507]{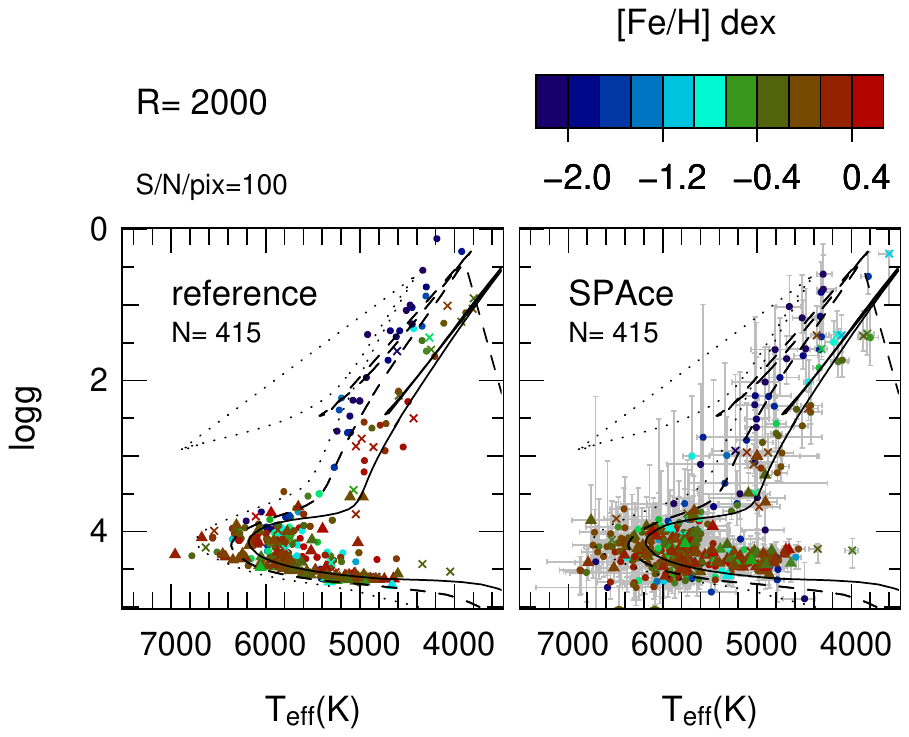}
\vfill
{\includegraphics[width=14cm,bb=76 281 532 530]{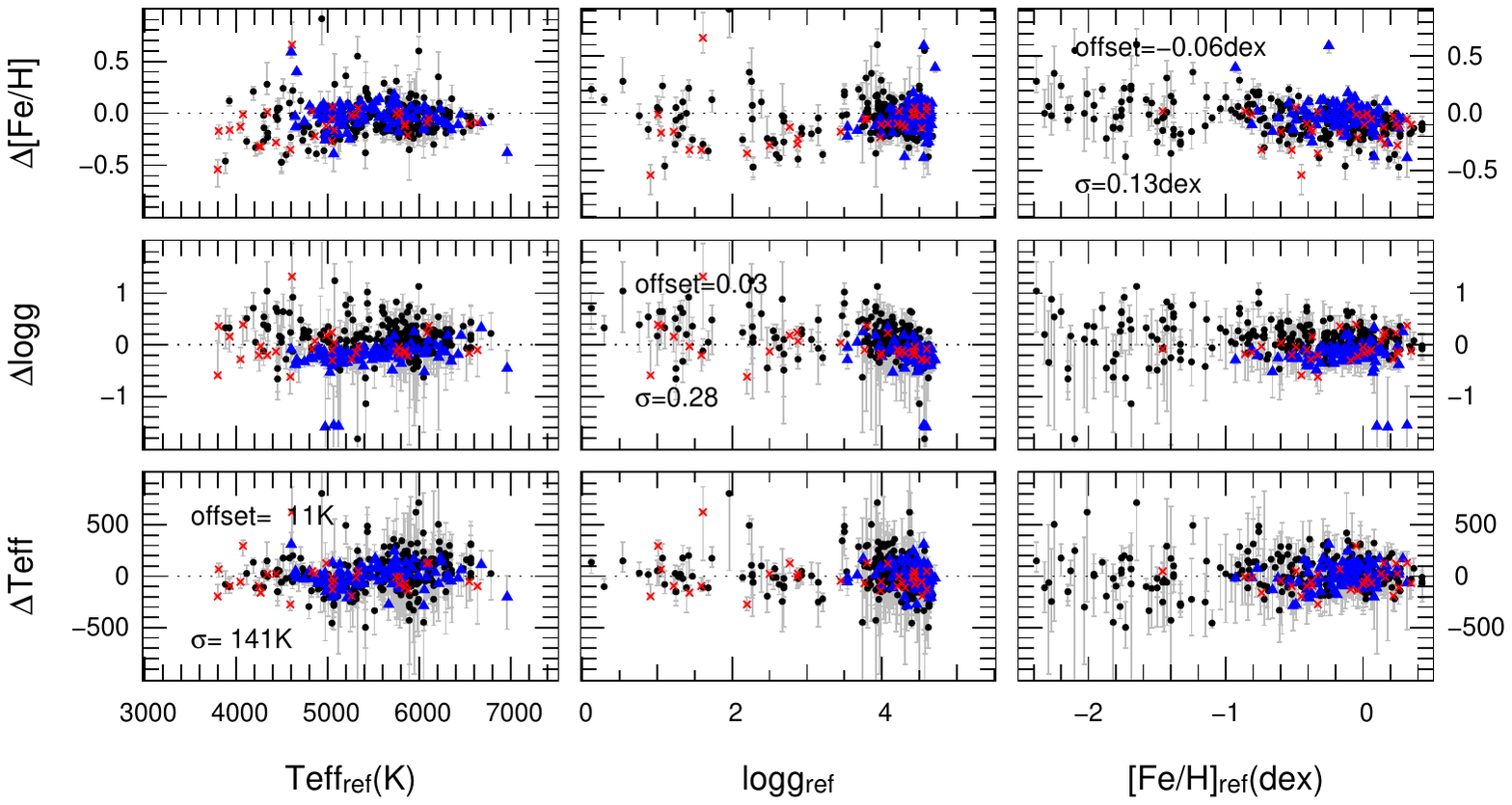}}
\vfill
{\includegraphics[width=14cm,bb=82 281 532 530]{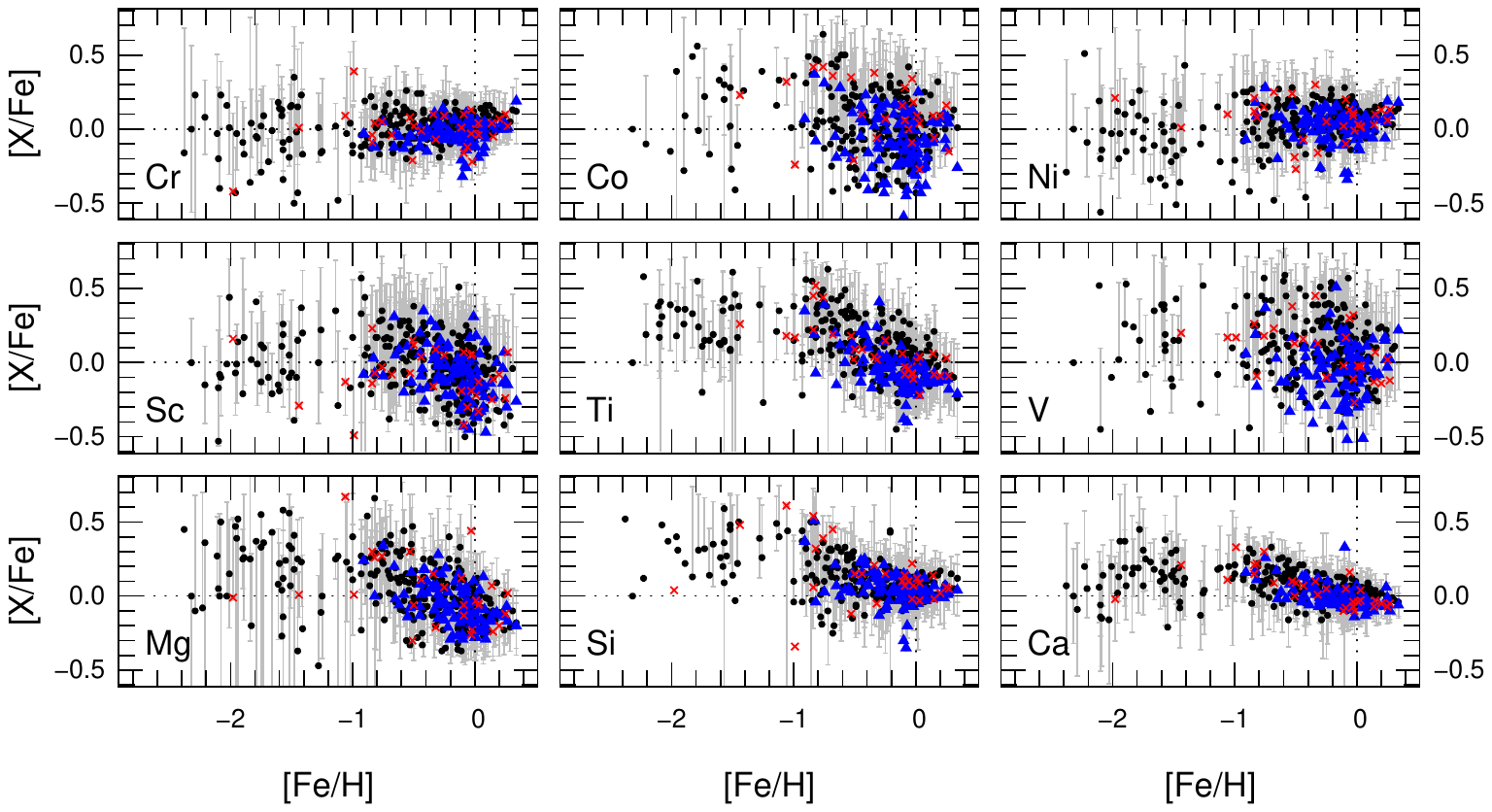}}
\caption{{\bf Top}: distribution of the reference \temp\ and \logg\ 
of real spectra (top-left panel) and the ones derived 
by \Space\ (top-right panel). 
{\bf Middle}: Residuals between derived and reference parameters (y-axis) as a
function of the reference parameters (x-axis). {\bf Bottom}: Chemical abundances
derived by \Space\ for the same spectra. These spectra have a resolution of
R=2000 and S/N/pixel=100. The symbols are as in
Fig.~\ref{correl_R12_SN100_real_ABDloop}. The color version of this plot is
available in the electronic edition.}
\label{R02_SN100_real_noABDloop}
\end{figure*}
\clearpage

\begin{figure*}[t]
\centering
\includegraphics[width=8cm,bb=95 287 363 507]{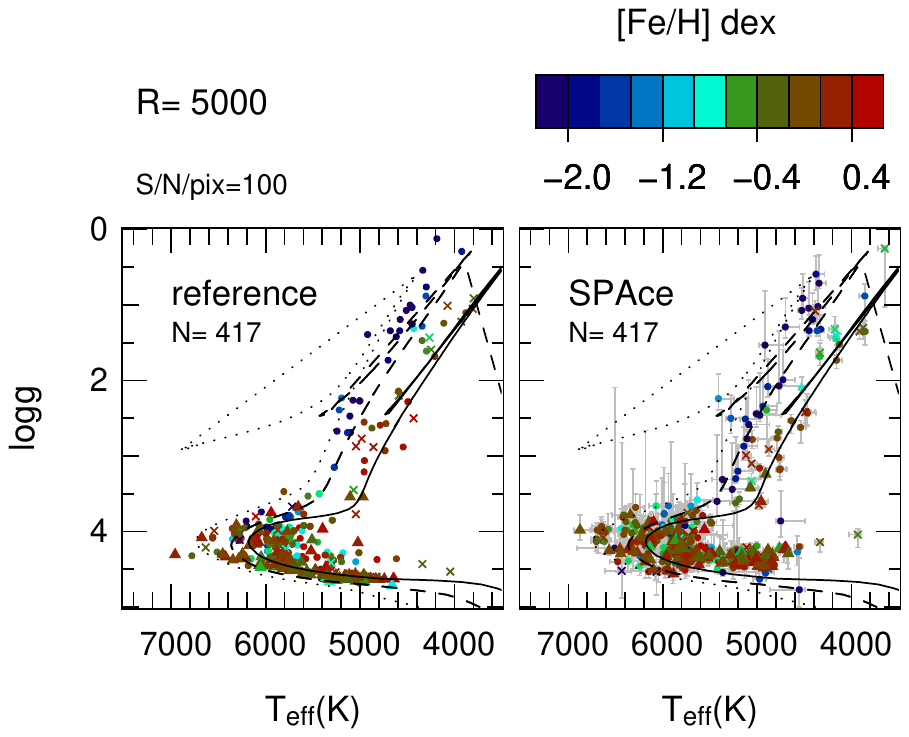}
\vfill
{\includegraphics[width=14cm,bb=76 281 532 530]{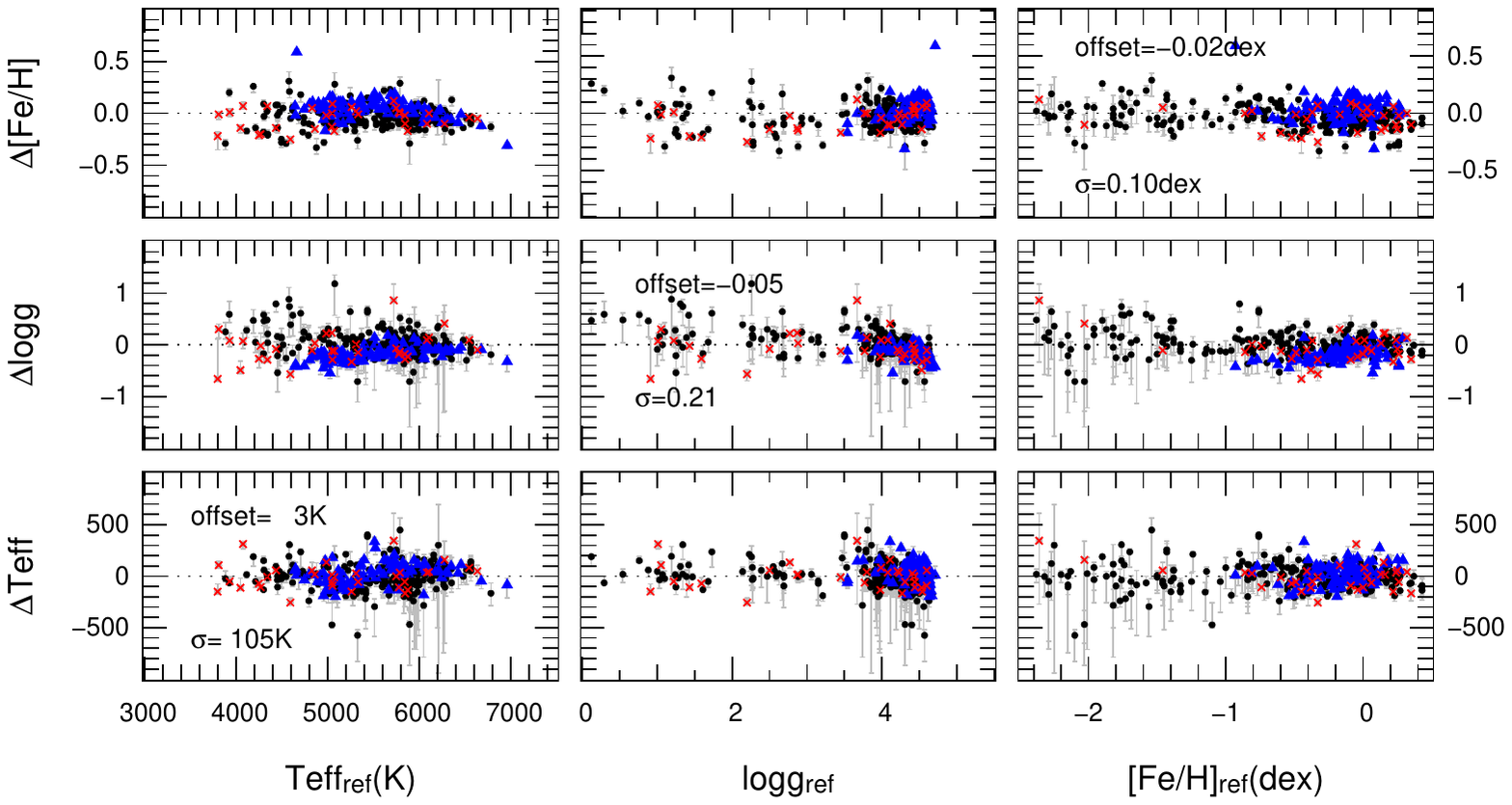}}
\vfill
{\includegraphics[width=14cm,bb=82 281 532 530]{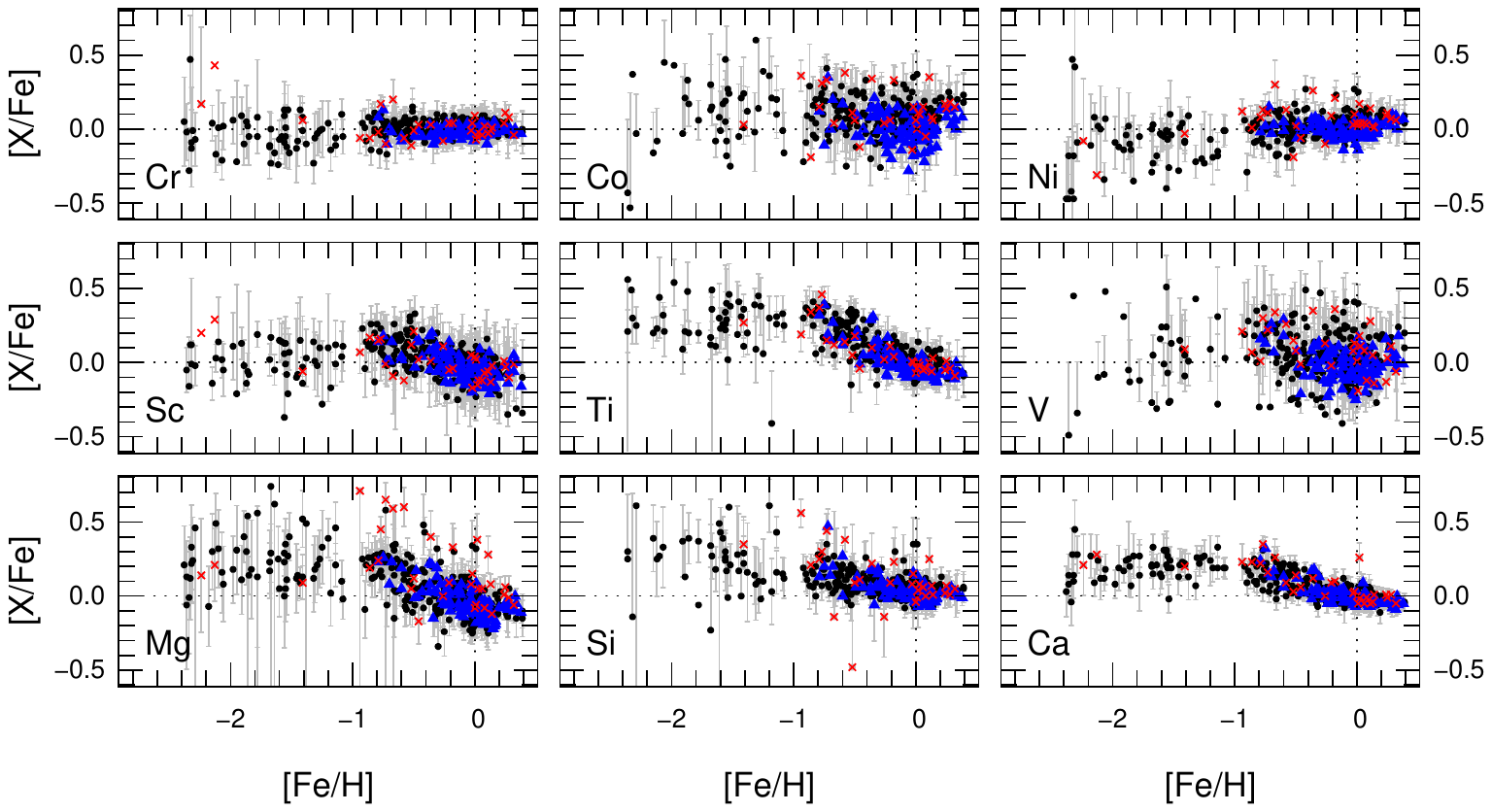}}
\caption{As in Fig.~\ref{R02_SN100_real_noABDloop} but for R=5\,000 and S/N/pixel=100.
The color version of this plot is available in the electronic edition.}
\label{R05_SN100_real_noABDloop}
\end{figure*}
\clearpage

\begin{figure*}[t]
\centering
\includegraphics[width=8cm,bb=95 287 363 507]{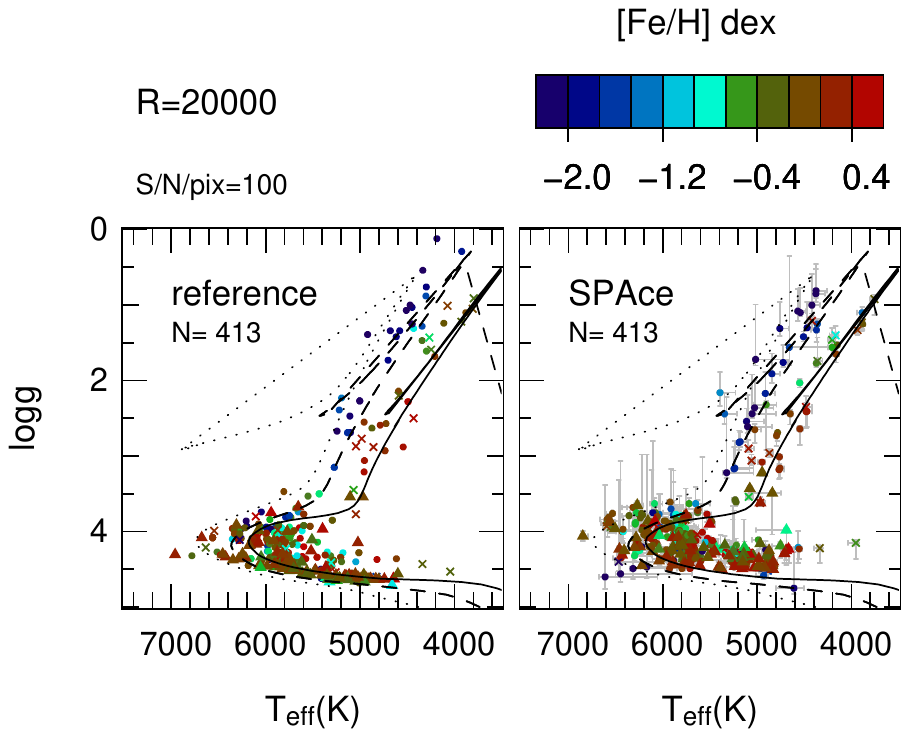}
\vfill
{\includegraphics[width=14cm,bb=76 281 532 530]{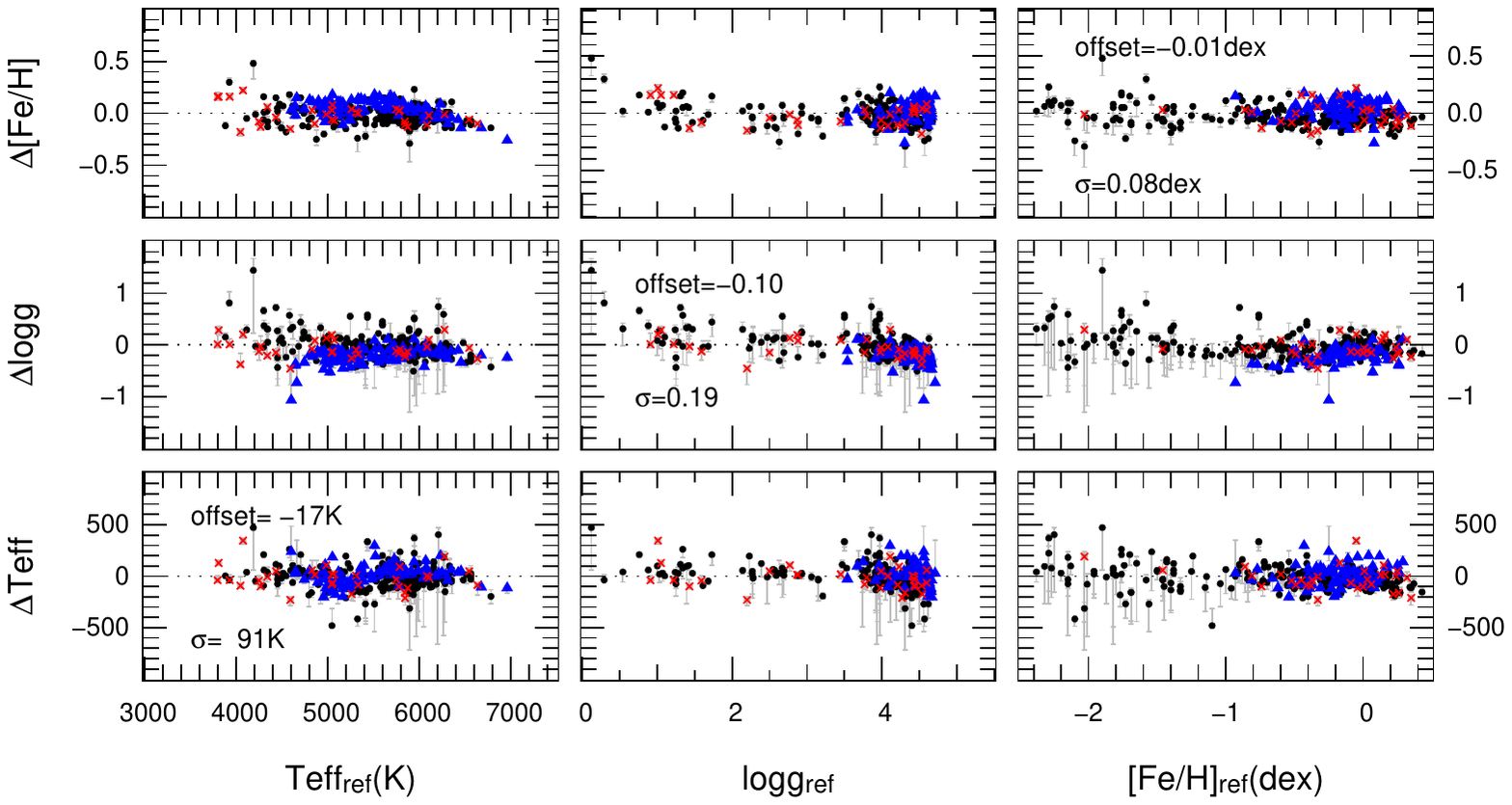}}
\vfill
{\includegraphics[width=14cm,bb=82 281 532 530]{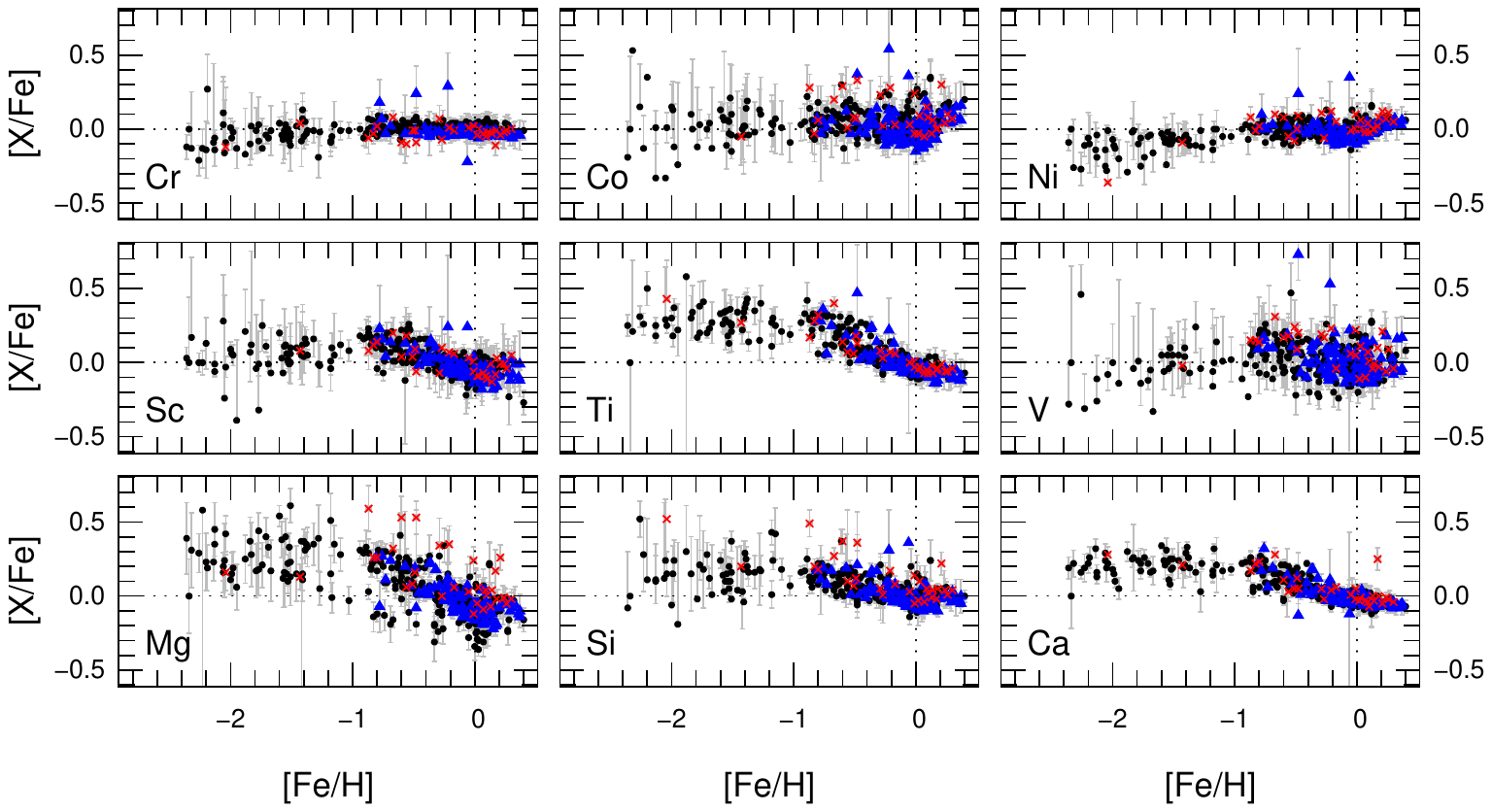}}
\caption{As in Fig.~\ref{R02_SN100_real_noABDloop} but for R=20\,000 and S/N/pixel=100.
The color version of this plot is available in the electronic edition.}
\label{R20_SN100_real_noABDloop}
\end{figure*}

\clearpage

\begin{figure*}[t]
\centering
\includegraphics[width=8cm,bb=20 22 290 240]{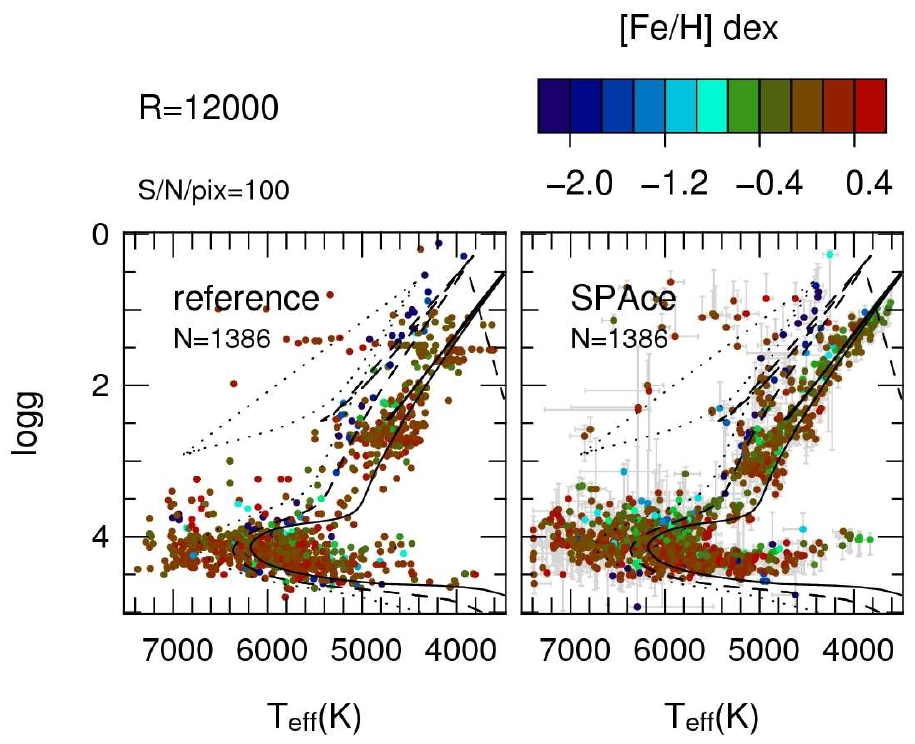}
\vfill
{\includegraphics[width=14cm,bb=15 25 473 270]{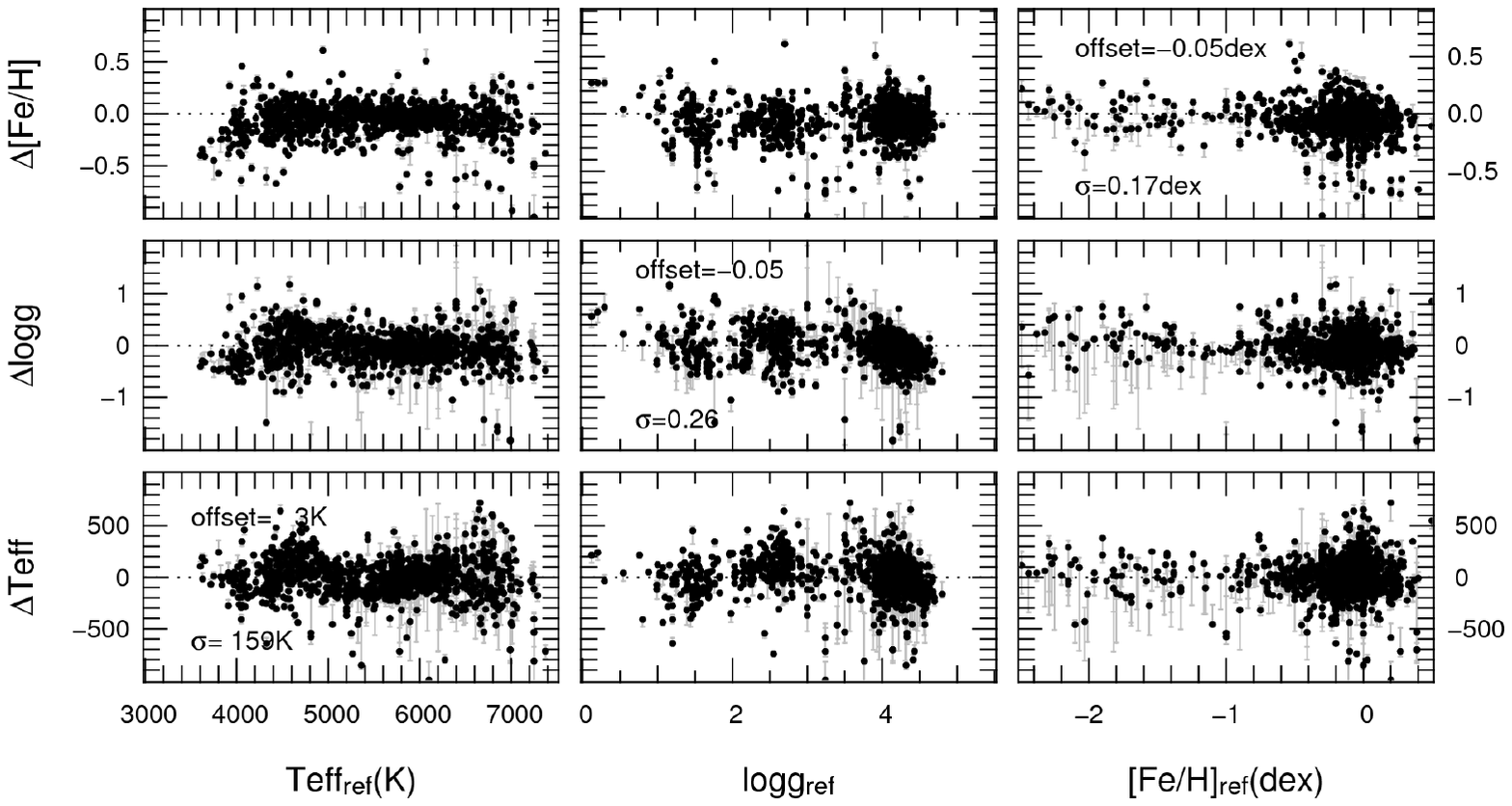}}
\vfill
{\includegraphics[width=14cm,bb=15 25 473 270]{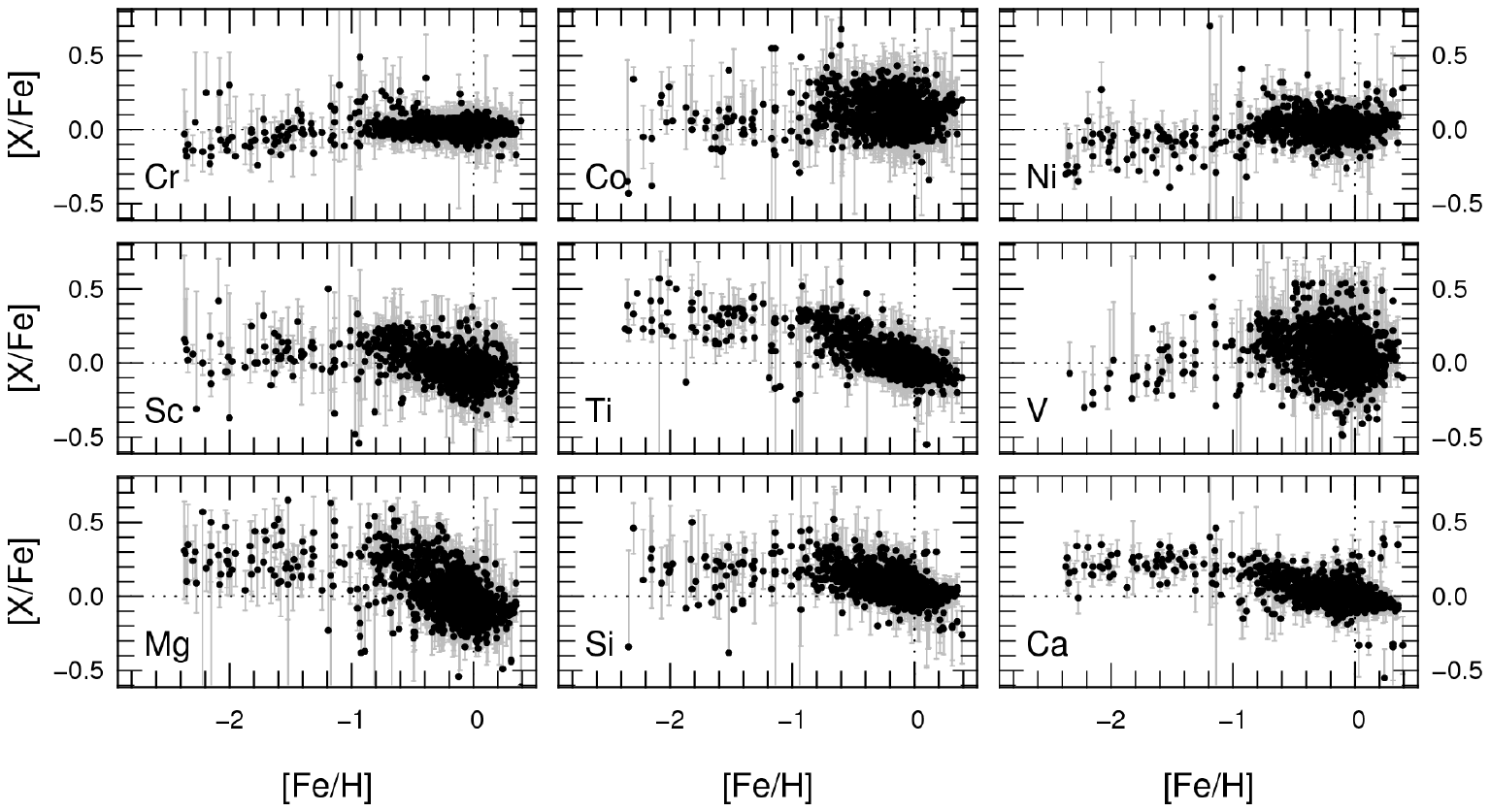}}
\caption{As in Fig.~\ref{R02_SN100_real_noABDloop} but for 1386 ELODIE
spectra (out of the total 1959 spectra of the whole ELODIE database) measured at R=12\,000 and S/N/pixel=100.
In this case all reference parameters are considered without quality
criteria selection. The color version of this plot is available in the
electronic edition.}
\label{R12_SN100_elodie_whole_noABD}
\end{figure*}
\clearpage


\end{document}